\documentclass[%
 reprint,
 amsmath,amssymb,
 aps,
rmp,
]{revtex4-2}

\usepackage{graphicx}
\usepackage{dcolumn}
\usepackage{bm}
\usepackage[colorlinks=true]{hyperref}
\usepackage[version=4]{mhchem} 

\usepackage[usenames,dvipsnames]{xcolor}
\usepackage{amsmath, amssymb}

\usepackage{lipsum}
\usepackage{longtable}

\begin{document}


\title{Kagome metals}





\author{Domenico Di Sante}\email{domenico.disante@unibo.it}
\affiliation{Department of Physics and Astronomy, University of Bologna, Bologna, Italy}

\author{Titus Neupert}\email{titus.neupert@physik.uzh.ch}
\affiliation{Physik-Institut, Universit\"{a}t Z\"{u}rich, Z\"{u}rich, Switzerland}

\author{Giorgio Sangiovanni}\email{giorgio.sangiovanni@uni-wuerzburg.de}
\affiliation{Institut f\"{u}r Theoretische Physik und Astrophysik \\
and W\"{u}rzburg-Dresden Cluster of Excellence ct.qmat, Universit\"{a}t W\"{u}rzburg, W\"{u}rzburg, Germany}

\author{Ronny Thomale}\email{rthomale@physik.uni-wuerzburg.de}
\affiliation{Institut f\"{u}r Theoretische Physik und Astrophysik \\
and W\"{u}rzburg-Dresden Cluster of Excellence ct.qmat, Universit\"{a}t W\"{u}rzburg, W\"{u}rzburg, Germany}

\author{Riccardo Comin}\email{rcomin@mit.edu}
\affiliation{Department of Physics, Massachusetts Institute of Technology, Cambridge, Massachusetts, USA}

\author{Ilija Zeljkovic}\email{ilija.zeljkovic@bc.edu}
\affiliation{Department of Physics, Boston College, Chestnut Hill, Massachusetts, USA}

\author{Joseph G. Checkelsky}\email{checkelsky@mit.edu}
\affiliation{Department of Physics, Massachusetts Institute of Technology, Cambridge, Massachusetts, USA}

\author{Stephen D. Wilson}\email{stephendwilson@ucsb.edu}
\affiliation{Materials Department, University of California Santa Barbara, Santa Barbara, California, USA}

\date{\today}

\begin{abstract}
Three important driving forces for creating qualitatively new phases in quantum materials are the topology of the materials’ electronic band structures, frustration in the electrons’ motion or magnetic interactions, and strong correlations between their charge, spin, and orbital degrees of freedom. In very few material systems do all of these aspects come together to contribute on an equal footing to stabilize new electronic states with unprecedented properties; however the search for such systems can be guided by models of configurational motifs or key sublattices that can host such physics. One of the most fascinating structural motifs for realizing this rich interplay of frustration, electronic topology, and electron correlation effects is the kagome lattice.  In this review, we provide an overview of the theoretical underpinnings driving the  physics of kagome lattices, and we then discuss experimental progress in realizing novel states enabled by kagome networks in crystalline materials.  Different material classes are discussed with an emphasis on the phenomenologies of their electronic states and how they map to interactions arising from their kagome lattices. 
\end{abstract}

\maketitle

\tableofcontents

\section{\label{sec:Intro}Introduction}

The kagome lattice is recognized as being composed of the iconic elementary motive of frustration, the triangle. This insight was translated into the physics of crystalline materials in two waves:   
 The first focus was on magnetic insulating compounds, such as Herbertsmithite, derived from strongly correlated oxides, and exhibit local magnetic $S = 1/2$ moments of Cu kagome nets~\cite{mendels2010quantum,normanRMP}. As the Coulomb interactions emanating from the Cu $d$-orbitals are strong, these systems are placed deep in the Mott limit where the kinetic electronic degrees of freedom are frozen out. Herbertsmithite, and polymorphs thereof, thus provided an excellent setting for geometrically frustrated quantum magnetism and topologically ordered spin liquids~\cite{lacroix2011introduction,iqbal2013gapless,bauer2014chiral,he2017signatures}.

Second came a focus on kagome \emph{metals}. Preceded by binary kagome metals tailored toward supporting Dirac and flat band electronic states~\cite{ye_massive_2018}, the field of many-body electronic order in kagome metals was ignited by the synthesis of layered AV$_3$Sb$_5$ (A = K, Cs, Rb) crystals~\cite{ortiz2019new} or the ``135" family of kagome compounds. The V$^{3+}$ constituents in this family form a kagome lattice, \textit{i.e.} a non-Bravais lattice of corner-sharing triangles.
Kagome metals harboring an unprecedentedly rich phenomenology offering itinerant magnetism, charge order, electronic topology, and superconductivity (SC) in a variety of compounds, has initiated a new era of kagome quantum materials~\cite{Neupert-review-2022}. Theoretical studies proposed the emergence of $\mathbb{Z}_{2}$ topological insulators via spin-orbit coupling (SOC) \cite{guo2009topological} and quantum anomalous Hall insulators through magnetism and SOC \cite{xu_intrinsic_2015,guterding2016prospect}, highlighting the potential for the kagome lattice to host phenomena at the nexus of topology and magnetism. The first experimental confirmation of topological (massive) Dirac fermions in a kagome material was achieved in ferromagnetic binary kagome compound Fe$_3$Sn$_2$ \cite{ye_massive_2018, yin_giant_2018}, soon followed by the direct observation of the characteristic kagome flat band in the same material \cite{zhiyong_lin_flatbands_2018}. 

\begin{figure}[!b]
\centering
\includegraphics[width=\columnwidth,angle=0,clip=true]{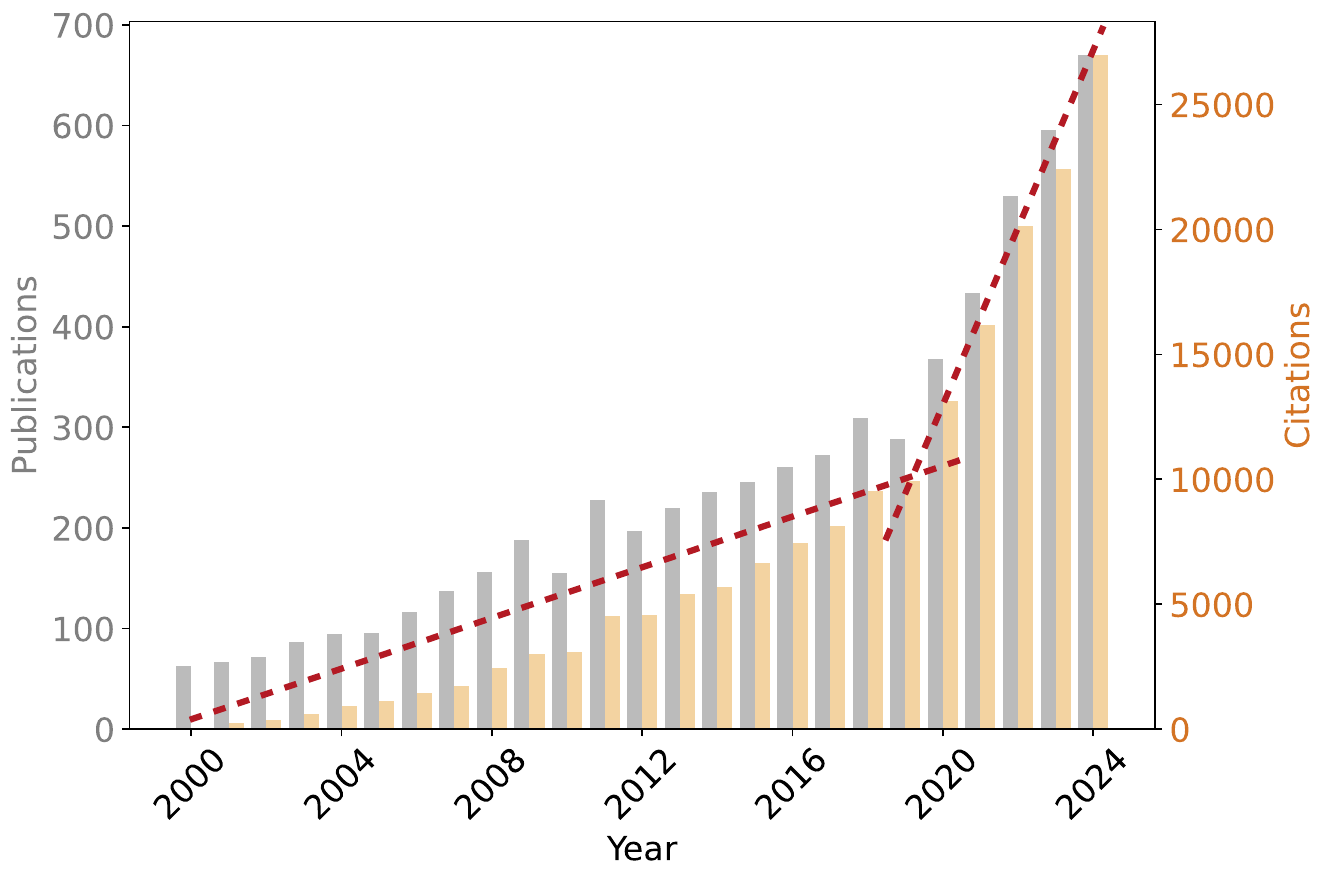}
\caption{Number of publications per year, and relative citations, from 2000 to 2024 in which the word \textit{kagome} is mentioned. Data have been retrieved from the Web of Science platform, searching within all fields. The red dashed lines are guide to the eye to emphasize the surge of interest by the scientific community around 2019 that we ascribe to the experimental discovery of binary and ternary kagome metals~\cite{ye_massive_2018, ortiz2019new}.}
\label{citations}
\end{figure}

Explicitly topological flat bands, possessing a $\mathbb{Z}_{2}$ invariant due to SOC, were experimentally verified in the kagome metal CoSn \cite{kang_topological_2020}. The first explorations of the 135 family of ternary kagome compounds AV$_3$Sb$_5$ led to the discovery of a high-temperature charge ordering (CO) transition and SC at ambient pressure~\cite{PhysRevMaterials.5.034801,chen2021roton,PhysRevLett.125.247002,kang_charge_2023}. The evidence for high-temperature electronic charge order alone, however, which has been previously witnessed in other materials such as NbSe$_2$~\cite{ugeda2016characterization}, would not provide a sufficient basis for the tremendous excitement surrounding kagome metals, and kagome materials in general. Rather, the excitement, as evident from the growing trend of publications and citations in Fig.~\ref{citations} in the last few years, is fueled by various aspects.

To begin with, the unconventional nature of charge order with respect to time reversal symmetry (TRS) breaking or nematicity, and the possible appearance of pair density wave (PDW) SC, provides a strong hint that electronic correlations are vital to the formation of quantum many-body phases in kagome metals, which are further enhanced through the presence of van Hove singularities (VHS) nearby the Fermi level~\cite{jiang_unconventional_2021,PhysRevLett.127.046401,mielke2022time,PhysRevLett.126.247001,Guo2022chiralTransport,PhysRevX.13.031030}. The possibility of orbital currents further suggests a largely unexplored facet of correlated electron systems, \textit{i.e.}, a TRS breaking mechanism not descending from local spins, but orbital moments. Surprisingly, within a relatively short time span, an astounding proliferation of new discoveries and kagome material classes suggests that kagome matter is a uniquely diverse host for exotic electronic order. As an added example, FeGe features an itinerant magnetic phase at high-temperatures before it develops an onset CDW, as well as the possibility of a spin density wave (SDW), yielding a fascinating intertwining of charge and spin fluctuations to be explored further~\cite{teng_discovery_2022,xiaokun_teng_magnetism_2022,yin_discovery_2022,lebing_chen_competing_2023,mason_klemm_vacancy-induced_2024,oh2025disentangling}.

Parallel to this development on symmetry-broken phases has been the exploration of the prototypical non-interacting kagome band structure features and their fate in real materials.  A central line of this research is the flat electronic band, long-expected from destructive interference on the hexagonal plaquette of the kagome network~\cite{bergman2008band}.  Previously explored in the context of metalizing insulating kagome materials~\cite{mazin2014theoretical, kelly2016electron}, significant progress has been made recently in kagome metals including spectroscopic identification of these flat bands and observation of interaction effects arising therefrom~\cite{kang_dirac_2020, kang_topological_2020, liu_orbital-selective_2020, gim2023fingerprints, ye_hopping_2024,xieCr135}.

To date, the field keeps expanding into novel kagome materials classes with entirely different ordering phenomena such as the 134 family (Eu,Yb)V$_3$Sb$_4$~\cite{ortiz2023ybv}, the 166 family like ScV$_6$Sn$_6$~\cite{pokharel_electronic_2021,hu_tunable_2022}, and new materials classes with varied kagome-site character such as Ti-based and Cr-based kagome metals~\cite{Liu2023CTB135_ARPES,liuCr135}. Additionally, with the advent of twisted hexagonal moir\'{e} systems~\cite{cao2018unconventional,andrei2020graphene,devakul2021magic,andrei2021marvels,kennes2021moire,bernevig2024twisted} and advances in quantum material heterostructure modeling by design, the possibility of emergent kagome lattices is starting to arise~\cite{claassen2022ultra}. While this branch of kagome matter is still at its very early stage, the prospects of such pursuits are promising for reaching the strong coupling regime of kagome electron systems where topological order and itinerant quasiparticle excitations face off. This would allow access to uncharted territory of strongly correlated electron systems.

A key motivation behind a recent surge of interest in twisted hexagonal moiré systems and heterostructure modeling is the pursuit of truly two-dimensional kagome lattices. Within many bulk kagome materials realized to date, out-of-plane electron hopping can disrupt the flat bands and electron localization essential for many-body correlated and topological phenomena. To address this challenge, several complementary approaches have been actively explored. First, van der Waals kagome metals provide an opportunity to realize atomically thin layers preserving the intrinsic kagome geometry~\cite{mantravadi2024van,fang2023polarization,park2020kagome,yun2025flat,sabin_regmi_observation_2023,yan2022evolution}. Second, thin-film growth and surface-engineering techniques enable the fabrication of ultrathin or artificially constructed kagome lattices, allowing precise control over dimensionality and electronic structure~\cite{shuyu_cheng_epitaxial_2023,shuyu_cheng_atomic_2022,ren2024persistent,mihalyuk20222d,kim2023realization,lee2024atomically,vekovshinin2024lifshitz}. Together, these strategies aim to create kagome systems where low dimensionality enhances electron correlation effects and topological features, providing a promising avenue for future research in the field.

As much as the experimental side of the field has flourished with new discoveries, kagome metals have also pushed the frontiers of theoretical understanding. In particular, the microscopic description and conceptualization of kagome metals posed a new level of intricacy in correlated electron systems due to three main reasons. 
\begin{enumerate}
    \item The kagome lattice exhibits three sublattices, and combined with the relevance of multiple orbitals in all kagome metals known to date, enforces a complicated kinetic electronic model even in its most reduced effective form~\cite{kiesel_sublattice_2012}. This presents a fundamental challenge to a large fraction of contemporary many-body methods beyond mean field, which are typically constrained to simpler kinetic models.
    \item Coulomb interactions have likewise been unveiled as a necessary ingredient already in the V-based kagome metals dominated by the vanadium $d$-orbitals~\cite{diSantePRR2023}, which is likely to be even strengthened in future Cr-~\cite{giorgioNV} or even Cu-based~\cite{anja_wenger_theory_2024} kagome metals with an increasingly filled $d$-shell valence. It implies that a many-body electronic ansatz for kagome metals will be largely indispensable.
    \item Phonons have also been shown to be a necessary component of any microscopic kagome modeling, as suggested by large electron-phonon coupling and charge orderings commensurate with dominant phonon modes~\cite{tuniz2023dynamics,tuniz2024strain,Hu2023optical_spec_Sc166}. It suggests that any all-electronic modeling of kagome matter will miss out on essential aspects and that phonons have to be considered at all levels of effective kagome metal modeling.
\end{enumerate}

As a consequence, the actual theoretical perspective on kagome metals turned out to be highly fragmented, as most approaches cannot treat the impact of all relevant degrees of freedom on equal footing. As experimental evidence suggests, however, this is needed to account for the astounding richness of electronic orders in kagome matter. The same holds for the typical experimental observables provided by bulk measurements, surface probes, transport signatures, and spectroscopic evidence, where a quantitative correspondence is only enabled by starting from a complex material modeling.

This has put kagome metals at a new level of complexity as compared to previous paradigms of correlated electron systems such as high-temperature cuprates. There, as intricate as the nature of the cuprate phase diagram may be and as many questions it may pose to this date, the single band Hubbard model provides an extremely powerful effective description. Such a simple yet universal effective model does not exist for kagome materials. Instead, understanding kagome matter at a deeper level has necessitated methods and models with a more inclusive ansatz with respect to different low-energy degrees of freedom.

Within such a diversified and complex landscape of theoretical and experimental efforts, the goal of this review is to provide a comprehensive overview of kagome metals, beginning with an analysis of their single-particle electronic and phonon properties in Sec.~\ref{sec:chapII}, including the electronic structure of spinless electron models in~\ref{sec:chapIIA}, spin-orbit interactions in~\ref{sec:chapIIB}, and lattice vibrations in~\ref{sec:chapIIC}. We then explore in Sec.~\ref{sec:chapIII} many-body phenomena such as CDWs in~\ref{sec:chapIIIA}, SC in~\ref{sec:chapIIIB}, PDWs in~\ref{sec:chapIIIC}, electron-phonon interactions in~\ref{sec:chapIIID}, and the role of correlations and fluctuating local moments in~\ref{sec:chapIIIE}. A detailed discussion follows on the various materials classes in Sec.~\ref{sec:chapIV}, encompassing binary and ternary kagome compounds in~\ref{sec:chapIVA} and~\ref{sec:chapIVB} respectively, with representative examples across different elemental compositions. Finally in Sec.~\ref{sec:Conc}, we summarize key insights and outline future research directions in the study of kagome metals.

\section{\label{sec:chapII}Single-particle electronic and phonon properties}

The kagome lattice geometrically consists of a two-dimensional network of corner-sharing triangles, as shown in Fig.~\ref{figure1}(a). The unit cell includes three sublattice sites, conventionally dubbed as the A, B and C sites. These sites result in a high degree of geometrical frustration for the hopping of electrons between the different atoms. In the following, in Sections~\ref{sec:chapIIA} and \ref{sec:chapIIB}, we discuss the properties of several single-particle models on the kagome lattice, without accounting for any many-body contribution that could potentially arise from electron-electron and electron-phonon interactions. These will be extensively reviewed in Section~\ref{sec:chapIII}.

\subsection{\label{sec:chapIIA}Electronic structure of spinless electrons}

At the single-particle level, the first important differentiation is the number of orbital degrees of freedom per site. This influences the size of the associated Hilbert space, and the number of electronic states. We start here in ~\ref{sec:chapIIA1} by revising the famous and widely used model of a single orbital per site, and more specifically, simple isotropic $s$-orbital-like electrons, and we later in ~\ref{sec:chapIIA2} and ~\ref{sec:chapIIA3} extend it by the inclusion of two or more orbitals. This extension and increase in the complexity of the model turns out to be crucial for its expressivity in relation to real-world kagome materials.

\subsubsection{\label{sec:chapIIA1}One orbital per site}

The simplest kagome lattice model is constructed by assuming a single orbital per lattice site, most commonly of 
$s$-like character (such as $s$, $p_z$, or $d_{z^2}$) because of their trivial transformation under the action of point group symmetry operations. This simplification, long adopted in theoretical studies, captures the essential lattice connectivity responsible for the characteristic Kagome band structure. In real materials, however, the spatial orientation and nature of the relevant orbitals with respect to the underlying lattice play a crucial role in realizing the Kagome electronic bands, and we return to this point in later sections.

The starting point is then the tight-binding (TB) model of spinless $s$-orbital electrons hopping on the lattice, given by the single-particle Hamiltonian
\begin{equation}
\label{eq:H0}
    H_0 = -t \sum_{\langle (i,s) (j,l) \rangle} c^{\dag}_{i,s} c^{}_{j,l} \, ,
\end{equation}
\noindent where $t$ is the hopping amplitude, $c^{\dag}_{i,s}$ $(c^{}_{i,s})$ creates (destroys) an electron in the unit cell $i$ and sublattice site $s$ of the kagome lattice, and $\langle (i,s) (j,l) \rangle$ denotes all pairs of nearest neighbor sites. Transforming Eq.~(\ref{eq:H0}) in momentum space, it becomes $H_0 = \sum_{\textbf{k}}\Psi^{\dag}_{\textbf{k}}\mathcal{H}^0_{\textbf{k}}\Psi^{}_{\textbf{k}}$ with $\Psi^{\dag}_{\textbf{k}} = (c^{\dag}_{\textbf{k}, 1},c^{\dag}_{\textbf{k},2},c^{\dag}_{\textbf{k},3})$ and
\begin{equation}
\label{eq:H0k}
    \mathcal{H}^0_{\textbf{k}} =  
\begin{bmatrix}
0 & \cos{\textbf{k}\cdot\textbf{a}_1} & \cos{\textbf{k}\cdot\textbf{a}_2}\\
\cos{\textbf{k}\cdot\textbf{a}_1} & 0 & \cos{\textbf{k}\cdot\textbf{a}_3}\\
\cos{\textbf{k}\cdot\textbf{a}_2} & \cos{\textbf{k}\cdot\textbf{a}_3} & 0
\end{bmatrix}
\, .
\end{equation}
%
%
\begin{figure*}[!ht]
\centering
\includegraphics[width=\textwidth,angle=0,clip=true]{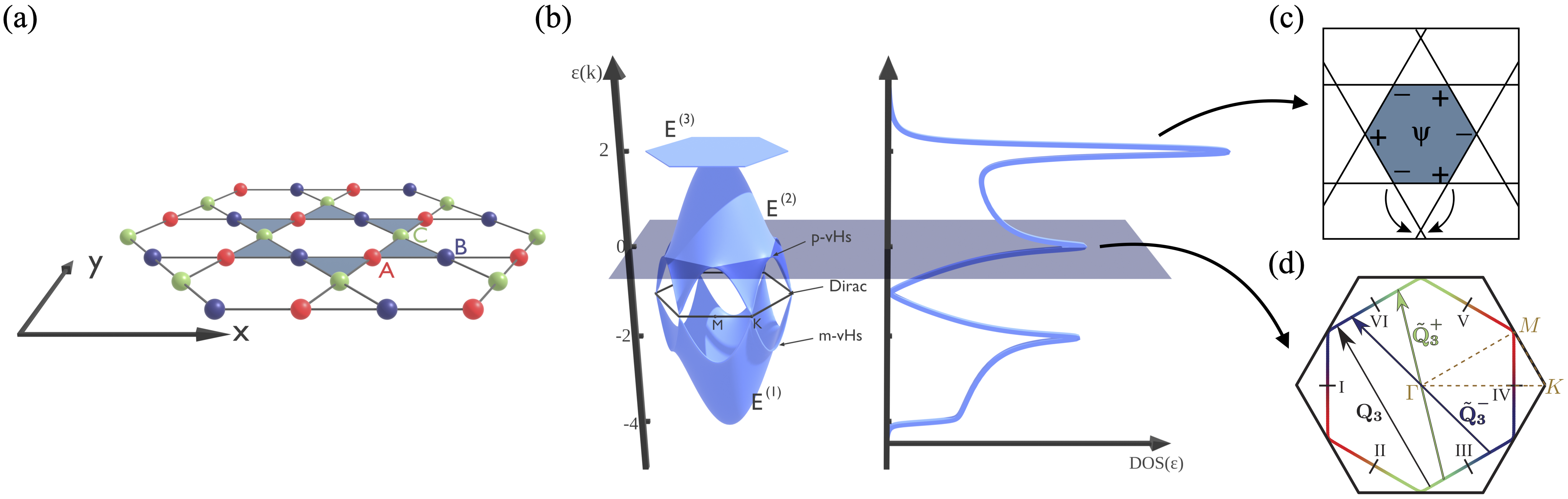}
\caption{(a) Structure of the ideal kagome lattice. A, B and C are the three sites in the unit cell. The blue triangles highlight the corner sharing configuration. (b) Band structure and corresponding density of states (DOS) in units of
the hopping integral $t$ resulting from the three-sublattice structure of the kagome lattice shown in (a). The electron fillings $n = 3/12$ and $n = 5/12$ (grey horizontal plane) are
located at VHS, as visible in the density of state plot from the two divergences symmetric to the location of the Dirac point. (c) Confinement of the electron wavefunction $\Psi$. Plus and minus signs indicate the phase of flat band eigenstate at neighboring sublattices. Any hoppings outside the hexagon (arrows) are canceled by destructive quantum interference, resulting in the perfect localization of electron in the blue-colored hexagon. (d) Fermi-surface distribution of the sublattice weight for the pure type VHS at filling $n = 5/12$. The Fermi surface touches the \textbf{M} point of the hexagonal BZ where the DOS diverges, as shown in (b). Its topology allows for three nesting vectors connecting parallel opposite sides of the Fermi surface, one of which is $\textbf{Q}_3 = (-\pi/2, \sqrt{3}\pi/2)$. The colors red, blue, and green label the major sublattice occupation of the Fermi surface states, in agreement with the sublattice coloring scheme of panel (a). $\tilde{\textbf{Q}}^{\pm}_3$ originate from opposite shifts of $\textbf{Q}_3$ and link states of similar sublattice weights. The labels I-VI are guides to the eye to help reading off the modulation of sublattice occupation weights encoded in $|u_{s,n=2}(\textbf{k})|^2$ for band $E^{(2)}$ and momenta $\textbf{k}$ along the Fermi surface.
Panel (c) adapted from~\textcite{kang_topological_2020}. Panel (d) adapted from~\textcite{kiesel_sublattice_2012}.}
\label{figure1}
\end{figure*}
%
%
\noindent The index $s=1,2,3$ in $c^{\dag}_{\textbf{k},s}$ refers to any of the three basis sites A, B and C in the triangular unit cell, whereas $\textbf{a}_1 = a\hat{x}$, $\textbf{a}_2 = a(\hat{x}+\sqrt{3}\hat{y})/2$ and $\textbf{a}_3 = \textbf{a}_2 - \textbf{a}_1$ identify the three nearest-neighbor vectors, $a$ being the lattice parameter.

The eigenvalue spectrum of $\mathcal{H}^0_{\textbf{k}}$, shown in Fig.~\ref{figure1}(b) with the corresponding DOS, consists of two dispersive bands
\begin{equation}
\label{eq:Edirac}
    E^{(1,2)}_{\textbf{k}} = -t[1 \pm \sqrt{4A_{\textbf{k}}-3}] \, ,
\end{equation}
\noindent with $A_{\textbf{k}} = \cos^2{\textbf{k}\cdot\textbf{a}_1} + \cos^2{\textbf{k}\cdot\textbf{a}_2} + \cos^2{\textbf{k}\cdot\textbf{a}_3}$, and one flat band at energy $E^{(3)}_{\textbf{k}} = 2t$. The dispersive bands $E^{(1,2)}$ touch at the corner of the Brillouin zone (BZ) $\textbf{K} = (2\pi/3a,0)$ and $\textbf{K}' = (-2\pi/3a,0)$, in proximity of which they exhibit a linear energy–momentum dispersion relationship analogous to the case of the honeycomb lattice. In fact, by linearizing $\mathcal{H}^0_{\textbf{k}}$ near $\textbf{K}$ and $\textbf{K}'$ and subsequently projecting onto the subspace spanned by the bands $E^{(1,2)}$, one obtains the Dirac Hamiltonian $h_{\textbf{k}} = v_F(\tau_3 k_x + \tau_1 k_y)$, where $v_F = \sqrt{3}t$ is the Fermi velocity and $\tau_i$ are Pauli matrices~\cite{guo2009topological}.

In addition to the dispersive bands, a flat band $E^{(3)}$ exists independent of the electron wave vector $\textbf{k}$. This flat band is degenerate with one of the two dispersive bands at the center of the BZ $\Gamma$, where the band touching is quadratic in momentum. The presence of a flat band can be understood from a mechanism relying on the destructive quantum phase interference of fermion hopping paths in certain two-dimensional networks. Interestingly, besides the kagome lattice, these include, for instance, also the dice~\cite{sutherland1986localization}, Lieb~\cite{lieb1989two} and decorated square~\cite{tasaki1992ferromagnetism} lattices. With specific focus on the nearest-neighbor electronic hopping model of Eq.~(\ref{eq:H0}), one can in fact build the real-space eigenfunction $\Psi$ with alternating phases at neighboring corners of the hexagon shown in Fig.~\ref{figure1}(c). Such an electronic state is geometrically confined within a single hexagon because any hopping to neighboring cells is canceled out by destructive phase interference. This real-space electronic localization translates into flat momentum–space eigenfunctions with no energy dispersion.

It is important to emphasize that this mechanism strictly relies on the nearest-neighbor hopping approximation; further-neighbor terms generally introduce a finite dispersion into the nominally flat band. Moreover, the formation of perfectly flat bands in real kagome materials is nontrivial because the relevant $p$- and 
$d$-orbitals have distinct symmetry properties compared to idealized
$s$-like orbitals. Achieving an ideal kagome band structure thus requires that the orbital character is properly aligned with the lattice geometry, a condition recently clarified in \cite{kim2023realization}, where it was shown that only certain orbital configurations (e.g., $d_{z^2}$) or rotated orbital bases under large crystal-field splitting can host true flat bands.

A second instructive way to understand the appearance of a flat band on the kagome lattice takes root in the theory of graphs~\cite{mielke1991ferromagnetic,mielke1991ferromagnetism,roychowdhury2022supersymmetry}. This method also explains why the dispersion of bands $E^{(1,2)}$ resembles that of nearest-neighbor hopping models on the honeycomb lattice. If $\textbf{A}$ is a generic matrix of arbitrary dimensions, then the two matrices $\textbf{A}\textbf{A}^{\dag}$ and $\textbf{A}^{\dag}\textbf{A}$ are isospectral except for zero modes which result from a potential dimension mismatch between the kernel of $\textbf{A}$ and $\textbf{A}^{\dag}$ if $\textbf{A}$ is not a square matrix. Identifying $\textbf{A}\textbf{A}^{\dag}$ as the adjacency matrix of the honeycomb lattice graph, $\textbf{A}^{\dag}\textbf{A}$ turns out to be the adjacency matrix of its incident or line graph, \textit{i.e.} the graph originating from the promotion of edges to vertices. 

For the honeycomb lattice, the incidence graph is the kagome lattice~\cite{roychowdhury2022supersymmetry}. The adjacency matrix can be thought of as a hopping-1 nearest-neighbor Hamiltonian similar to Eq.~(\ref{eq:H0}). The adjacency matrix of a line graph can also be shown to have a zero mode as its lowest eigenvalue~\cite{cvetkovic2004spectral,chiu2022line}. The flat band, or zero mode, in the kagome spectrum can then be traced back to the difference in the dimensions of the two graph adjacency matrices, which is simply the difference in number of sites (three) in the kagome lattice unit cell versus the two sites of the honeycomb lattice unit cell. This analysis explains why, besides the flat band, the two lattices are identical from a spectral point of view, both featuring Dirac-like linear dispersions at the $\textbf{K}$ and $\textbf{K}'$ valleys and VHS at the $\textbf{M}$ points. 

However, different from the honeycomb lattice, the lack of particle-hole symmetry in the kagome spectrum emerges prominently when one looks at the redistribution of the real space A, B and C sublattice characters in the momentum-dependent wavefunction. In fact, the transition from real space to momentum space upon Fourier transform reads
\begin{eqnarray}
\label{eq:transform}
    c^{\dag}_{i,s} &=& \sum_{\textbf{k}} c^{\dag}_{\textbf{k},s} \exp{[-i \textbf{k}(\textbf{R}_i + \textbf{r}_s)]} = \nonumber \\
    &=& \sum_{\textbf{k},n} u^{*}_{s,n}(\textbf{k})c^{\dag}_{\textbf{k},n} \exp{[-i \textbf{k}(\textbf{R}_i + \textbf{r}_s)]},
\end{eqnarray}
\noindent where $\textbf{R}_i$ denotes the unit cell location and $\textbf{r}_s$ the sublattice
location within the unit cell. The core information is encoded in the transformation coefficients $u_{s,n}(\textbf{k})$, namely the eigenvectors of $\mathcal{H}^0_{\textbf{k}}$ in Eq.~(\ref{eq:H0k}), known as sublattice weights~\cite{kiesel_sublattice_2012}. For a given band $n$ that identifies any of the $E^{(1,2,3)}$ and momentum point $\textbf{k}$ in the BZ, the coefficients obey $\sum_{s}|u_{s,n}(\textbf{k})|^2 = 1$. 

Noteworthy is the distribution of the sublattice weights at the VHS. At variance with the bipartite honeycomb lattice, the kagome bands host two different types of VHS which are commonly labeled as sublattice mixed ($m$--type) and sublattice pure ($p$--type), characterized by odd and even parity at the $\textbf{M}$ point, respectively~\cite{kiesel_sublattice_2012,wang2013competing,kiesel2013unconventional,kang2022twofold,jonas_b_profe_kagome_2024}. Figure~\ref{figure1}(d) shows the evolution of $|u_{s,n=2}(\textbf{k})|^2$ on the Fermi surface at the $p$--type VHS, where the eigenstates in the vicinity of the three $\textbf{M}$ points are localized on mutually different sublattices. By contrast, the lower VHS has mixed sublattice character, with the eigenstates equally distributed over mutually different sets of two sublattices for each $\textbf{M}$ point.

Independently of the sublattice decoration, the Fermi surface topology at both VHS suggests $\textbf{Q}_1 = (-\pi/2, -\sqrt{3}\pi/2)$, $\textbf{Q}_2 = (\pi, 0)$ and $\textbf{Q}_3 = (-\pi/2, \sqrt{3}\pi/2)$ as nesting vectors. However, for the $p$--type VHS, the different sublattice character of states at the three distinct $\textbf{M}$ points reduces the effectiveness of the nesting vectors in the particle-hole fluctuation channels compared to a Fermi surface with identical topology and constant orbital makeup. In fact, because of sublattice interference~\cite{kiesel_sublattice_2012,kiesel2013unconventional}, the particle-hole fluctuation channels split into six different nesting vectors that connect Fermi surface regions of equal sublattice weights. For instance, for the case of $\textbf{Q}_3$, this amounts to a shift to $\tilde{\textbf{Q}}^{\pm}_3 = \textbf{Q}_3 \pm (\pi/4, \pi/4\sqrt{3})$ as shown in Fig.~\ref{figure1}(d).


\begin{figure}[!t]
\centering
\includegraphics[width=\columnwidth,angle=0,clip=true]{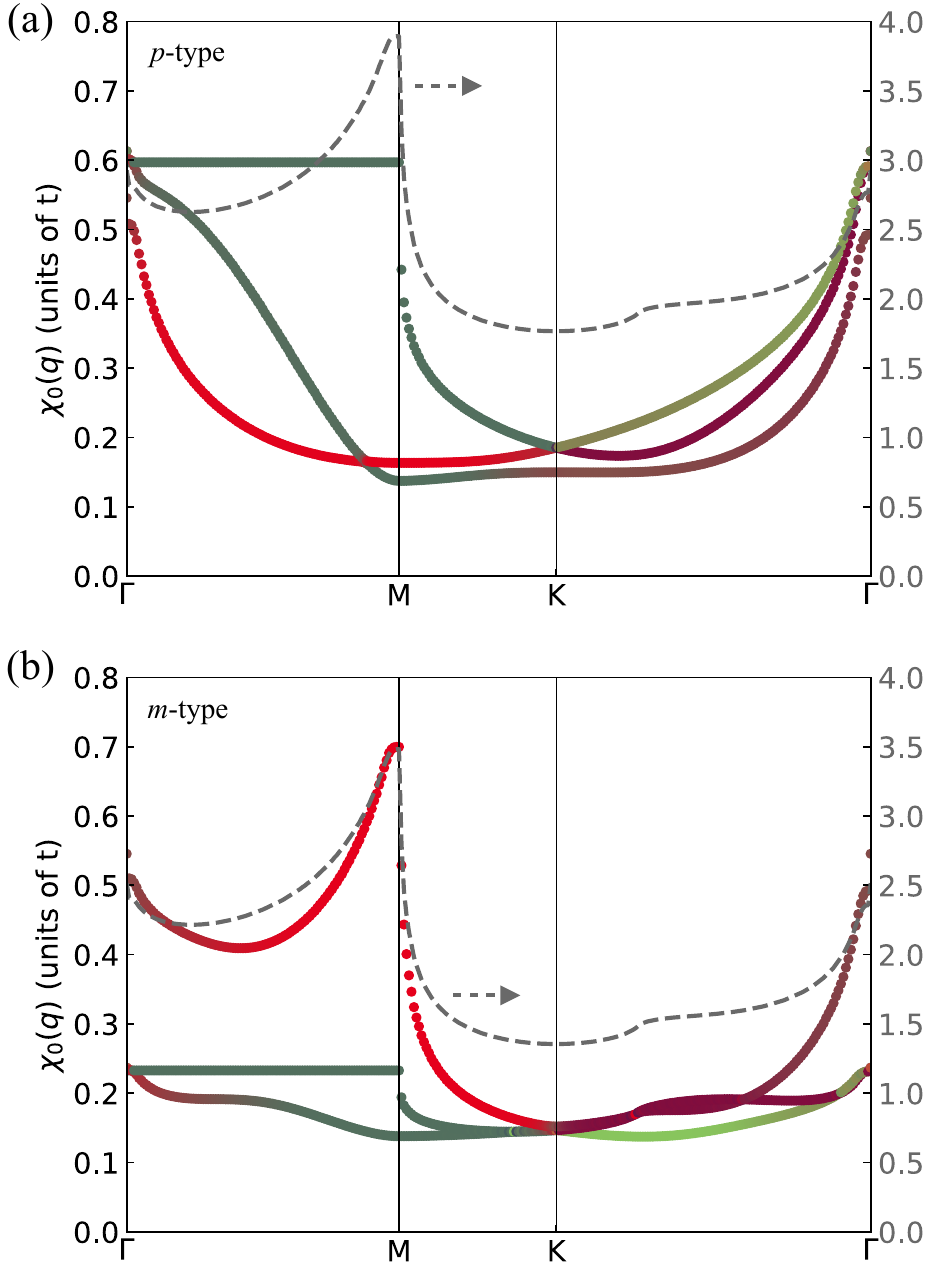}
\caption{(a) Eigenvalues of the static non-interacting susceptibility $\chi^{0}_{s_1s_1s_3s_3}(\textbf{q},\omega = 0)$ along a high-symmetry path of the kagome BZ for the chemical potential set at the $p$--type VHS. The colors refer to the sublattice contributions in the susceptibility eigenvectors, following the color scheme given in Fig.~\ref{figure1}(a) for the A, B and C sublattices. The dashed grey line, whose y-axis is at the right, reports the Lindhard function. (b) Same as panel (a) but for the chemical potential set at the $m$--type VHS.}
\label{figure2}
\end{figure}


The distribution of sublattice characters of VHS in kagome systems are important for possible Fermi-surface instabilities, which are modulated by the matrix-element effects of the sublattice textures. As a matter of fact, the non-interacting two-particle susceptibility in momentum-frequency space is defined as
\begin{eqnarray}
\label{eq:chi0}
    &&\chi^{0}_{s_1s_2s_3s_4}(\textbf{q},\omega) = -\Omega \int \frac{d\textbf{k}}{4\pi^2} \times \nonumber \\
    &&\sum_{n,m}
    u_{s_4,n}(\textbf{k})
    u^{*}_{s_2,n}(\textbf{k})
    u_{s_1,m}(\textbf{k}+\textbf{q})
    u^{*}_{s_3,m}(\textbf{k}+\textbf{q}) \times \nonumber
    \\
    && \frac{ n_{\text{F}}[E^{(n)}_{\textbf{k}}] - n_{\text{F}}[E^{(m)}_{\textbf{k} + \textbf{q}}] }{\omega + i\delta + E^{(n)}_{\textbf{k}} - E^{(m)}_{\textbf{k} + \textbf{q}}},
\end{eqnarray}
\noindent where $\Omega$ is the unit cell volume, $n/m$ the band indices, $n_{\text{F}}(E)$ is the Fermi-Dirac distribution, $\delta$ an infinitesimally small broadening, and the integral is carried out over the whole BZ. 

It is clear how the presence of the weighting coefficients $u_{s,n}(\textbf{k})$ has a sizable influence on the particle-hole fluctuations. The physical susceptibility is given by the susceptibility matrix elements $\chi^{0}_{s_1s_1s_3s_3}(\textbf{q},\omega)$, whereas the Lindhard function neglects the contribution from the coefficients $u_{s,n}$ in Eq.~(\ref{eq:chi0}). Figure~\ref{figure2}(a) displays the sublattice-resolved eigenvalues of the static physical bare susceptibility $\chi^{0}_{s_1s_1s_3s_3}(\textbf{q},\omega=0)$ (solid colored points) and the Lindhard function (dashed grey line) along a high-symmetry path at the $p$--type VHS. The former does not reveal any peak around the $\textbf{M} = \textbf{Q}_3$ point and the eigenvector of the largest eigenvalue is mainly attributed to the B and C sublattices, as the nesting vector $\textbf{Q}_3$ connects parts of the Fermi surface which are dominated by those sublattices [Fig.~\ref{figure1}(d)]. 

As a result, the susceptibility along the line $\Gamma$--$\textbf{M}$ is almost constant, because the opposite edges of the hexagonal Fermi surface connected by $\textbf{Q}_3$ are likewise entirely dominated by the B and C sublattices. However, the Lindhard function, that discards the orbital makeup, displays a pronounced peak at the $\textbf{M}$ point as a consequence of the Fermi-surface nesting. On the other hand, at the $m$--type VHS [Fig.~\ref{figure2}(b)], the physical susceptibility shows a noticeable peak at the $\textbf{M}$ point attributed to the A sublattice, that is the common sublattice shared by the Fermi-surface segments connected by the $\textbf{Q}_3$ nesting vector. The Lindhard function exhibits a similarly pronounced feature. A comparison between Fig.~\ref{figure2}(a) and Fig.~\ref{figure2}(b) demonstrates the clear effect of the sublattice interference that originates from the modulation of sublattice weights on the Fermi surfaces at the VHS~\cite{kiesel_sublattice_2012,kiesel2013unconventional}.

\subsubsection{\label{sec:chapIIA2}Two and three orbitals per site}

The description in terms of itinerant electrons as given by the one $s$-like orbital per site model can be adopted as a basis to study the phenomenology of some realistic kagome systems that are located at intermediate coupling and not in the Mott-type local spin regime.  Herbertsmithites~\cite{normanRMP}, like ZnCu$_3$(OH)$_6$Cl$_2$, are notable candidates in the latter regime~\cite{mazin2014theoretical,di2020turbulent}. Promising alternative platforms have also been proposed, such as optical kagome lattices of ultracold fermionic atomic gases, including isotopes like $^6$Li and $^{40}$K~\cite{jo2012ultracold}.

In the weaker coupling, itinerant regime, the discovery of $A$V$_3$Sb$_5$ Kagome metals compounds~\cite{ortiz2019new} with their large number of contributing orbitals from both V and Sb in the vicinity of the Fermi level, has required the development of more sophisticated multi-orbital effective models as a prerequisite to any analysis of many-body instabilities. We briefly use this material as one concrete example to illustrate the complexity that arises in this regime.  In this context, \textcite{wu2021nature} started from the assumption that the \textit{ab initio} bandstructure of AV$_3$Sb$_5$ matches closely angle-resolved photoemission spectroscopy (ARPES) measurements taken below the CDW transition temperature, despite the density functional theory (DFT) calculations being done without considering the star-of-David-type structural distortion~\cite{PhysRevLett.125.247002,PhysRevLett.127.046401}. 

The layered structure of AV$_3$Sb$_5$, combined with the large anisotropy in the resistivity $\rho_c/\rho_{ab}$ ($\approx 600$ for CsV$_3$Sb$_5$~\cite{PhysRevLett.125.247002}) motivated~\textcite{wu2021nature} to constrain themselves to the two-dimensional V--Sb kagome plane. Setting $k_z=0$, DFT predicts three distinct Fermi surfaces in AV$_3$Sb$_5$: (i) a pocket formed by vanadium $d_{xy}$, $d_{x^2-y^2}$, and $d_{z^2}$ orbitals near a $p$--type VHS, (ii) two additional pockets created by vanadium $d_{xz}$ and $d_{yz}$ orbitals close to another $p$--type VHS below the Fermi level and an $m$--type VHS above it, and (iii) a circular pocket around $\Gamma$ comprised of antimony $p_z$ orbitals. Due to opposite mirror $M_z$ eigenvalues, electronic states in pockets (i) and (ii) do not hybridize, while the allowed hybridization terms between pockets (ii) and (iii) are small.


\begin{figure}[!t]
\centering
\includegraphics[width=\columnwidth,angle=0,clip=true]{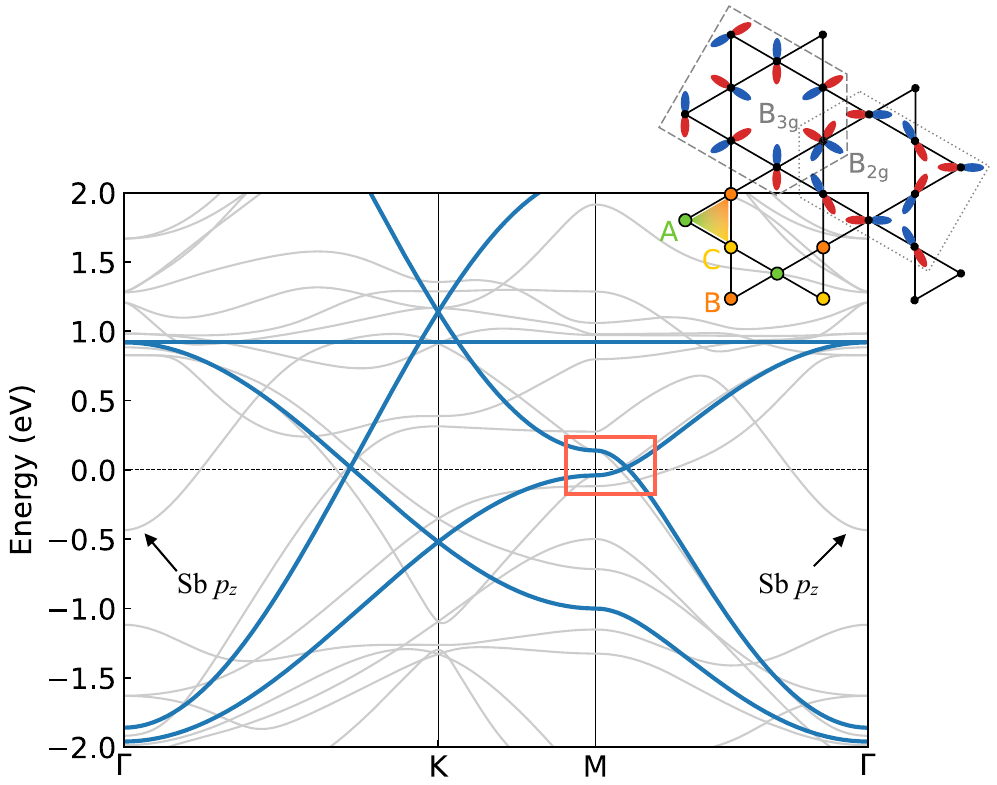}
\caption{Eigenvalues of the two orbitals per site model of Eq.~(\ref{eq:H0_2orb}) (blue lines) overlaying the band structure as obtained from the first-principles calculations in the absence of SOC (grey lines) for KV$_3$Sb$_5$. The red box highlights the energy-momentum region where the multiple VHS of both $p$-type and $m$-type character emerge below and above the Fermi level, respectively. The top right corner inset reports the real-space structure of the kagome V plane where the sign structure (blue or red lobes) and spatial orientation of the $B_{2g}$ and $B_{3g}$ orbitals is shown on the different lattice sites. Around the $\Gamma$ point, the arrows identify the electron pocket originating from the $p_z$ orbital of the Sb atom located at the center of the hexagon of the kagome lattice. Inset adapted from~\textcite{wu2021nature}.}
\label{figure3}
\end{figure}


For their two orbitals per site effective model, \textcite{wu2021nature} considered only the Fermi pockets (ii) because those carry the dominant DOS at the Fermi levels, and preserve the complexity of multiple VHS of both $p$--type and $m$--type character. More importantly, the effective model captures all irreducible band representations at the high symmetry points in the BZ. The orbitals $d_{xz}$ and $d_{yz}$ that constitute the Fermi pockets (ii) transform, respectively, under the $B_{2g}$ and $B_{3g}$ irreducible representations of the site symmetry group $D_{2h}$ for the $3f$ atomic Wyckoff positions (shown in Fig.~\ref{figure3}). As such, they form two sets of three bands with opposite mirror eigenvalues and a mirror symmetry-protected Dirac cone along the $\Gamma$--$\textbf{M}$ line, hence giving rise to an upper and lower Van Hove filling (emphasized by the red box in Fig.~\ref{figure3}) with opposite sublattice parity. The proposed effective TB Hamiltonian reads~\cite{wu2021nature}
\begin{eqnarray}
\label{eq:H0_2orb}
    H_0 && = \sum_{\textbf{k}s\alpha}\epsilon_{\alpha}c^{\dag}_{\textbf{k},s, \alpha}c_{\textbf{k},s, \alpha} - \sum_{\textbf{k}sl\alpha}t_{\alpha}\Phi_{sl}(\textbf{k})c^{\dag}_{\textbf{k},s, \alpha}c_{\textbf{k},l, \alpha} \nonumber \\
    && - t'\sum_{\textbf{k}sl}\Phi_{sl}(\textbf{k})\nu_{sl}(c^{\dag}_{\textbf{k},l, xz}c_{\textbf{k},s, yz} - c^{\dag}_{\textbf{k},l, yz}c_{\textbf{k},s, xz})
\end{eqnarray}
\noindent with $s,l = A, B, C$ and $\alpha = xz, yz$. Here, $\epsilon_{\alpha}$ denotes the crystal field splitting and the operator $c^{\dag}_{\textbf{k},s, \alpha}$ ($c_{\textbf{k},s, \alpha}$) creates (annihilates) an electron on orbital $\alpha$ on sublattice site $s$ with momentum $\textbf{k}$. Further, $\Phi_{sl}(\textbf{k}) = 1 + e^{-2 i \textbf{k} \cdot \textbf{d}_{sl}}$ are the lattice structure factors, with $\textbf{d}_{sl}$ being the sublattice-connecting vectors. They obey  $\Phi_{sl}(\textbf{k}) = (1-\delta_{sl})\Phi_{ls}^{*}(\textbf{k})$. The transformation of the $B_{2g}$ and $B_{3g}$ orbitals under the site symmetry group demands that the sign-structure of the third term in Eq.~(\ref{eq:H0_2orb}) is ensured by $\nu_{AC} = \nu_{CB} = -\nu_{AB} = 1$ and $\nu_{sl} = -\nu_{ls}$. Figure~\ref{figure3} shows that the model is capable of faithfully capturing the electronic behavior of AV$_3$Sb$_5$ Kagome metals in the energy region around the chemical potential, reproducing the nature and dispersion of the two VHS at the $\textbf{M}$ point as well as the mirror symmetry-protected Dirac crossing along the $\Gamma$--$\textbf{M}$ line.

This model has been used as the non-interacting platform for refined random phase approximation (RPA) analysis of superconducting instabilities (see Section~\ref{sec:chapIIIC} for details), to interpret ARPES experiments on the twofold nature of VHS~\cite{kang2022twofold} and to predict the existence of possible chiral excitonic order~\cite{scammell2023chiral}. An extension of the model to include the electron pocket at the $\Gamma$ point, that derives from the central Sb $p_z$ orbital and is clearly detected in photoemission measurements~\cite{PhysRevLett.125.247002,kang2022twofold}, has also been proposed~\cite{wu2021nature}. However, numerical calculations suggest that its contribution to the superconducting properties of AV$_3$Sb$_5$ Kagome metals is marginal, while measurements under non-hydrostatic pressure point towards a strong correlation between a monoclinic structural transition, the termination of the second high-pressure superconducting dome and the removal of the Sb $p_z$ electron pocket from the Fermi surface~\cite{PhysRevB.107.174107}.


\begin{figure}[!t]
\centering
\includegraphics[width=\columnwidth,angle=0,clip=true]{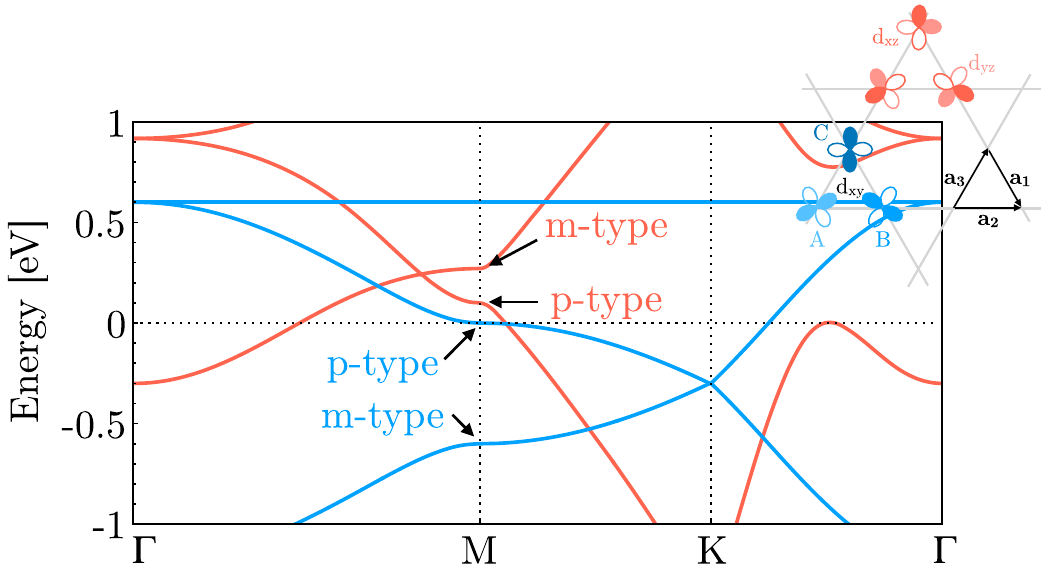}
\caption{Band structure induced by $d_{xy}$ (blue) and $d_{xz}/d_{yz}$ (red) orbitals on the kagome lattice (see top right corner inset) and corresponding pure and mixed nature of VHS. Adapted from~\textcite{denner2021analysis}.}
\label{figure4}
\end{figure}


An extension of the above model includes an additional V $d_{xy}$ orbital represented by a Wannier state in the $A_g$ irreducible representation of $D_{2h}$ formed by a linear combination of $d_{xy}$, $d_{x^2-y^2}$ and $d_{z^2}$ orbitals~\cite{denner2021analysis}. Figure~\ref{figure4} shows the band dispersion of such a model, and crossings between the $d_{xy}$ (blue) and the $d_{xz}/d_{yz}$ (red) bands are protected by mirror symmetry. The coexistence of two $p$--type and one $m$--type VHS near the Fermi level has been confirmed by recent ARPES measurements of the CsV$_3$Sb$_5$ kagome metal~\cite{kang2022twofold}. The three bands forming these states at the $\textbf{M}$ point have been dubbed K1, K2 and K2$^\prime$, and their location in the proximity of the Fermi level could set the stage for electronic symmetry breaking~\cite{denner2021analysis,kang2022twofold}.

One further model for the AV$_3$Sb$_5$ kagome metals worth mentioning was proposed by~\textcite{deng2023two} based on the identification of elementary band representations (EBRs)~\cite{bradlyn2017topological,cano2018building,elcoro2021magnetic}. This model is built considering an orbital with $A_g$ symmetry on each kagome site, \textit{i.e.}, V $d_{x^2-y^2}$, and an orbital with $A''_2$ symmetry, \textit{i.e.} $p_z$, on all Sb atoms on the hexagonal ``graphene-like" sublattice. This model, called two-EBR graphene-Kagome model, captures the dispersion of the two low-energy VHS, and correctly describes the overall nontrivial band topology of AV$_3$Sb$_5$ materials. 


\subsubsection{\label{sec:chapIIA3}Extended models}

The electronic structures of realistic kagome materials are complicated and involve many orbitals and bands at the Fermi level. The strongly simplified models such as, for instance, those described in previous Sections~\ref{sec:chapIIA1} and~\ref{sec:chapIIA2}, while capable of describing general band structure features in a qualitative way, fail to give quantitative descriptions. To overcome this drawback and to bring the faithfulness of TB approaches closer to that of first-principles calculations, more realistic models have been developed, where not only the multi-orbital nature of kagome-like atoms is accounted for, but also the atomic environment. In fact, the importance of inter-orbital hybridization and of grouping the $d_{x^2-y^2}/d_{xy}$, $d_{xz}/d_{yz}$, and $d_{z^2}$ orbitals has been recognized in earlier studies of kagome metals~\cite{kang_topological_2020,okamoto2022topological,kim2023realization}, which highlights how orbital symmetry and hybridization strongly influence the emergence of flat bands and VHS.

As another material specific example, a comprehensive understanding of the complicated electronic properties of the XY  family of kagome metals (discussed in Section ~\ref{sec:chapIVA}), where X is a $3d$ transition metal such as Fe and Co and Y is a main group element such as Sn or Ge, has been proposed by~\textcite{jiang2023kagome} on a strategy based upon band decomposition and orbital grouping. Each of the three groups is formed by separating the $d$ orbitals of X in combination with specific orbitals of Y based on chemical and symmetry analyses. Within each group, an analytical understanding of the band structures can be obtained. The resulting three decoupled effective TB models quantitatively reproduce the quasi-flat bands, VHS, and Dirac points of XY kagome metals~\cite{kang_dirac_2020,kang_topological_2020}.


\begin{figure*}[!ht]
\centering
\includegraphics[width=\textwidth,angle=0,clip=true]{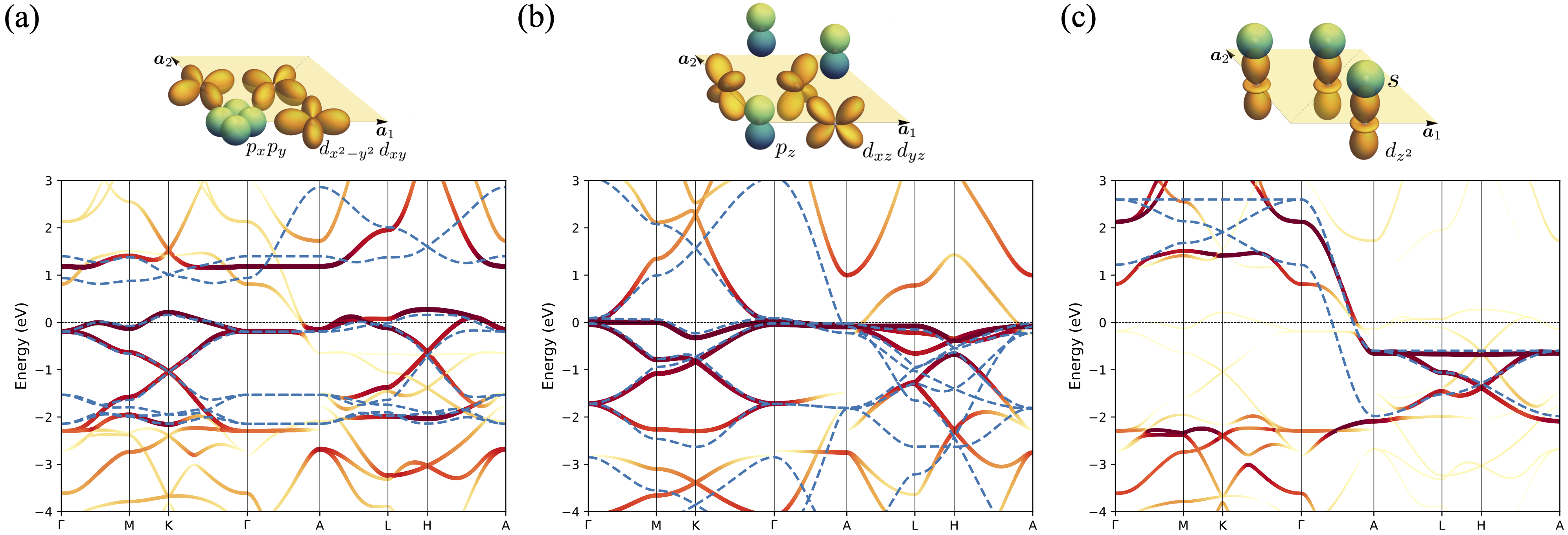}
\caption{Band structures of first-principles calculations (solid lines) and of TB model in Eq.~(\ref{eq:H0_bernevig}) (dashed blue lines) for the specific case of FeGe. The \textit{ab initio} band structure has been computed in the paramagnetic phase of FeGe. The three panels (a), (b) and (c) refer to the three groups of Fe $(d_{xy},d_{x^2-y^2})$, $(d_{xz},d_{yz})$ and $d_{z^2}$ orbitals, respectively. The darker the color of the bands, the stronger is the character of the respective set of orbitals shown in the top insets (adapted from~\textcite{jiang2023kagome}).} 
\label{figure5}
\end{figure*}


More specifically, \textcite{jiang2023kagome} split the five kagome X $d$ orbitals into the three groups $(d_{xy},d_{x^2-y^2})$, $(d_{xz},d_{yz})$ and $d_{z^2}$, because the $M_z$ mirror symmetry prevents any planar hybridization between the orbitals of the first two groups, whereas the hopping terms between the $(d_{xy},d_{x^2-y^2})$ and $d_{z^2}$ orbitals, although allowed by symmetry, can be neglected. These three groups are then combined with specific orbitals of Y atoms, as depicted in the top insets of Fig.~\ref{figure5}. Indeed, first-principles calculations confirm: (1) large overlap and $\sigma$--like bonding between the X $(d_{xy},d_{x^2-y^2})$ and trigonal Y $(p_x,p_y)$ orbitals [Fig.~\ref{figure5}(a)] (2) large overlap and $\pi$--like bonding between the X $(d_{xz},d_{yz})$ and trigonal as well as hexagonal Y $p_z$ orbitals [Fig.~\ref{figure5}(b)] and (3) sizable overlap between X $d_{z^2}$ orbitals and $sp^2$ bonding state of honeycomb Y atoms [Fig.~\ref{figure5}(c)]. 

\begin{widetext}
In matrix form, the total TB Hamiltonian becomes~\cite{jiang2023kagome}
\begin{eqnarray}
\label{eq:H0_bernevig}
&&\mathcal{H}^0_{\textbf{k}} = H_{d_{z^2}}(\textbf{k}) \oplus\\  
&&\begin{bmatrix}
    H_{p_{xy}^t}(\textbf{k}) & S_{p_{xy}^t, d_{xy}}(\textbf{k}) & S_{p_{xy}^t, d_{x^2-y^2}}(\textbf{k}) \\
     & H_{d_{xy}}(\textbf{k})  & S_{d_{xy},d_{x^2-y^2}}(\textbf{k}) \\
     h.c. &  & H_{d_{x^2-y^2}}(\textbf{k})+H_{d_{x^2-y^2}, p_z^h}^{(2)}(\textbf{k}) \\
\end{bmatrix}
\oplus
\begin{bmatrix}
    H_{p_z^h}(\textbf{k})+H_{p_z^h,d_{x^2-y^2}}^{(2)}(\textbf{k}) & \bm{0} & S_{p_z^h,d_{yz}}(\textbf{k})\\
    & H_{d_{xz}}(\textbf{k})+H_{d_{xz}, p_z^t}^{(2)}(\textbf{k}) & S_{d_{xz},d_{yz}}(\textbf{k})\\
    h.c. & & H_{d_{yz}}(\textbf{k})\\
\end{bmatrix} \nonumber
\end{eqnarray}
\end{widetext}
\noindent where the blocks $H_i(\textbf{k})$ along the diagonals derive from short-range hopping contributions of the corresponding orbital $i$ ($p_{...}^t$ and $p_{...}^h$ refer to orbitals on the trigonal and honeycomb Y atoms respectively), while the off-diagonal blocks $S_{i, j}(\textbf{k})$
account for hybridization terms between different orbitals. The blocks $H_{i,j}^{(2)}(\textbf{k})$ arise as corrections from second-order perturbation theory and are required to keep the Hilbert space of the model still manageable and to write Eq.~(\ref{eq:H0_bernevig}) as the direct sum $\oplus$ of three independent terms. Their eigenvalues, shown in Fig.~\ref{figure5} as dashed blue lines overlaid to the \textit{ab initio} band structures for the specific case of paramagnetic FeGe, quantitatively reproduces the quasi-flat bands, VHS and Dirac points close to the Fermi level. Disagreements between the TB model and bands from first-principles calculations appear at high energies, putting forward Eq.~(\ref{eq:H0_bernevig}) as a useful quantitative model to study low-energy physics in kagome metals.

Interestingly, the model in Eq.~(\ref{eq:H0_bernevig}) for the XY kagome metals can be considered as a LEGO-like building block for the large family of RX$_6$Y$_6$ or ``166" kagome materials in the paramagnetic phase, such as Y(Mn,V)$_6$Sn$_6$ ~\cite{PhysRevLett.127.266401,pokharel_electronic_2021,Arachchige2022Sc166}. The Hamiltonian of these 166 members can be obtained by doubling and perturbing the model in Eq.~(\ref{eq:H0_bernevig}), because the 166 class materials RX$_6$Y$_6$ can be seen as a doubled X$_3$Y$_3$ material with an R atom inserted as spacer in the middle honeycomb layer.

In addition to binary kagome metals, simplified models of ternary compounds such as AV$_3$Sb$_5$ also struggle to quantitatively capture several key aspects revealed by first-principles calculations and ARPES~\cite{kang2022twofold}.  These include the sublattice character of the two VHS close to the Fermi energy, the correct Fermi surface orientation, and the multiorbital nature of the low-energy bands. To address these shortcomings, an extended Slater–Koster multiorbital formalism has been developed by~\textcite{zeng2025electronic}. This approach systematically incorporates the crystal-field environment of Sb atoms, employs a symmetry-adapted $C_3$-respecting orbital basis, and allows hybridization between orbitals of the same symmetry.

Importantly, the Slater–Koster framework is based on six independent hopping parameters, effectively renormalizing the influence of the surrounding crystal environment while preserving orbital symmetries. A central advance of the Slater-Koster method over the simplified models of Sections~\ref{sec:chapIIA1} and~\ref{sec:chapIIA2} is the explicit treatment of interorbital hopping, particularly between mirror-even and mirror-odd $d$-orbitals. This mechanism explains the emergence of ``mirror interorbital flat bands"~\cite{zeng2024interorbital} and accounts for the anomalous double \textit{p}-type VHS observed in AV$_3$Sb$_5$. When intraorbital hoppings are included, these interorbital flat bands evolve into dispersive kagome-like bands that remain consistent with the double \textit{p}-VHS scenario. The resulting Slater-Koster multiorbital models achieve excellent agreement with DFT and ARPES by reproducing the correct Fermi surface shapes and orientations, the sublattice and orbital character of the VHS, the multiorbital composition of the low-energy dispersions across the BZ, and higher-order VHS dispersions consistent with experimental fits~\cite{zeng2025electronic}.

\subsection{\label{sec:chapIIB}Effect of spin-orbit interaction}

\begin{figure*}[!ht]
\centering
\includegraphics[width=\textwidth,angle=0,clip=true]{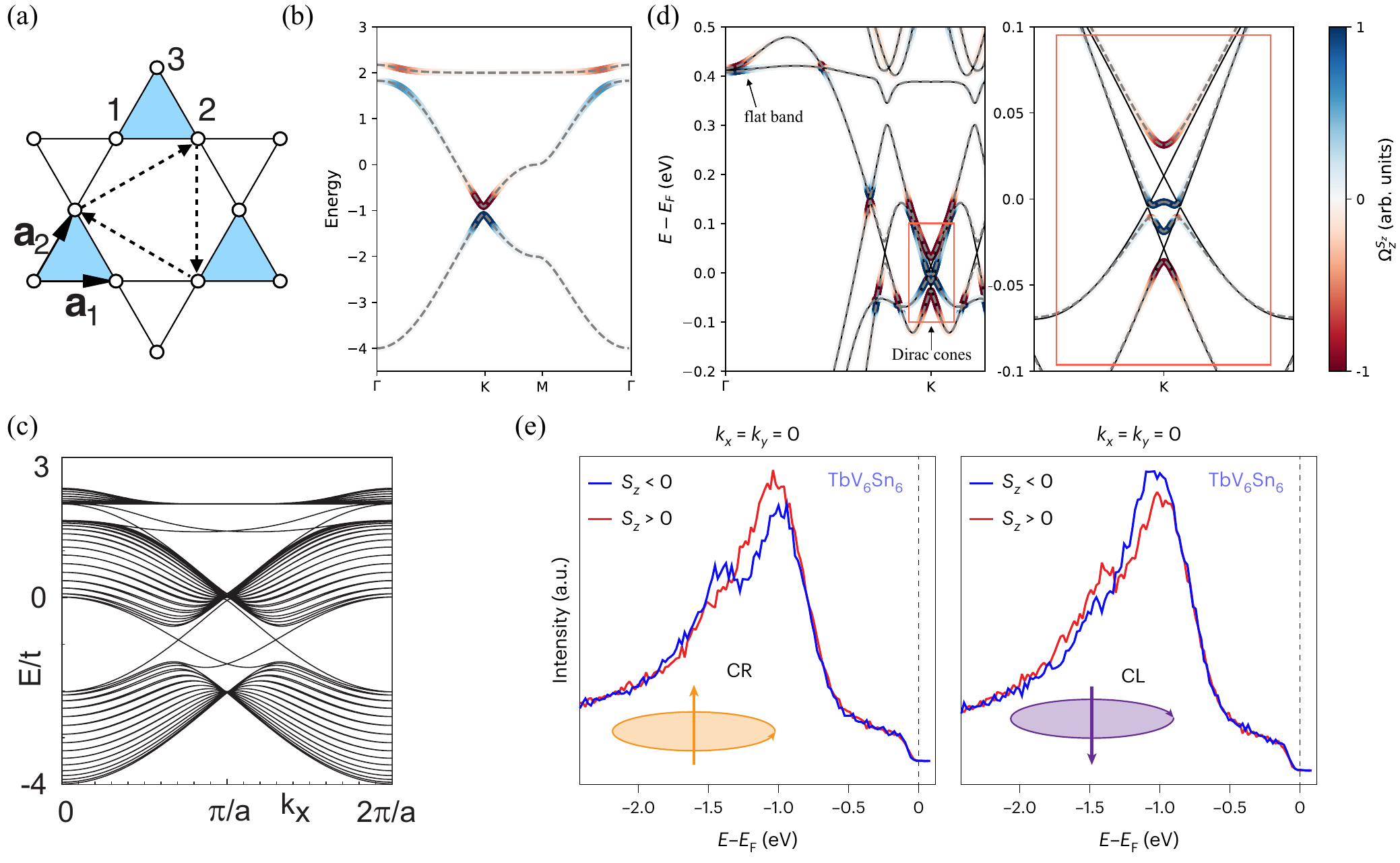}
\caption{(a) In the presence of SOC, spin up electrons hop between next nearest neighbor sites with amplitude $i\lambda_{\text{NNN}}$ when moving in accordance to the dashed arrows, $-i\lambda_{\text{NNN}}$ when moving against. For spin down electrons, the arrows are reversed. (b) The SOC opens a gap at the Dirac cone and at the touching point between the dispersing and flat bands. Concomitantly, a finite spin Berry curvature $\Omega_z^{S_z}$ arises at those gaps. (c) Band structure of the kagome model in open boundary conditions featuring the appearance of topological edge states at both gaps. (d) First-principles band structure of TbV$_6$Sn$_6$ kagome metal without (solid black line) and with (dashed grey line) SOC. The red and blue colors highlight $\Omega_z^{S_z}$ as in (b). The right panel is a close-up view of the Dirac cone region.
(e) Circular right (CR) and circular left (CL) spin-resolved ARPES energy distribution curves (EDCs) measured for the spin-up ($S_z > 0$, red) and spin-down ($S_z < 0$, blue) channels in TbV$_6$Sn$_6$. The EDCs were acquired at the center of the BZ with the incident light entirely confined to the sample’s mirror plane. Under these conditions, geometric contributions to the circular dichroism (CD) can be reliably ruled out. Panels (a) and (c) adapted from~\textcite{guo2009topological}. Panel (e) adapted from~\textcite{di2023flat}.} 
\label{figure6}
\end{figure*}

To study the physics of kagome lattice in the presence of SOC, one needs to expand the spinless basis of the model in Eq.~(\ref{eq:H0}) to $\Psi^{\dag}_{\textbf{k}\sigma} = (c^{\dag}_{\textbf{k}, 1},c^{\dag}_{\textbf{k},2},c^{\dag}_{\textbf{k},3}) \otimes (\uparrow,\downarrow)$ to include the spin degree of freedom. In the absence of SOC, at $\frac{1}{3}$ filling, the lowest band $E_{\textbf{k}}^{(1)}$ in Fig.~\ref{figure1}(b) is filled and the low-energy electronic excitations of $H_0$ resemble those of graphene. A perturbation term bilinear in the fermionic operators that breaks the SU(2) spin symmetry while preserving both the translational and TRS of $H_0$ is the spin-orbit interaction that induces hopping between next nearest neighbors (NNN) sites~\cite{guo2009topological}. It takes the form
\begin{eqnarray}
\label{eq:SOCNNN}
H_{\text{SO}} =\ && \imath\frac{2\lambda_{\text{NNN}}}{\sqrt{3}}\\
&&\times \sum_{\langle\langle (i,s)(j,l) \rangle\rangle \sigma\sigma'} (\mathbf{d}_{ij}^{1} \times \mathbf{d}_{ij}^{2}) \cdot c^{\dag}_{i,s,\sigma} \boldsymbol{\sigma}_{\sigma\sigma'} c_{j,l,\sigma'} \, , \nonumber
\end{eqnarray}
\noindent where $\lambda_{\text{NNN}}$ is the NNN SOC strength, $c^{\dag}_{i,s,\sigma}$ $(c_{i,s,\sigma})$ creates (annihilates) an electron in the unit cell $i$, sublattice site $s$ and spin $\sigma$, and $\langle\langle (i,s) (j,l) \rangle\rangle$ denotes all next nearest neighbors. $\mathbf{d}_{ij}^{1,2}$ are nearest neighbor vectors traversed between NNN pairs $(i,s)$ and $(j,l)$, and $\boldsymbol{\sigma}$ is the vector of Pauli spin matrices. Because the vectors $\mathbf{d}_{ij}^{1,2}$ all lie within the $xy$ plane, only the $\sigma^z$ Pauli matrix appears in Eq.~(\ref{eq:SOCNNN}) and the spin-orbit Hamiltonian $H_{\text{SO}}$ decouples for the two spin projections along the $z$ axis. As such, the motif of NNN hoppings that are induced by the SOC [Fig.~\ref{figure6}(a)] resembles that of Haldane or Kane and Mele models of graphene~\cite{haldane1988model,kane2005z,kane2005quantum}.

Transforming Eq.~(\ref{eq:SOCNNN}) in momentum space, the spin-orbit Hamiltonian reads
\begin{eqnarray}
\label{eq:SOCNNNk}
\mathcal{H}^{\text{SO}}_{\textbf{k}} =\ && \pm \imath 2\lambda_{\text{NNN}}\\
&&\times
\begin{bmatrix}
0 & \cos{\textbf{k}\cdot(\textbf{a}_2+\textbf{a}_3)} & -\cos{\textbf{k}\cdot(\textbf{a}_3-\textbf{a}_1})\\
 & 0 & \cos{\textbf{k}\cdot(\textbf{a}_1+\textbf{a}_2})\\
 & & 0
\end{bmatrix}
\nonumber \,,
\end{eqnarray}
%
%
%
%
\noindent where the $+$ ($-$) sign refers to spin up (spin down) electrons. As shown in Fig.~\ref{figure6}(b), the SOC opens up a gap at the Dirac point. The magnitude of such a gap scales linearly with the strength of SOC, amounting to $\Delta_{\text{SO}} = 4\sqrt{3}|\lambda_{\text{NNN}}|$. Interestingly, $H_{\text{SO}}$ opens up a gap also at the touching point between the flat band $E^{(3)}$ and the dispersing band $E^{(2)}$ coming upwards from the Dirac point (Fig.~\ref{figure6}b). 

The Hamiltonian that governs the low-energy excitation in the vicinity of the $\textbf{K}$ point acquires a mass term $m_{\pm} = \pm 2\sqrt{3}\lambda_{\text{NNN}}$, taking the form $h_{\textbf{k}\pm} = v_{\text{F}}(\tau_3 k_x + \tau_1 k_y) + \tau_2 m_{\pm}$. Furthermore, performing a numerical diagonalization of the lattice Hamiltonian $\mathcal{H}^0_{\textbf{k}}+\mathcal{H}^{\text{SO}}_{\textbf{k}}$ in a strip geometry, one obtains a spectrum that is characterized by the presence of spin-filtered gapless states associated with each edge, as reported in Fig.~\ref{figure6}(c). These states traverse both bulk gaps that have been opened up by the SOC, and are hallmarks of a topologically nontrivial groundstate both at $\frac{1}{3}$ and $\frac{2}{3}$ filling. In this respect, with SOC, the electronic structure cannot be smoothly deformed into the one of a trivial insulator and, explicit calculations show that it possesses a nontrivial $\mathbb{Z}_{2}$ invariant~\cite{guo2009topological,kane2005z,fu2007topological}.

The block diagonal form of the spin-orbit Hamiltonian in Eq.~(\ref{eq:SOCNNN}), as far as the spin projection along the $z$ component is concerned, allows one to compute the $\mathbb{Z}_{2}$ invariant as the parity of the spin Chern number $C_{\text{s}}$ of the bands of interest
\begin{eqnarray}
\label{eq:SBC}
C_{\text{s}} = \frac{C_{\uparrow}-C_{\downarrow}}{2} = \frac{1}{2\pi}\int_{\text{BZ}} d\textbf{k}\,\Omega_z^{S_z}(\textbf{k})
\end{eqnarray}
\noindent where $C_{\sigma}, \sigma=\uparrow,\downarrow$, is the Chern number for the specific spin block of Eq.~(\ref{eq:SOCNNN}) and $\Omega_z^{S_z}(\textbf{k})$ is the spin Berry curvature (SBC)~\cite{xiao2010berry}. Figure~\ref{figure6}(b) demonstrates how the momentum space regions in proximity of the gaps opened up by SOC are characterized by a sizable SBC. Numerical integration of $\Omega_z^{S_z}(\textbf{k})$ as in Eq.~(\ref{eq:SBC}) gives $C_{\text{s}} = 1, 0, -1$ for bands $E^{(1)}$, $E^{(2)}$ and $E^{(3)}$, respectively. This analysis proves the topological nature of the ground state at $\frac{1}{3}$ and $\frac{2}{3}$ filling.

Beyond the simplified model, the argument given above is also valid for realistic kagome metals. To give one example, Fig.~\ref{figure6}(d) reports the first-principles band structure of TbV$_6$Sn$_6$, a member of the 166 family~\cite{rosenberg2022uniaxial,PhysRevLett.127.266401}. SOC opens up several gaps throughout the entire spectrum and, relevant to our discussion, specifically in the regions of the flat band at the \textbf{$\Gamma$} point and Dirac cones at the \textbf{K} point. The bands in the vicinity of those gaps feature a pronounced spin Berry curvature $\Omega_z^{S_z}(\textbf{k})$, even if the underlying electronic description is more complicated than that provided by the simple TB model of Eqs.~(\ref{eq:H0}) and (\ref{eq:SOCNNN}). Recently, the fingerprints of a finite spin Berry curvature in TbV$_6$Sn$_6$ kagome metal have been addressed experimentally by looking at spin resolved circular dichroism in the wide energy region of the occupied flat band, about 1~eV below the chemical potential~\cite{di2023flat}. The spin-asymmetry in both right- and left-polarized dichroic signals, shown in Fig.~\ref{figure6}(e), can be taken as indication of a finite spin Berry curvature~\cite{schuler_local_nodate}. As for the Dirac cone region, a spin-dependent orbital Zeeman shift of bands observed in quasiparticle interference (QPI) scanning tunneling microscopy (STM) was also attributed to a non-vanishing $\Omega_z^{S_z}$~\cite{li2024spin}.

Differing from the lattice model on the hexagonal lattice, electrons on the kagome lattice can also experience a first nearest neighbor (NN) spin-orbit interaction~\cite{tang2011high,ma2020spin}. The reason is that the electric field from neighboring ions felt during the nearest neighbor hopping on the kagome lattice is no longer perfectly compensated. Nevertheless, its presence does not qualitatively change the physical picture addressed above. Both NN and NNN SOC terms open up a gap at the Dirac point as well as at the flat band region, giving rise to a finite spin Berry curvature. 

The main quantitative difference between the two types of spin-orbit interactions appears when one looks at the symmetries of Bloch states at the high-symmetry points for the flat band, as dictated by its topological quantum chemistry~\cite{bradlyn2017topological,cano2018building}. Those are $\{ \bar{\Gamma}_7, \bar{\text{M}}_6, \bar{\text{K}}_9 \}$ and $\{ \bar{\Gamma}_8, \bar{\text{M}}_6, \bar{\text{K}}_9 \}$, depending on whether NN or NNN SOC is considered~\cite{ma2020spin}. However, in both cases, the symmetry-data vectors cannot be decomposed into a linear combination of EBRs which, being topologically equivalent to atomic orbitals, is topologically trivial. In this respect, both forms of spin-orbit interaction lead to topological flat bands, characterized by a nontivial $\mathbb{Z}_{2}$ invariant.  The coexistence of these topological flat bands and Dirac bands elevates the kagome lattice to one of the most interesting playgrounds in the study
of some fundamental physical phenomena, such as high-T$_c$ SC~\cite{zhao2021cascade,chen2021roton} and the fractional quantum Hall eﬀect~\cite{tang2011high,neupert2011fractional,sun2011nearly}.

\subsection{\label{sec:chapIIC}Phonons in the kagome lattice}

The fundamental equations that govern the physics of periodic lattices, whether they describe electronic, phononic or photonic degrees of freedom, show striking similarities, hinting at the possibility of a correspondence between phenomena occurring on a given lattice across different classes of systems from ionic crystals to meta materials~\cite{kane2014topological,huber2016topological,lu2014topological,yang2015topological,xue2022topological}. To quote few specific examples, the analogy between the TB Hamiltonians of electrons in a crystal, such as those introduced in the previous Section~\ref{sec:chapIIA}, and the stiffness (or dynamical) matrix of lattice vibrations, has led to the observation of mechanical topological helical states at the edge of a large lattice of coupled pendula~\cite{susstrunk2015observation}. The analogy between the electron Hamiltonian and the Laplacian of an electric circuit has fostered the prediction and discovery of topoelectrical circuits~\cite{lee2018topolectrical,imhof2018topolectrical}. In this framework, as for their electronic counterparts, mechanical vibrations of atoms, or phonons, in the kagome lattice feature rich physics. For instance, spring-mass models, consisting of a periodic arrangement of springs and point masses, serve as a simple platform to realize phenomena of topological nature governed by Newtonian equations of motion~\cite{kane2014topological}.

Within this description, the motion of point masses is described by the dynamics of a variable $\vec{u}_{i,s}$, which identifies the displacement of the mass in unit cell $i$ and sublattice $s=1,2,3$. Indeed, the notation used here is equivalent to that of TB Hamiltonians in Eqs.~(\ref{eq:H0}) and~(\ref{eq:SOCNNN}). For simplicity, assuming that all atoms in the lattice have the same unitary mass, the Lagrangian that governs the motion of point masses is that of coupled harmonic oscillators, reading
\begin{eqnarray}
\label{eq:Lphon}
\mathcal{L} = \ &&\frac{1}{2}\sum_{i,s,\mu} (\dot{u}_{i,s}^{\mu})^2 \\ 
&&-\frac{1}{2}\sum_{\langle (i,s)(j,l) \rangle}\sum_{\mu,\nu}(u_{i,s}^{\mu}-u_{j,l}^{\mu}) \mathcal{K}_{is,jl}^{\mu\nu} (u_{i,s}^{\nu}-u_{j,l}^{\nu}),  \nonumber
\end{eqnarray}
\noindent where, as usual, $\langle (i,s)(j,l) \rangle$ refers to the pairs of first-nearest-neighbors, and $\mu,\nu = x,y$ are cartesian coordinates in a two-dimensional space. $\mathcal{K}_{is,jl}^{\mu\nu}$ is the spring-constant matrix and encodes the geometry of the lattice. The Euler-Lagrange equations for $u_{i,s}^{\mu}$, \textit{i.e.} $d/dt(\partial \mathcal L/\partial \dot{u}_{i,s}^{\mu}) - \partial \mathcal L/\partial u_{i,s}^{\mu}=0$, give the equation of motion
\begin{eqnarray}
\label{eq:eqofmot}
\ddot{\vec{\boldsymbol{u}}} + \mathcal{D}\vec{\boldsymbol{u}} = 0
\end{eqnarray}
\noindent where $\vec{\boldsymbol{u}}$ is the column vector formed by $\vec{u}_{i,s}$ and $\mathcal{D}$ is the real-space dynamical matrix
\begin{eqnarray}
\label{eq:dynmat}
\mathcal{D}_{is,jl}^{\mu\nu} = -\mathcal{K}_{is,jl}^{\mu\nu} + \sum_{\langle (i,s)(k,o) \rangle}\mathcal{K}_{is,ko}^{\mu\nu}\delta_{ij}\delta_{sl} \,,
\end{eqnarray}
\noindent which is proportional to the spring-constant matrix and where the second term, clearly diagonal in the sublattice index, is needed to enforce the acoustic sum rule~\cite{grosso2013solid}. $\mathcal{D}$ is a real and symmetric matrix. If we assume that each mass point oscillates at frequency $\omega$, \textit{i.e.} $u_{i,s}^{\mu} = e^{i\omega t}\epsilon_{i,s}^{\mu}$, Eq.~(\ref{eq:eqofmot}) reads
\begin{eqnarray}
\label{eq:eqofmot2}
-\omega^2 \vec{\boldsymbol{\epsilon}} + \mathcal{D}\vec{\boldsymbol{\epsilon}} = 0
\end{eqnarray}
\noindent giving a real-space eigenvalue problem for the matrix $\mathcal{D}$, whose eigenbasis is $\vec{\boldsymbol{\epsilon}}$ and eigenvlues $\omega^2$. Further imposing translational invariance, $\epsilon_{i,s}^{\mu} = 1/N_{\textbf{q}}\sum_{\textbf{q}}e^{i\textbf{q}\cdot\textbf{R}_i}\epsilon_{\textbf{q},s}^{\mu}$, where $\textbf{q}$ belongs to the first BZ, Eq.~(\ref{eq:eqofmot2}) becomes
\begin{eqnarray}
\label{eq:eqofmot3}
-\omega_{\textbf{q}}^2 \epsilon_{\textbf{q},s}^{\mu} + \sum_{l} \Gamma_{sl}^{\mu\nu}(\textbf{q})\epsilon_{\textbf{q},l}^{\nu}= 0 \, ,
\end{eqnarray}
\noindent whose solution gives the phonon dispersion relation $\omega_{\textbf{q}}$ and phonon vibrational modes $\vec{\epsilon}_{\textbf{q},s}$. The matrix $\Gamma$ is named the momentum-space dynamical matrix, and appears as the lattice Fourier trasform of $\mathcal{D}$
\begin{eqnarray}
\label{eq:dynmat2}
\Gamma_{sl}^{\mu\nu}(\textbf{q}) = \sum_j \mathcal{D}_{0s,jl}^{\mu\nu}e^{i\textbf{q}\cdot\textbf{R}_j} \, ,
\end{eqnarray}
%
%
\begin{figure*}[!ht]
\centering
\includegraphics[width=\textwidth,angle=0,clip=true]{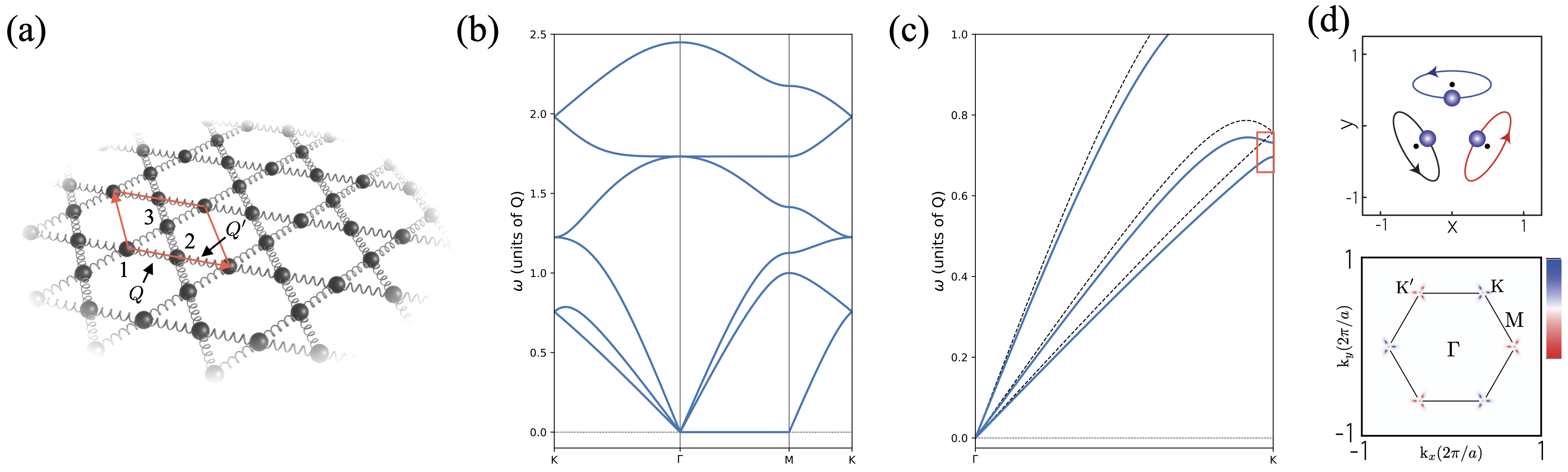}
\caption{(a) Nearest-neighbor spring mass model in the kagome lattice. The red lines identify the kagome unit cell with the three sublattice sites. $Q$ and $Q'$ set the strengths of the spring constant. (b) Phonon band dispersion in the $\mathcal{PT}-$symmetric $Q=Q'$ limit, featuring Dirac cones between the $1^{st}$ and $2^{nd}$ bands, as well as between the $5^{th}$ and $6^{th}$ bands. The simplified nearest-neighbor model also features a flat band along the $\Gamma$-\textbf{M} direction. (c) Zoom in the phonon dispersion when inversion symmetry $\mathcal{P}$ is broken by differentiating the intra-unit-cell $Q$ and inter-unit-cell $Q'$ spring constants. In the specific case, $Q'/Q = 0.8$. The dashed black line refers to the $\mathcal{P}-$symmetric case $Q'/Q = 1$ as in panel (b). (d) The top panel shows the phonon eigenmode $\vec{\boldsymbol{\epsilon}}_{\textbf{K},s}$ at the \textbf{K} point for each of the three $s=1,2,3$ sublattice sites for the $1^{st}$ phonon branch, highlighting the chiral nature of the atomic displacements. The bottom panel is the calculated phonon Berry curvature in momentum space for the same phonon branch. Panels (c) and (d) adapted from~\textcite{chen2019chiral}.}  
\label{figure7}
\end{figure*}
%
%
\noindent whose matrix-dimension depends only on the number of atoms in the unit-cell and it is independent cartesian coordinates. 

For the explicit case of interest here, \textit{i.e.}\ the spring-mass model on the nearest-neighbor kagome lattice depicted in Fig.~\ref{figure7}(a), the momentum-space dynamical matrix is a $6 \times 6$ matrix, owing to the three sublattices and the two spatial coordinates $x$ and $y$. The complete form of $\Gamma$ is
\begin{eqnarray}
\label{eq:Gamma}
\Gamma_{sl}^{\mu\nu}(\textbf{q}) = 
\begin{bmatrix}
D_1 & -\gamma_{12}(\textbf{q}) & -\gamma_{13}(\textbf{q})\\
 & D_2 & -\gamma_{23}(\textbf{q})\\
 & & D_3
\end{bmatrix}
\,,
\end{eqnarray}
\noindent with
\begin{eqnarray}
\label{eq:Gammapar}
&&\gamma_{12}(\textbf{q}) = Q
\begin{bmatrix}
1 & 0\\
0 & 0
\end{bmatrix}
(1+e^{-i\textbf{q}\cdot\textbf{a}_1}) \nonumber \\
&&\gamma_{13}(\textbf{q}) = Q
\begin{bmatrix}
\frac{1}{4} & \frac{\sqrt{3}}{4}\\
\frac{\sqrt{3}}{4} & \frac{1}{4}
\end{bmatrix}
(1+e^{-i\textbf{q}\cdot(\textbf{a}_1+\textbf{a}_2)}) \nonumber \\
&&\gamma_{23}(\textbf{q}) = Q
\begin{bmatrix}
\frac{1}{4} & -\frac{\sqrt{3}}{4}\\
-\frac{\sqrt{3}}{4} & \frac{1}{4}
\end{bmatrix}
(1+e^{-i\textbf{q}\cdot\textbf{a}_2}) \nonumber \, .
\end{eqnarray}
\noindent Denoting the above $2 \times 2$ matrices in $\gamma_{ij}(\textbf{q})$ as $\boldsymbol{A}$, $\boldsymbol{B}$ and $\boldsymbol{C}$ respectively, the diagonal blocks of $\Gamma$ read $D_1 = 2Q(\boldsymbol{A} + \boldsymbol{B})$, $D_2 = 2Q(\boldsymbol{B} + \boldsymbol{C})$ and $D_3 = 2Q(\boldsymbol{A} + \boldsymbol{C})$, in order to enforce the acoustic sum rules. The entries in $\boldsymbol{A}$, $\boldsymbol{B}$ and $\boldsymbol{C}$ originate from the projections of the Cartesian atomic displacements along the nearest-neighbor bonds of the kagome lattice. The parameter $Q$ determines the strength of the spring constant, as in Fig.~\ref{figure7}(a), and sets the overall energy scale (of lattice vibrations) in complete analogy to the hopping parameter $t$ in the electronic Hamiltonian of Eq.~(\ref{eq:H0}). The resulting phonon band structure is shown in Fig.~\ref{figure7}(b). It features a Dirac-like dispersion at the \textbf{K} point and VHS at the \textbf{M} point.

Despite its simplicity, this simple model of lattice vibrations on the kagome geometry can account for interesting nontrivial topological effects when specific symmetry breaking terms are included. For instance, differentiating between intra- $(Q)$ and inter-unit-cell $(Q')$ spring constants, the model breaks inversion symmetry $\mathcal{P}$, while maintaining the overall $C_3$ rotational symmetry. This has noticeable effects on the topological properties of the model. In fact, the Berry curvature is even under the action of $\mathcal{P}$, while it is odd under TRS $\mathcal{T}$. 

When both $\mathcal{P}$ and $\mathcal{T}$ are present, the Berry curvature vanishes. Breaking $\mathcal{P}$ via $Q \neq Q'$, a consequence of a weak trimerization seen in some kagome metals~\cite{nakatsuji2015large,zhang2023visualizing}, results in a gap opening of the Dirac cones at the \textbf{K} valleys shown by the red box in Fig.~\ref{figure7}(c). Interestingly, the vibration eigenmodes describe chiral phonons [top panel in Fig.~\ref{figure7}(d)]~\cite{chen2019chiral}, a concept theoretically predicted in the graphene-type honeycomb lattice~\cite{zhang2015chiral}, and experimentally verified in a monolayer of the transition metal dichalcogenide WSe$_2$~\cite{zhu2018observation}, where the phonon eigenmodes at the Brillouin
zone corner have a well-defined sense of chirality and quantized phonon pseudoangular momentum. 

In contrast to the honeycomb lattice, however, chiral phonons on the kagome lattice include motion of all three sublattices with the same chirality, and the vibration orbit can take an elliptical shape. Moreover, these chiral phonons possess a valley-contrasting Berry curvature that is sharply peaked at the \textbf{K} and \textbf{K$^\prime$} points, as depicted in the bottom panel of Fig.~\ref{figure7}(d). This is the basis of topological phonon Hall effect~\cite{strohm2005phenomenological,sheng2006theory,kagan2008anomalous,wang2009phonon,zhang2010topological}, \textit{i.e.}, the phonon analogue to the anomalous Hall effect (AHE) in electron systems. Besides chiral phonons, variants of the simple nearest-neighbor spring-mass model can also support mechanical analogues to the quantum spin Hall effect~\cite{chen2018elastic} and higher-order topological phases~\cite{wakao2020higher}.

When it comes to the study of phonon dispersions in actual kagome materials, quantum mechanical first-principles methods are preferred to simplified spring-mass models. To substantiate this, we briefly discuss two notable cases of $A$V$_3$Sb$_5$ and ScV$_6$Sn$_6$ kagome metals, for which an extensive \textit{ab initio} investigation has been performed to unveil the origin of their CDW instabilities.


\begin{figure*}[!ht]
\centering
\includegraphics[width=\textwidth,angle=0,clip=true]{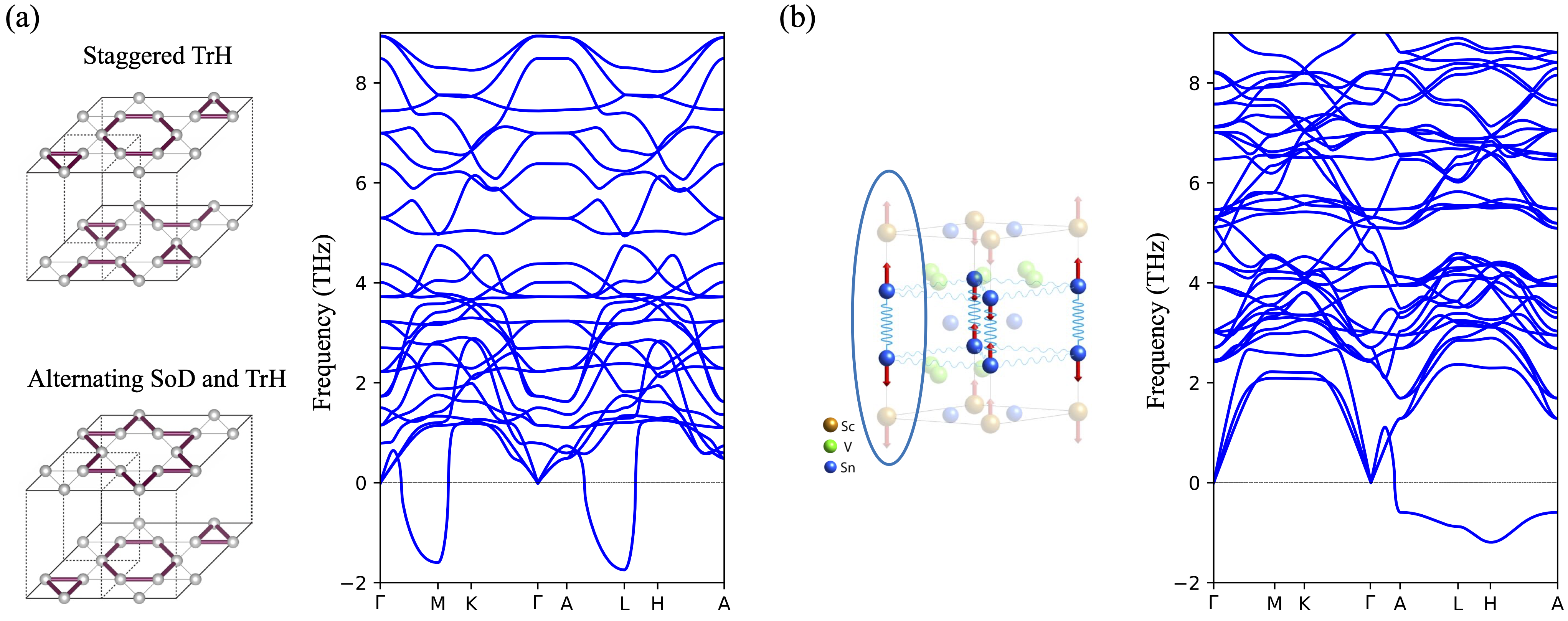}
\caption{(a) \textit{Ab initio} phonon band dispersion of CsV$_3$Sb$_5$, characterized by imaginary phonon frequencies at the \textbf{M} and \textbf{L} points. The crystal structures show the expected in-plane modulation of vanadium atoms for (K,Rb)V$_3$Sb$_3$ (staggered TrH) and CsV$_3$Sb$_3$ (alternating SoD and TrH). (b) \textit{Ab initio} phonon band structure of ScV$_6$Sn$_6$, characterized by a flat imaginary phonon band whose eigenmode primarily affects the out-of-plane vibration of Sn and Sc atoms. Panels adapted from~\textcite{kang_charge_2023,PhysRevB.111.054113}.}
\label{figure8}
\end{figure*}


For the first example, an in-plane lattice distortion of the kagome lattice is often energetically favorable and characterized via a multi-$q$ distortion at inequivalent \textbf{M}-points.  Interactions between the lattice planes are possible in three-dimensional compounds, leading to a rich spectrum of possible distortion modes.  As reviewed in more detail in section~\ref{sec:chapIVB2}, $A$V$_3$Sb$_5$ kagome compounds exhibit such a CDW instability ~\cite{PhysRevLett.125.247002,PhysRevX.11.041030,PhysRevX.11.031050} embodied by these energetically favored breathing modes in the kagome plane~\cite{PhysRevLett.127.046401,consiglio2022van}. The in-plane distortion is identified by a breathing mode into a Star-of-David (SoD) or tri-hexagonal (TrH) modulation. \textit{Ab initio} phonon modeling fully captures both SoD and TrH patterns, as shown in Fig.~\ref{figure8}(a), with the preferred distortion modes including in-plane \textbf{M}-point modes, along with \textbf{L}-point modes that add an out-of-plane modulation to the CDW structure~\cite{PhysRevLett.127.046401,PhysRevB.104.214513}. 

The energy differences between various distortion types are small, with the predicted distortion mode being the staggered TrH arrangement, which involves 3\textbf{q}=(\textbf{M}, \textbf{L}, \textbf{L}) modes for (K,Rb)V$_3$Sb$_5$, and an alternating SoD and TrH arrangement, which comprises 3\textbf{q}=(\textbf{L}, \textbf{L}, \textbf{L})+(\textbf{M}, \textbf{M}, \textbf{M}) modes for CsV$_3$Sb$_5$~\cite{kang_charge_2023} [inset of Fig.~\ref{figure8}(a)]. First-principles calculations of the phonon dispersion of $A$V$_3$Sb$_5$ kagome metals under external pressure~\cite{consiglio2022van,zhang2021first} were also able to faithfully reproduce the observed reduction of the CDW distortion~\cite{wang2021competition,PhysRevB.103.L220504,PhysRevLett.126.247001,Yu2021}, diagnosed in the calculation by a decrease, and eventual disappearance, of the imaginary phonon softening at the \textbf{M} and \textbf{L} points.

As a second example, another common distortion type is an out-of-plane chain instability that can couple to the kagome plane in a secondary manner. Discussed in greater depth in section~\ref{sec:chapIVB3}, ScV$_6$Sn$_6$ is an experimental realization of this distortion-type~\cite{Arachchige2022Sc166}. Here, distinct from the previous example of AV$_3$Sb$_5$, ScV$_6$Sn$_6$'s distortion is primarily characterized by a displacement of out-of-plane Sc and Sn atoms, accompanied by a weak modulation of the kagome-network V atoms. This is driven by the electron-phonon interaction and the softening of a flat phonon mode related to an out-of-plane vibration of Sn atoms~\cite{Lee2024Sc166_arpes,Korshunov2023flat_phonon_Sc166,cao2023competing,Hu2024MingShi_Sc166,tuniz2023dynamics}. As shown in Fig.~\ref{figure8}(b), the first-principles calculation of the phonon spectrum of ScV$_6$Sn$_6$ agrees well with the experimental observations, predicting the presence of an unusually flat imaginary phonon band, whose eigenmode weight comprised of the out-of-plane vibration of Sn and Sc atoms~\cite{tuniz2023dynamics,tuniz2024strain,PhysRevB.111.054113}.


\section{\label{sec:chapIII}Many-body phenomena}

The kagome lattice, with its inherent geometric frustration and intriguing electronic structure, provides an exceptional playground for many-body interactions that give rise to a diverse landscape of emergent quantum phenomena. Beyond the single-particle band structure features explored in the previous sections, kagome metals exhibit rich physics driven by electron-electron and electron-lattice interactions. These many-body effects not only modify the electronic states but can also stabilize novel ordered phases that are highly sensitive to doping, temperature, and external perturbations.

Here we discuss the theoretical aspects behind key many-body phenomena observed in kagome metals. We begin with a discussion on CDWs in Section~\ref{sec:chapIIIA}, which often emerge in response to Fermi surface nesting and lattice instabilities, manifesting as periodic modulations in the electronic density. Following this, in Section~\ref{sec:chapIIIB} we delve into SC, a central theme in the study of kagome systems, particularly in compounds like $A$V$_3$Sb$_5$, where unconventional pairing mechanisms and nontrivial topology are believed to play a role.

In Section~\ref{sec:chapIIIC}, we then examine the more exotic concept of a PDW, wherein the superconducting order parameter itself exhibits spatial modulation, an intriguing possibility in systems with competing orders and strong correlations. The role of electron-phonon interactions is also scrutinized in Section~\ref{sec:chapIIID}, especially given their relevance to both CDW formation and SC. Finally, in Section~\ref{sec:chapIIIE} we consider electronic correlations and fluctuating local moments, which underpin much of the emergent behavior in kagome systems, contributing to magnetic fluctuations and possible non-Fermi liquid behavior.


\subsection{\label{sec:chapIIIA}Charge orders}

We start with an imperative comment on the terminology of different orders. The systems of interest are often multi-band (with electron and hole bands), multi-orbital, and moderately spin-orbit coupled. Therefore, CDWs and SDWs, flux orders, and even exciton condensates are not sharply differentiated concepts. We will therefore use the umbrella term charge orders here. All of them are electron-hole-condensates, \textit{i.e.}, the order parameter takes the general form
\begin{equation}
    \Delta_{\alpha,\beta; \textbf{q}}(\textbf{k})\sim\left\langle
c^\dagger_{\textbf{k}+\textbf{q},\alpha}
c_{\textbf{k},\beta}
    \right\rangle,
\end{equation}
where $\alpha$ and $\beta$ are compound indices including spin, orbital and sublattice degrees of freedom. The center of mass momentum $\textbf{q}$ is set subscript to indicate that the order parameter is typically nonzero only for a small set of discrete $\textbf{q}$, while it has a continuous dependence on the relative momentum $\textbf{k}$ \cite{Nayak00}.

In general charge orders can also appear at $\textbf{q}=0$, as long as they break other symmetries besides translation (e.g., in nematic phases if the flux order in the Haldane model is spontaneously generated).~\footnote{This distinction of $\textbf{q}=0$ and finite $\textbf{q}$ phases is more widespread in SC, where a finite $\textbf{q}$ order is a Fulde–Ferrell–Larkin–Ovchinnikov (FFLO) \cite{Fulde64,osti_4653415} phase or PDW \cite{Agterberg20}, while a $\textbf{k}$-dependence that breaks lattice symmetries is often referred to as unconventional superconductors with nonzero angular momentum (`\textit{p}-wave', `\textit{d}-wave', etc.)} 
The $\textbf{k}$-dependence, if breaking lattice symmetries by itself, would justify calling the charge order unconventional and may be categorized by a finite `angular momentum' that it carries (which of course is only meaningful modulo the rotation symmetry of the crystal). A nontrivial $\textbf{k}$-dependence is typically associated with an order parameter that renormalizes -- in a TB picture -- the hopping integrals, rather than only onsite-terms, and is therefore also referred to as a bond order.  

Another important symmetry distinction is whether the order breaks TRS. Time-reversal symmetry breaking through complex hopping amplitudes will induce spontaneous ring currents, and is therefore also dubbed ``flux phase''. Such orbital magnetism would, due to SOC, in general also induce spin magnetism and therefore a spin-density wave. Whether a phase should be termed flux order or spin-density wave depends on the primary driver for the time-reversal breaking order, but this is not always a well-defined question. 

Among the charge ordering phenomena, those occurring in the $A$V$_3$Sb$_5$ family are most studied and perhaps the most puzzling~\cite{PhysRevLett.127.046401, PhysRevB.104.214513, PhysRevB.106.144504, PhysRevB.107.205131, enzner2025phononfluctuationdiagnosticsorigin}. We will therefore focus our description primarily on effective theories applying to the situation found there, namely two-dimensional orders on the Kagome lattice with a $2\times 2$ enlarged unit cell \cite{Grandi23,Wagner23}. Building on these, we can comment on how these orders couple into three-dimensional structures. Such a symmetry-based description can be connected to microscopic models~\cite{PhysRevB.106.144504}. 

We restrict our discussion to the spinless case, focusing on the point group $C_{6v}$ instead of the double groups. We focus on ordering momenta $\textbf{q}\in\{\textbf{M}_1,\textbf{M}_2,\textbf{M}_3\}$ corresponding to the three \textbf{M} points. This enlarges the point group by four translations acting within the enlarged unit cell. We denote the new group by $C_{6v}'''$, its character table is listed in Table~\ref{tab:CT}. Any (spinless) charge order with translation symmetry breaking according to a $2\times 2$ period in the two-dimensional plane will fall into one of the irreducible representations $F_i,\ i=1,2,3,4$, which are all of dimension three. We further denote by $F'_i,\ i=1,2,3,4$ TRS breaking cousins of these orders. Determining which of $F_i$ or $F'_i$ is realized in a given system is thus the primary goal in understanding it. 

\begin{table}[t]
\begin{tabular}{c|c|c|c|c|c|c|c|c|c|c}
                               & $I$ & $t_i$ & $C_2$ & $t_iC_2$ & $C_3$ & $C_6$ & $\sigma_v$ & $t_i\sigma_v$ & $\sigma_d$ & $t_i\sigma_d$ \\ \hline
$|\mathcal{C}|$ & 1   & 3     & 1     & 3        & 8     & 8     & 6          & 6             & 6          & 6             \\ \hline
$A_1$                          & 1   & 1     & 1     & 1        & 1     & 1     & 1          & 1             & 1          & 1             \\
$A_2$                          & 1   & 1     & 1     & 1        & 1     & 1     & -1         & -1            & -1         & -1            \\
$B_1$                          & 1   & 1     & -1    & -1       & 1     & -1    & 1          & 1             & -1         & -1            \\
$B_2$                          & 1   & 1     & -1    & -1       & 1     & -1    & -1         & -1            & 1          & 1             \\
$E_1$                          & 2   & 2     & -2    & -2       & -1    & 1     & 0          & 0             & 0          & 0             \\
$E_2$                          & 2   & 2     & 2     & 2        & -1    & -1    & 0          & 0             & 0          & 0             \\\hline
$F_1$                          & 3   & -1    & 3     & -1       & 0     & 0     & 1          & -1            & 1          & -1            \\
$F_2$                          & 3   & -1    & 3     & -1       & 0     & 0     & -1         & 1             & -1         & 1             \\
$F_3$                          & 3   & -1    & -3    & 1        & 0     & 0     & 1          & -1            & -1         & 1             \\ 
$F_4$                          & 3   & -1    & -3    & 1        & 0     & 0     & -1         & 1             & 1          & -1           
\end{tabular}
\caption{Character table of the group $C_{6v}'''$ \cite{Venderbos}. The one- and two-dimensional irreducible representations preserve the translation symmetry of the original kagome lattice, while the three-dimensional irreducible representations lead to a $2\times2$ increase in the unit cell.}
\label{tab:CT}
\end{table}

\begin{figure}[t]
\includegraphics[width=\columnwidth]{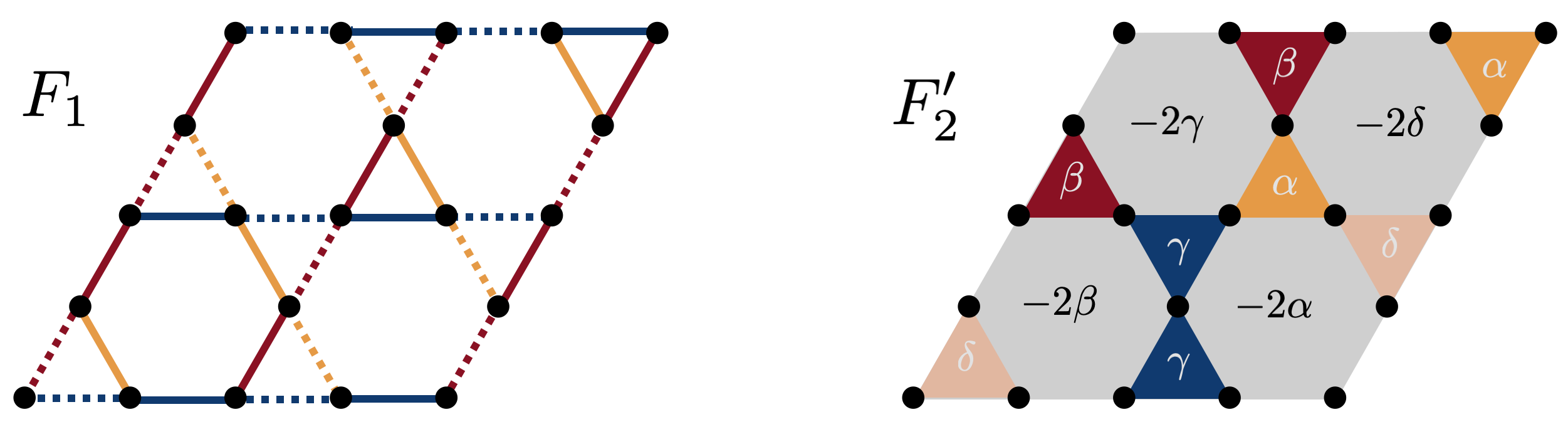}
\caption{ Examples of orders with $2\times2$ enlarged unit cell. (left) Bond order pattern in irreducible representation $F_1$. Straight and dotted lines represent strong and weak bonds, respectively. The three colors represent different amplitudes of the bond modulations that together represent the three-dimensional order parameter of $F_1$. If all amplitudes are chosen equal, one obtains the tri-hexagonal (TrH) and Star-of-David (SoD) orders, for positive and negative value of the order parameter, respectively. (right) Flux order pattern in irreducible representation $F_2'$, where $\alpha,\ \beta,\ \gamma, \delta$ represent fluxes subject to the constraint $\alpha+\beta+\gamma+\delta=0$. With this, they represent the three-dimensional order parameter space  of $F_2'$. Note that, even though the magnetic flux through the $2\times2$ unit cell vanishes, this does not imply that the (orbital) magnetic moment of the state vanishes (a famous example where this is not the case is the Haldane model \cite{haldane1988model}).}  
\label{figure9}
\end{figure}

Group-theory analysis \cite{Wagner23, Venderbos} shows that on-site density orders on the kagome lattice sites can realize the irreps $F_1,\ F_3$, and $F_4$, while bond orders on the nearest-neighbor kagome bonds can realize any $F_i$, $i=1,2,3,4$. Finally, flux orders on the plaquettes outlined by the nearest-neighbour kagome bonds can realize $F_2^\prime$ and $F_4^\prime$. See Fig.~\ref{figure9} for exemplary pictorial representations of these orders. 

The effective free energy governing these orders has two remarkable features. First, it may contain third-order terms in the order parameters. This is enabled by the fact that the three momenta corresponding to the three components of the irreducible representations $F^{(\prime)}_i$, $i=1,2,3,4$, add up to $0=\textbf{M}_1+\textbf{M}_2+\textbf{M}_3$, allowing for terms that transform trivial under translation. Considering the other symmetry constraints, such third-order terms exist only for $F_1$ individually, reading
\begin{equation}
\Delta_1\Delta_2\Delta_3,
\label{eq: delta cube}
\end{equation}
where $(\Delta_1,\Delta_2,\Delta_3)$ is the three-dimensional vector order parameter associated with the three-dimensional irreducible representation $F_1$. The three components carry momentum $\textbf{M}_1,\ \textbf{M}_2,\ \textbf{M}_3$, respectively. The term Eq.~\eqref{eq: delta cube} in the free energy renders the transition first order, irrespective of whether the microscopic origin of the order is phononic or electronic. In addition, the following coupling terms between $F_1$ order and $F_2^\prime$ or $F_4^\prime$ are allowed
\begin{equation}
\Delta_1\Delta'_2\Delta'_3
+
\Delta'_1\Delta_2\Delta'_3
+
\Delta'_1\Delta'_2\Delta_3,
\label{eq: delta delta prime}
\end{equation}
where $(\Delta'_1,\Delta'_2,\Delta'_3)$ is the three-dimensional order parameter of either $F_2'$ or $F_4'$. Consider a scenario where the $F_1$ bond order is the primary instability, and $F_2'$ or $F_4'$ have a comparably lower transition temperature. The effect of Eq.~\eqref{eq: delta delta prime} in the free energy is to renormalize the transition temperature to higher values -- the bond order thus promotes a secondary flux order. This is not the case for bond orders $F_2$, $F_3$, and $F_4$.

Some general conclusions can be drawn for the coupling of these orders to magnetic fields and in-plane strain. The lowest order coupling term between an out-of-plane magnetic field  $B$ and the mentioned bond and flux orders is one that mixes a bond and a flux order, namely
\begin{equation}
B(\Delta_1\Delta'_1
+
\Delta_2\Delta'_2
+
\Delta_3\Delta'_3).
\label{eq: B delta delta prime}
\end{equation}
It is only symmetry allowed for the combination of orders $(F_1,F_2')$ and $(F_3,F_4')$. If we again consider the situation of a well-established $F_1$ order, for instance, the effect of Eq.~\eqref{eq: B delta delta prime} for $B\neq 0$ is to admix the $F_2'$ flux order, even though this might not have an instability by itself (since it effectively acts as a first order term for $F_2'$). This creates the intriguing phenomenology that a \emph{homogeneous} magnetic field can be used to induce a staggered flux order and reversal of the field orientation reverses the sign of the staggered flux. This is enabled by the presence of the $F_1$ order, which, due to its symmetry breaking, renders the combination of $F_1$ and $F_2'$ not perfectly ``antiferromagnetic" or free of net magnetization, but induces a finite magnetic moment in $F_2'$ to which $B$ can couple. A second feature of a phase with coexisting $F_1$ and $F_2'$ orders is that it becomes anisotropic (in the kagome plane). This is true both for a mixed phase due to the coupling Eq.~\eqref{eq: delta delta prime} or one brought about by a magnetic field via Eq.~\eqref{eq: B delta delta prime}. The latter scenario leads to a counterintuitive response: one starts with an isotropic $F_1$ order and applies a magnetic field that also has the full rotation symmetry of the kagome lattice (and more). Yet, the resulting state is anisotropic.  


In-plane strain explicitly breaks the rotation symmetry (down to two-fold). However, the impact this explicit symmetry breaking has on the (electronic) response of the system depends on the susceptibility of the system towards such an isotropy-breaking deformation. Since anisotropy arises from a combination of $F_1$ and $F_2'$ orders as explained above, an enhanced strain response of the system appears if it is proximate to a phase where  $F_1$ and $F_2'$ orders coexist. 

In this way, responses of the system to external perturbations can be used to systematically argue about the irreducible representation(s) that its symmetry-breaking pattern realizes. They are independent of whether phonons or electronic interactions are the main driver of the instability.   
In this discussion, we ignored the three-dimensional nature of the charge order for simplicity. Considering orders with nonvanishing momentum (translation symmetry breaking) perpendicular to the kagome planes will induce further constraints on possible terms in the free energy. For instance, if one only considers order parameters with $\pi$ momentum in this direction (and the same in-plane momenta, \textit{i.e.}, $L$-point orders), one obtains no third-order terms in the free energy as they would violate momentum conservation. The unusual coupling of bond and flux orders to the magnetic field remains, however.

For instance, in the the $A$V$_3$Sb$_5$ family of compounds, experimental observations place clear constraints on the order that is realized. In \cite{Guo24}, by perturbing CsV$_3$Sb$_5$ in a controlled way by strain and magnetic field, it has been argued that the material is close to a phase where $F_1$ and $F_2'$ orders coexist, but in its pristine form realizes pure $F_1$ order. In \cite{Xing2024opticalM} RbV$_3$Sb$_5$ was argued to have a coexisting $F_1$ and $F_2'$ order, and the specific configuration of the respective three-dimensional order parameters has been determined.


\subsection{\label{sec:chapIIIB}Superconductivity}

\begin{figure*}[!t]
\centering
\includegraphics[width=0.95\textwidth,angle=0]{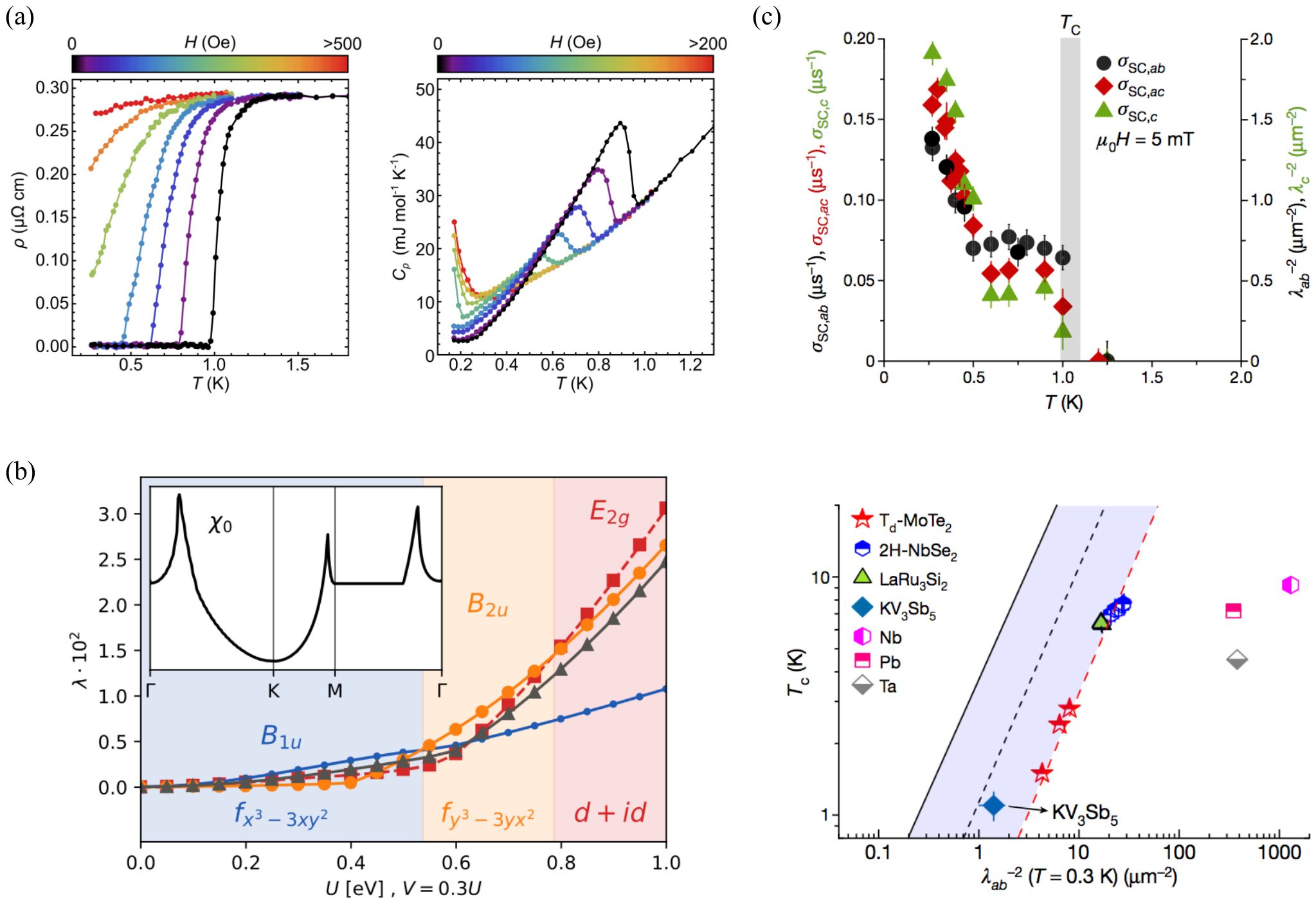}
\caption{SC in kagome metals. (a) Original resistivity and specific heat data obtained for KV$_3$Sb$_5$ featuring a T$_c$=0.94 K. (b) Theoretical predictions of $f$-wave and chiral $d$-wave for an all-electronic mechanism of SC in kagome metals. (c) up: $\mu$SR scattering rates and penetration depth measurements for CsV$_3$Sb$_5$ down: Uemura plot in comparison to other hexagonal superconductors, cuprates bounds (hole doping straight black line, electron doping dashed black line), and traditional Bardeen–Cooper–Schrieffer (BCS) superconductors. Panels adapted from~\textcite{wu2021nature,mielke2022time,PhysRevMaterials.5.034801}.}  
\label{sc-figure}
\end{figure*}

In the same way the community has tried to reconcile high-T$_c$ cuprates from the perspective of doping a magnetically ordered Mott insulator~\cite{RevModPhys.78.17}, kagome metals suggest a scenario where a high-temperature itinerant charge order may seed superconducting order~\cite{Neupert-review-2022}. A rather low T$_c$ has so far been recorded for kagome metals at ambient pressure~\cite{PhysRevLett.125.247002,PhysRevMaterials.5.034801,Yang2024CTB135_STM}, ranging between 0.8---4~K depending on the specific compound and the sample quality (Fig.~\ref{sc-figure}a). This explains why a majority of experimental activity has so far focused on exploring the nature of the high-temperature charge order rather than such low-T$_c$ phases that evade certain spectroscopic methods. Furthermore, the intricacy of charge ordering cascades as a function of temperature renders it difficult to isolate an effective electronic model that SC originates from.  

In the absence of strong magnetic fluctuations as observed in the majority of kagome metal compounds with local magnetic moments, phonon-mediated SC is a common suspect and cannot be excluded even though the electron-phonon coupling is estimated to be too small for the observed T$_c$ in the 135 kagome metals~\cite{PhysRevLett.127.046401}. From an electronically mediated microscopic mechanism of SC, kagome metals present themselves as weakly to intermediately correlated electron systems, and suggest a high relevance of the VHS nearby the Fermi level~\cite{wu2021nature,romer2022superconductivity}, combined with sublattice interference~\cite{kiesel_sublattice_2012}, to explain the nature of superconducting pairing (Fig.~\ref{sc-figure}b). Either way, the complicated nature of charge order and its concomitant Fermi surface reconstruction has rendered it a difficult task to accomplish studies of superconducting order starting from a microscopically accurate charge ordered itinerant parent state.  From an Uemura plot of T$_c$ versus penetration depth, $\mu$SR locates this kagome metal in the domain of unconventional SC (Fig.~\ref{sc-figure}c).


\begin{figure*}[!t]
\centering
\includegraphics[width=0.95\textwidth,angle=0,clip=true]{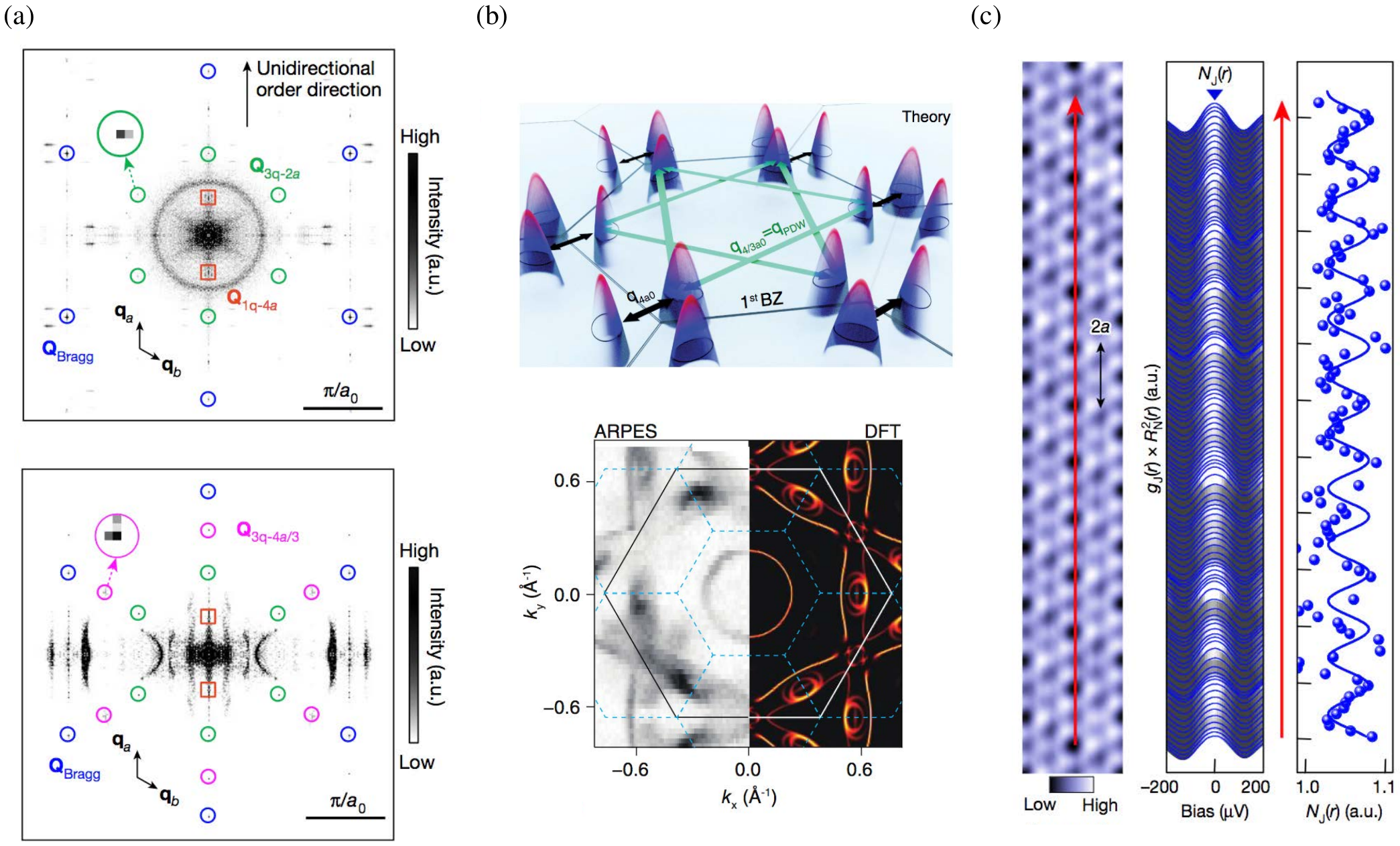}
\caption{Pair denstiy wave order in kagome metals. (a) Original observation of onset ordering vectors $Q_{3q-4a/3}$ in the superconducting phase of CsV$_3$Sb$_5$. (b) Fermi pocket reconstruction from charge order; the nesting features suggest similarities to the observed onset order below $T_c$. (c) Josephson tunneling spectroscopy along a clean Sb-terminated CsV$_3$Sb$_5$ crystal, where the spatial modulation of the SC pairing function becomes visible.  Panels adapted from~\textcite{PhysRevX.13.031030,Deng-Josephson-Nature-2024,chen2021roton}.} 
\label{pdw-figure}
\end{figure*}

From a synoptic viewpoint, the nature of SC in kagome metals appears to be strongly compound- and family-dependent, and to navigate in between an all-phononic and all-electronic pairing mechanism. It is likely that SC  can only be quantitatively accounted for through taking into account phonons and electronic correlations on a similar footing~\cite{PhysRevB.109.075130}. 


\subsection{\label{sec:chapIIIC}Pair Density Wave}
A PDW is a SC condensate that also breaks translation symmetry besides particle conservation. 
The principle found its original conception in the cuprates~\cite{Agterberg20}.
One may differentiate between a strict and a more inclusive definition of a PDW. In the strict reading, one would require the translation symmetry to be restored at temperatures right above the SC transition, while in the more inclusive picture, SC appearing within a translation breaking ordered phase (such as charge order) is also termed PDW, if a noticeable modulation of the SC order parameter is imprinted from the higher-temperature order. 
While PDWs had evaded unambiguous observation for a significant amount of time, recent discoveries in cuprates~\cite{Du-PDW-Nature2020}, UTe$_2$~\cite{Gu-PDW-Nature2023}, transition metal dichalcogenides such as NbSe$_2$~\cite{liu2021discovery}, and, in particular, the kagome metal CsV$_3$Sb$_5$~\cite{chen2021roton,Deng-Josephson-Nature-2024} have significantly broadened the empirical basis for alleged PDWs. 

A prototypical signature of a PDW results from the STM analysis of the pairing function above and below $T_c$, where the intertwined onset of SC and translation symmetry breaking would show up through additional Fourier moments at $T_c$. For CsV$_3$Sb$_5$, Fig.~\ref{pdw-figure}a depicts a comparison between the Fourier-tansformed STM topography (up) versus the Fourier-transformed dI/dV map at 300 mK (down), \textit{i.e.}, deep in the superconducting regime from~\cite{chen2021roton}. The additional purple peaks represent the additional translational symmetry breaking at $Q_{3q-4a/3}$ assigned to the superconducting condensate, in line with the Josephson tunneling profile along a clean Sb surface line depicted in Fig.~\ref{pdw-figure}c from~\cite{Deng-Josephson-Nature-2024}. 
From theory, sublattice interference~\cite{kiesel_sublattice_2012} has been claimed to be pivotal in explaining the system's propensity for the formation of PDW SC~\cite{PhysRevB.108.L081117}, while other mechanisms ascribe pivotal relevance to the tentative formation of loop current order~\cite{Zhou2022pockets}.

To date, a key challenge in the unambiguous experimental identification of a PDW -- in the strict sense mentioned above -- is the exclusion of a possible preceding reconstruction with the same ascribed Fourier moment above $T_c$. Ideally, Fourier data would be necessary as a function of temperature, scanned from above all the way to below $T_c$, and suggests the need for additional future exploration. In kagome metals, the charge-ordered parent state within which the PDW appears is particularly intricate, and exhibits a cascade of translation symmetry breaking through charge order. Notably, the reconstructed charge order Fermi pockets trigger a rather similar Fourier moment as to what has been ascribed to the PDW (Fig.~\ref{pdw-figure}b from~\cite{PhysRevX.13.031030}).

Instilled by the non-trivial unit cell of the kagome lattice, alternative proposals such as sublattice-modulated superconducting pairing have been proliferating~\cite{PhysRevB.110.024501}. There, the superconducting order experiences modulations within the unit cell; they accordingly do not break translation symmetry but still yield a site-modulated profile similar to a PDW breaking other crystalline symmetries.

\subsection{\label{sec:chapIIID}Electron-phonon Interaction}
In this section, we provide a brief discussion of the impact of phonons on many-body states in kagome metals with AV$_3$Sb$_5$ compounds used as an experimental touchstone. Although various superconducting pairing symmetries have been proposed, recent ARPES experiments have revealed distinct evidence of nodeless superconducting gaps in AV$_3$Sb$_5$ compounds~\cite{Zhong2023}, placing significant constraints on the nature of the pairing mechanism. In addition, multi-boson kinks in photoemission spectra were observed in both CsV$_3$Sb$_5$ and KV$_3$Sb$_5$ systems~\cite{zhong2023testing,luo2022electronic,wu2023unidirectional,luo2023unique}, showing contrasting behaviors across different electron bands, with some predominantly exhibiting a single kink and others displaying double kinks. These low-energy excitations are often attributed to strong electron-phonon $(e\text{-}ph)$ coupling~\cite{engelsberg1963coupled,hengsberger1999electron}, although the connection between photoemission kinks and emergent electronic orders is not always straightforward. A well-known example is provided by copper-oxide superconductors, where the $e\text{-}ph$ interaction accounts for the $\sim 70$ meV photoemission kinks but is generally regarded as too weak to explain the high $T_c$ and inconsistent with the nodal $d$-wave pairing symmetry~\cite{li2021unmasking}.


Initial first-principles calculations using DFT indicated a weak to moderate overall $e-ph$ coupling $\lambda \approx 0.25$ in CsV$_3$Sb$_5$ at both ambient and elevated pressures~\cite{zhang2021first,wang2023phonon,PhysRevLett.127.046401}, failing to support the superconducting transition temperature $T_c \approx 2.6$ K and hinting towards an unconventional electron-electron pairing mechanism for SC. Nevertheless, recent low-energy laser-based ARPES measurements reported evidence of anomalies (kinks) in the intensity and dispersion of the spectra for the Sb $5p$ and V $3d$ electronic bands near a binding energy of $\approx 32$ meV. 

\begin{figure*}[!t]
\centering
\includegraphics[width=0.75\textwidth,angle=0,clip=true]{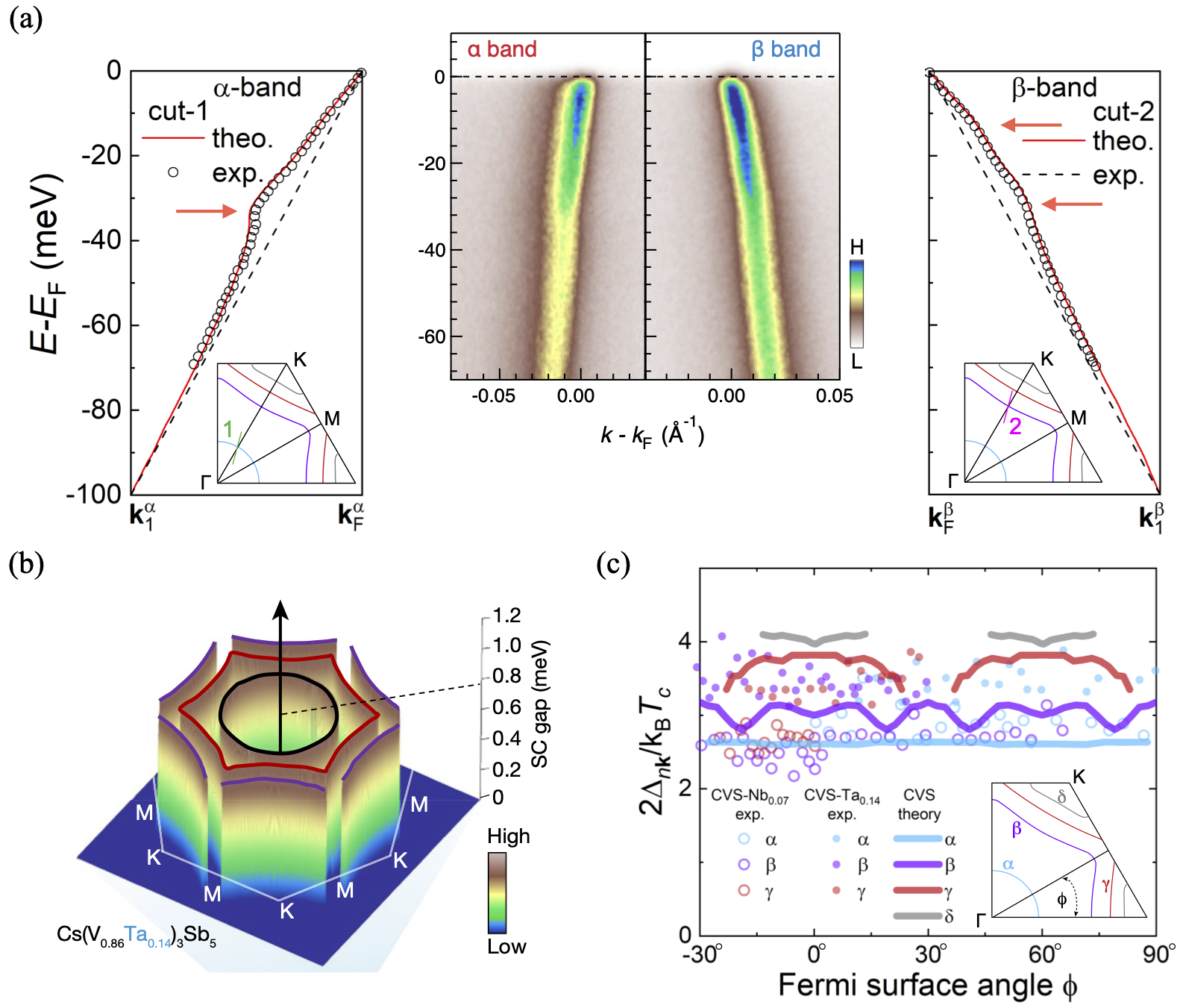}
\caption{(a) ARPES intensity plot in the central panel, and momentum distribution cut-derived dispersion relation between experiment (open circles) and $GW$ theory (red lines) in the left and right panels for the $\alpha$ and $\beta$ bands. The red arrows highlight the positions of the kinks in the photoemission data. (b)  Schematic momentum dependence of the superconducting gap amplitude of the Cs(V\(_{0.86}\)Ta\(_{0.14}\))\(_3\)Sb\(_5\) sample. (c) Calculated and experimental angle distribution of the rescaled superconducting gaps $2 \Delta /k_BT_c$ on multiple Fermi surface sheets of CsV\(_3\)Sb\(_5\)-derived compounds within the $k_z = 0$ plane. Panels adapted from~\textcite{zhong2023testing,Zhong2023,you2024diverse}.}  
\label{figure10}
\end{figure*}

Fig.~\ref{figure10}(a) (central panel) shows the ARPES intensity plots for the so-called $\alpha$ (Sb $5p$-derived) and $\beta$ (V $3d$-derived) bands, where $e-ph$ coupling-induced kinks are clearly visible. While the kink around $\approx 32$ meV is evident on both $\alpha$ and $\beta$ bands, an additional kink appears at a lower binding energy of $\approx 12$ meV only on the $\beta$ band. State-of-the-art $GW$-based many-body perturbation calculations demonstrated that $e-ph$ coupling is the universal origin of those multi-boson photoemission kinks~\cite{you2024diverse}. The simulated spectral functions shown in Fig.~\ref{figure10}(a) (left and right panels) reproduce the salient features of the experimental data with clear signatures of phonon-induced electron self-energy effects in the form of dispersion kinks (highlighted by the red arrows) and spectral width broadening. In fact, the $\alpha$-band displays a single kink around $-32$ meV, whereas the $\beta$-band shows double kinks at $-12$ and $-30$ meV, respectively, agreeing well with the experimental measurements. The calculated atom-vibration resolved real and imaginary parts of the $e-ph$ self-energy attribute the kinks at $\approx -30$ meV and $-12$ meV primarily to V and Sb atom vibrations.


Ultra high-resolution, low-temperature ARPES experiments have also directly observed a nodeless, nearly isotropic, and orbital-independent superconducting gap in momentum space for two representative CsV\(_3\)Sb\(_5\)-derived kagome superconductors: Cs(V\(_{0.93}\)Nb\(_{0.07}\))\(_3\)Sb\(_5\) and Cs(V\(_{0.86}\)Ta\(_{0.14}\))\(_3\)Sb\(_5\)~\cite{Zhong2023}. Fig.~\ref{figure10}(b) gives a schematic momentum dependence of the superconducting gap amplitude. Notably, this gap structure remains unaffected by the presence or absence of charge order in the normal state~\cite{Zhong2023}. In agreement with the observed band dispersion kinks, the robust isotropic superconducting gaps with small $2 \Delta /k_BT_c$ in the presence or absence of the CDW seem to be consistent with a conventional $s$-wave and $e-ph$ coupling-driven pairing. This interpretation is further reinforced by the \( GW \)-based calculations reported in Fig.~\ref{figure10}(c)~\cite{you2024diverse}, which, within the framework of anisotropic Eliashberg theory, predict an "almost isotropic" angular distribution of gap functions consistent with experimental observations. Overall, ARPES experiments and state-of-the-art calculations support a conventional $e-ph$ scenario of SC.

That said, in spite of supporting a fully gapped superconducting state, recent STM and muon spin spectroscopy ($\mu$SR) results bring evidence for TRS-breaking SC in Cs(V, Ta)$_3$Sb$_5$ and magnetism-induced modulation of Cooper pairs~\cite{deng2024evidence}. It is however important to also point out that the $e-ph$ driven SC is not incompatible with both the recently observed PDW in CsV$_3$Sb$_5$~\cite{chen2021roton} and TRS-breaking orders~\cite{mielke2022time}. Indeed, PDW has been also observed in the conventional superconductor NbSe$_2$, where the pair density modulation is due to the real space charge density modulation~\cite{liu2021discovery}.

Besides SC, strong $e-ph$ coupling is potentially responsible for the appearance of a CDW order, specifically when it is momentum dependent~\cite{varma1983strong}. Several experimental and theoretical studies support this scenario as a mechanism for the formation of the various charge orders in the AV$_3$Sb$_5$ family. In fact, although bulk sensitive hard-X-ray and neutron scattering showed the absence of soft phonons~\cite{PhysRevX.11.031050,xie2022electron}, \textit{i.e.} phonon modes whose frequency softens upon cooling toward a phase transition, Raman scattering measurements on CsV$_3$Sb$_5$ reported high-frequency amplitude modes   that hybridize significantly with other lattice modes, indicating strong $e-ph$ coupling within the CDW state~\cite{liu2022observation}. To reconcile the experimental evidence with the potential lack of phonon softening, it was suggested that the CDW transition in the AV$_3$Sb$_5$ family is a weakly first-order transition without a continuous change of the lattice dynamics~\cite{miao2021geometry}.

 At the theoretical level, DFT calculations revealed unstable phonon modes at the \textbf{M} and \textbf{L} points of the hexagonal BZ, with the softening of these modes proposed to be the driving mechanism for the CDW formation~\cite{PhysRevLett.127.046401,consiglio2022van,PhysRevB.104.214513}, as shown in Fig.~\ref{figure8}(a). Moreover, variational Monte Carlo simulations for the Hubbard model on the kagome lattice pointed towards the pivotal role that electron-phonon coupling plays to stabilize the experimentally observed CDW phases in AV$_3$Sb$_5$~\cite{Ferrari2022}.

\subsection{\label{sec:chapIIIE}Correlations and fluctuating local moments}

This section discusses the degree of electronic correlation in kagome metals, especially of the 135 family.
An unbiased approach to quantify the strength of the electron-electron interaction is to calculate the Coulomb tensor from first-principles. Operatively, this can be done by defining a localized model spanning a given energy window of the band structure around the Fermi level and performing a constrained random phase approximation (cRPA) calculation. This excludes the screening effects within the so-called ``target'' manifold when calculating the polarization of the system and yields the interaction parameters specific to that low-energy model (for an overview on cRPA, see \cite{ferdiBook}).

cRPA results for the kagome metals have been reported for KV$_3$Sb$_5$ \cite{diSantePRR2023} and for CsV$_3$Sb$_5$ \cite{PhysRevB.105.235145}.
In both cases, the average values for the Kanamori-type intraorbital repulsion $U$ and Hund’s exchange $J_\text{Hund}$ are approximately 6.1\,eV and 0.5\,eV, respectively, when screening is removed over the entire V-3$d$ and Sb-5$p$ manifolds. These values are significantly reduced — to about 1.6\,eV and 0.4\,eV, respectively — when the target space is restricted to the V-3d orbitals only.
Compared to oxides, such as for instance SrVO$_3$ or LaVO$_3$, the $d$-bandwidth in the 135-kagome is larger and the energy distance to the $p$-bands is a factor 3 to 4 smaller, resulting in a much more pronounced covalency.  Taking other kagome materials, such as Ni$_3$In, FeSn and Co$_3$Sn$_2$S$_2$, as a reference, the interaction values for KV$_3$Sb$_5$ are roughly a factor of two smaller. This all indicates a rather weak degree of correlation in the V-compounds of the 135-family. 

\begin{figure}[!t]
\centering
\includegraphics[width=\columnwidth,angle=0,clip=true]{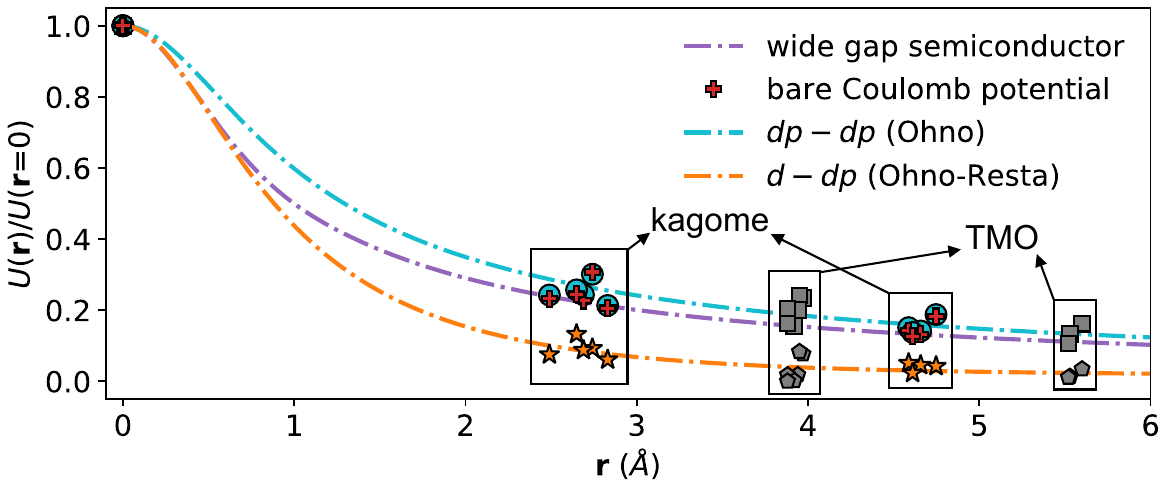}
\caption{Decay with the interatomic distance of the intraorbital Coulomb repulsion $U$ normalized to the on-site strength and compared between two different choices for the target subspaces in cRPA calculations. The representative of the transition-metal oxide (TMO) family are LaVO$_3$, LaTiO$_3$ and LaCrO$_3$ while KV$_3$Sb$_5$, Ni$_3$In, FeSn and Co$_3$Sn$_2$S$_2$ are considered for the kagome class. Figure taken from \cite{diSantePRR2023}.}  
\label{figureUcRPA}
\end{figure}

In \cite{diSantePRR2023}, the decay of the Coulomb repulsion with the interatomic distance has been compared for the two different target spaces against other classes of materials. 
The results shown in Fig.~\ref{figureUcRPA} indicate a good degree of universality, despite the different absolute values of the interaction strengths.

In addition to cRPA calculations, one can estimate the degree of correlation from a comparison between experimentally measured quantities and theoretical calculations. Possibilities are offered by optical measurements, tunneling or photoemission spectroscopies which can be compared to corresponding first-principles calculations. In the cases where the experimental spectrum displays a gap that is not predicted by DFT or GW -- and if long-range order can be ruled out at the temperatures of the measurements -- the material qualifies as a candidate strongly correlated Mott insulator \cite{imadaRMP}. According to this criterion, for instance, high-temperature superconducting cuprates are classified as Mott systems as the pseudogap extends to zero doping and persists above the N\'eel temperature. Such a strong-coupling scenario can be immediately ruled out for both V-based 135 and 166 kagome materials which display metallic bands in photoemission and well-defined Fermi surfaces. 
Using the ratio of the kinetic energy obtained from optics experiments and the kinetic energy from band theory as a proxy to the degree of correlation also puts the V-based 135 compounds in the same category as other weakly correlated itinerant systems \cite{zhouPRB2023}.

Therefore, more than one criterion points towards a rather small degree of correlation in the V-based 135 compounds. Yet, because of their multi-orbital nature, one could still ask if the V-based kagome metals could be classified as ``Hund's metals'' \cite{georgesPhysToday2024}, extending the definition given to iron-based superconductors and some transition-metal oxides.  Hund's metals are also characterized by metallic single-particle spectra. Yet, their spin-spin correlation function reveals sizeable instantaneous local magnetic moments that are long-lived \cite{hauleNJP,aichhornFeSe,yin2011,QimiaoReview,DaiReview} and fluctuate in a way characteristic of systems close to a Mott transition \cite{hansmannPRL2010,watzenboeckPRL}. Moreover, they display strong differentiation of the band renormalizations due to the orbital decoupling induced by the presence of $J_\text{Hund}$~\cite{HundReview,janusPRL}.


Analogously to many prototypical Hund's metals, the majority of the kagome materials discussed here comprise transition-metal elements of the fourth period (principal quantum number $n$=$3$) and, as discussed in Sec.~\ref{sec:chapIII}, the kagome bands close to the Fermi level involve partially filled atomic shells with angular momentum $l$=$2$. 
This situation promotes in general the formation of large local magnetic moments. From an atomic point of view, electrons are indeed significantly confined because of the nodeless radial wave function with $l$=$n-1$ \cite{georgesLecture}.

\begin{figure}[!b]
\centering
\includegraphics[width=\columnwidth,angle=0,clip=true]{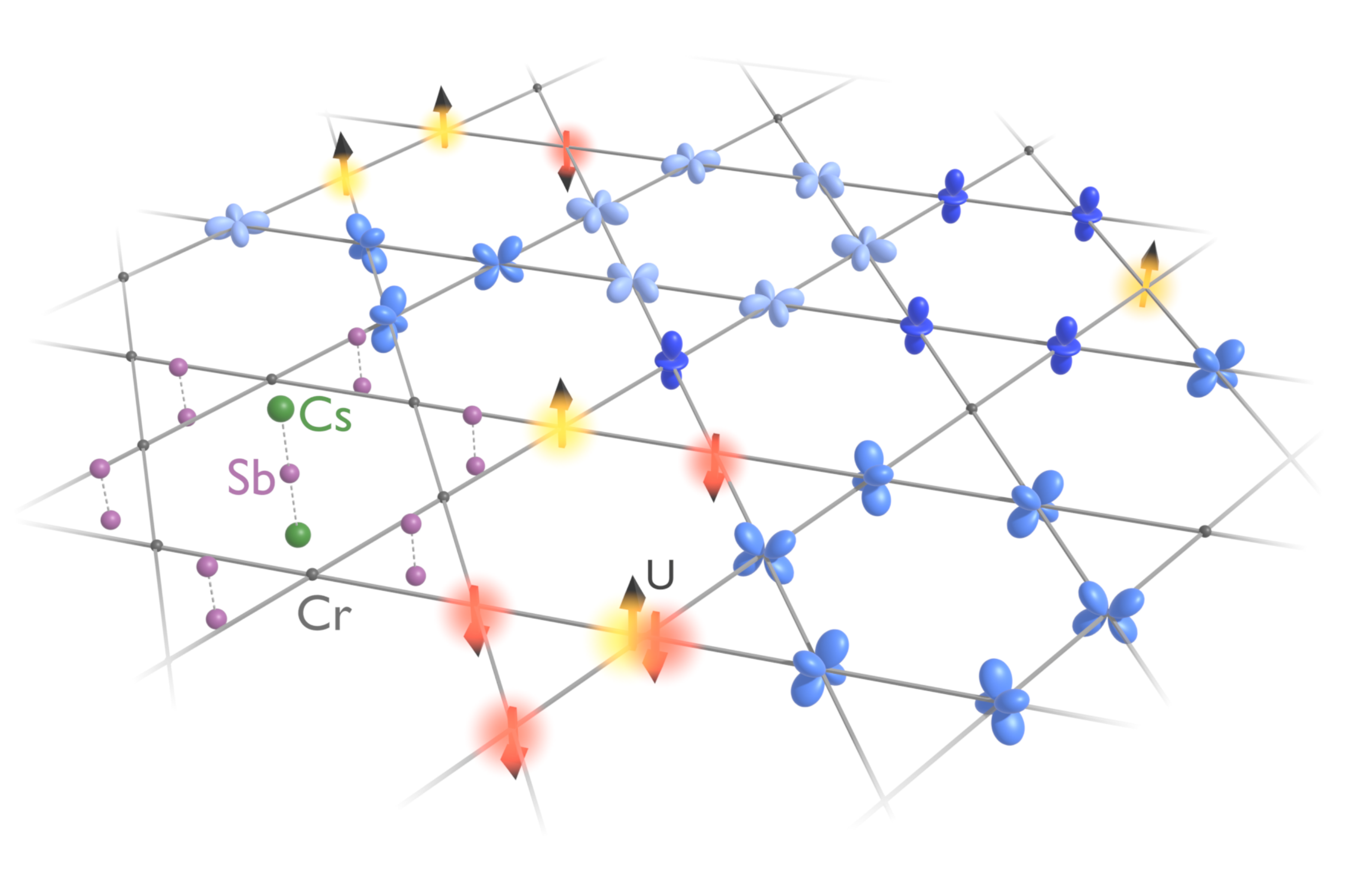}
\caption{Sketch of a kagome plane of Cr atoms (small grey sphere). The bigger green and purple balls represent Cs and Sb, respectively, in such a way to portray the CsCr$_3$Sb$_5$ compound. On randomly selected kagome sites, some of the Cr 3$d$-orbitals are drawn. Spin-up and spin-down electrons hop through the lattice and when two of them occupy the same site, they pay an energy cost. Its value depends on whether the sit on they same or on different orbitals. In the figure, readapted from \cite{giorgioNV}, we exemplify the case in which they occupy the same orbital, which means they have opposite spins and repulsion $U$. This demotes electron's delocalization through the lattice and favors the formation of (fluctuating) local moments.}  
\label{figureCr}
\end{figure}

However, when screening effects in the solid are included via cRPA, the intraorbital repulsion gets fairly suppressed, at least for V-based 135 materials.
Yet, independently from the absolute value of $U$ parameter, we can pictorially sketch the main physical ingredients of a 135 kagome metal using in the language of a microscopic Hubbard-like model, as in Fig.~\ref{figureCr}. This neglects for simplicity non-local interactions (see Fig.~\ref{figureUcRPA}) and also assumes no long-range magnetic order.

The illustration in Fig.~\ref{figureCr} highlights the specificity of the kagome lattice, which plays a role in determining the degree of correlation in these materials.
As discussed extensively already in Sec.~\ref{sec:chapIIA1} and Sec.~\ref{sec:chapIII}, the geometric frustration of single-particle wave functions influences the electrons' localization on the kagome lattice.
The resulting flat band with the associated singular DOS increases the impact of correlations. 
In fact, because the band velocity is proportional to the first derivative of the energy $E_{\textbf{k}}$ with respect to momentum $\textbf{k}$, bands that are very flat host Bloch electrons that are enormously slowed down, even without considering the kinetic energy reduction effect driven by $U$  \cite{vonsovskii1993,mravljePRL2011,Hausoel2017}.

One can therefore view the kagome lattice as a platform in which the level of itinerancy can be suppressed by one- and two-particle processes synergistically. The former can be roughly thought to be set by the bandwidth $W$ of the electronic manifold, or by better indicators sensitive to the presence of flat bands in the DOS -- such as e.g. the second moment of the DOS \cite{potthoff2mom}. 
Instead at the two-particle level the strength of the interaction matters in general, but the associated vertex corrections are crucially dependent on the filling of the correlated bands. This indeed determines the size of the phase space for the relevant many-body processes directly influencing the behavior of charge and spin fluctuations: the closer the localized orbitals are to half-filling, the stronger is the effective correlation. Borrowing jargon from meteorology, the filling has a similar role as humidity for the perceived temperature, \textit{i.e.} it amplifies the effects of correlation.

There is a simple consequence to this: if it is possible to increase the number of $d$-electrons with respect to known metals like V-based 135 and 166 compounds, there is a chance to reach a higher level of correlation. 
One way is to substitute V with Cr in the 135 compounds. This increases the nominal $d$-occupation by 3 per formula unit, though the cRPA estimates would be obviously influenced as well. At the present stage, the first indications point indeed towards more relevant many-body correlations: For the recently synthesized CsCr$_3$Sb$_5$, \textcite{liuCr135} report signatures of local magnetic moments and sizeable spin fluctuations. 
Using a parton approach based on subsidiary spins, \textcite{xieCr135} have stressed the tendency towards orbital selectivity in which the $d_{xz}$-orbital is the most correlated one and reaches quasiparticle renormalizations betwen 2 and 10, depending on the interaction used. 

This is in striking contrast with the V-compounds for which both experiments \cite{kenney2021muon} and theory \cite{zhaoDMFT2021} have reported no local moments. 
The latter DFT+DMFT calculations for V-based 135, in agreement with \textcite{liuPRB2022}, also indicate a rather weak electronic mass renormalization (smaller than 2) for KV$_3$Sb$_5$. 
This is true even if one focuses only on the $d_{xz}$-orbital, which is the one that responds most sensitively to the electron-electron interaction also in the V-compounds.
DFT+DMFT also indicates vanishing local moments and small mass renormalizations in the related ScV$_6$Sn$_6$ kagome metal \cite{yuPRB2024}.

\section{\label{sec:chapIV}Materials classes}

In this section, we provide an overview of the properties of key members of some of the more common classes of kagome metals.  The phenomenology associated with each kagome compound is governed by the corresponding lattice frameworks, the nature of the atoms comprising the kagome sublattice, the electron filling of the kagome bands, and the presence of other instabilities (native to other sublattices) that couple to the kagome framework.  This results in a rich spectrum of phase behaviors across numerous classes of kagome metals.

We discuss key kagome metal classes by breaking them down via their chemical complexities, which govern the phenomena and tunability possible in each.  As two elements are required to drive the formation of a kagome plane via distortions in close-packed hexagonal layers of single-element compositions \cite{kolli2021six}, the simplest class of kagome metals are binary compounds.  Although this phase space is seemingly simple, binary kagome networks harbor a remarkable range of complexity. This is governed by a combination of a stacking degree of freedom between kagome and non-kagome layers, which, in principle, can tune the dimensionality and the filling of the kagome network.  The character of the kagome ion can further tune magnetism within the kagome network and gapping within the electronic band structure.

Additional elements introduced within the crystal network, for example in ternary compounds, allow for new structural motifs that support interactions between the kagome network and other cation sublattices. Examples include, for instance, added magnetic layers distinct from the kagome network that can couple to carriers and break symmetries within the kagome planes.  Other examples include structural subunits that host bond order or steric instabilities that couple to and distort the kagome networks.  Some of the more common ternary kagome compounds hosting some of these features are discussed in the second part of this section.   

\subsection{\label{sec:chapIVA}Binary}

\begin{figure*}[!ht]
\centering
\includegraphics[width=\textwidth,angle=0,clip=true]{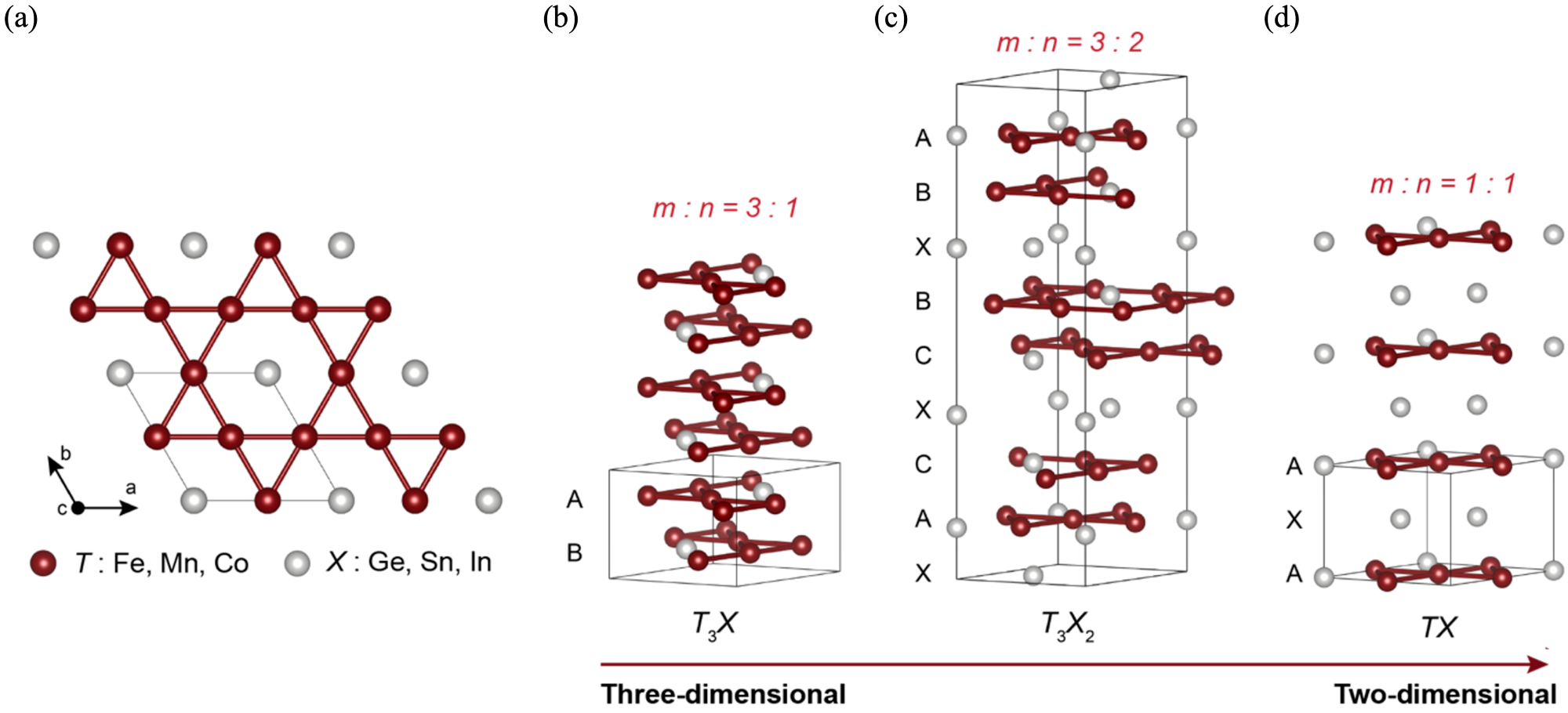}
\caption{(a) Top view of the kagome layer in the T$_m$X$_n$ series. (b-d) Stacking sequences of the T$_m$X$_n$ series with m:n = 3:1 (b), 3:2 (c), and 1:1 (d). Kagome layers (A–B-C) exhibit different in-plane lattice offsets. The spacing layers (S) consist of X atoms. The dimensionality increases with the X:T ratio. In the 1:1 structure (d), kagome layers are separated by the S layers, while in the 3:1 structure (b), neighboring kagome layers are stacked on top of ech other. The 3:2 structure (c) exhibits a mixture of both stacking types. Panels adapted from \cite{kang_dirac_2020}.}
\label{binary_kagome_stacking}
\end{figure*}

The binary kagome metal series, T$_m$X$_n$ (T = $3d$ transition metal, X = Sn, In, Ge), has been instrumental in the early exploration of the physics of kagome metals. This family of compounds offers a versatile platform to study the interplay between the unique electronic structure of the kagome lattice and magnetism. A key feature of these materials is the presence of a structural layer hosting a kagome sublattice that is occupied by 3\textit{d} transition metal elements. The stacking arrangements of these kagome layers, interspersed with spacer layers (S) of X atoms, can be varied depending on the specific composition ($m:n$ ratio in T$_m$X$_n$). This structural flexibility allows the tuning of interlayer interactions and the effective dimensionality of the electronic structure (Fig.~\ref{binary_kagome_stacking}).

For instance, in the 3:1 structure (T$_3$X), the kagome layers exhibit an AB type stacking, where each layer is shifted laterally with respect to the neighboring one by half unit cell (Fig.~\ref{binary_kagome_stacking}(b)). Such a configuration, where downward-pointing triangles face upward-pointing triangles, naturally leads to strong interlayer interactions and a 3D electronic structure. Consequently, some of the characteristic electronic features of an ideal 2D kagome lattice, such as Dirac fermions, are typically not observed in these materials. However, flat bands can partially survive in these structures even when interlayer hopping is allowed. Also, some compounds from the 3:1 family defy conventional understanding by demonstrating a large intrinsic AHE without net magnetization, arising from a noncollinear antiferromagnetic spin structure~\cite{nakatsuji2015large}. 

In the 3:2 structure (T$_3$X$_2$), interlayer interactions are partially reduced, allowing some characteristics of the 2D kagome electronic structure to emerge (Fig.~\ref{binary_kagome_stacking}(c)). In this stacking motif, kagome bilayers units are stacked between the spacer layers, with a stacking sequence A-B-S-B-C-S-C-A-S. Fe$_3$Sn$_2$ is a prime example, being the first material where kagome-derived Dirac fermions were experimentally realized \cite{ye_massive_2018, yin_giant_2018}. The material also exhibits other interesting properties such as room-temperature skyrmion bubbles, and a topological Hall effect \cite{hou_observation_2017}.

Lastly, the 1:1 structure (TX) represents a configuration where kagome layers are most strongly decoupled, offering the closest environment to realize the prototypical electronic structure of a 2D kagome lattice (Fig.~\ref{binary_kagome_stacking}(d)). CoSn, a non-magnetic kagome metal, displays ideal kagome flat bands with suppressed dispersion in all momentum directions \cite{kang_topological_2020}. The intrinsic SOC lifts the degeneracy between the flat and Dirac bands, leading to topologically non-trivial flat bands. In contrast, the antiferromagnetic kagome metal FeSn displays a richer electronic structure with coexisting surface and bulk Dirac fermions alongside a magnetic flat band. The interplay between Dirac fermions and the unique magnetic structure in FeSn leads to symmetry-protected Dirac nodes that are robust against SOC \cite{kang_dirac_2020}. FeGe, closely related to FeSn, exhibits a different magnetic structure and features a 2$\times$2 charge order coexisting with antiferromagnetism (AFM) \cite{teng_discovery_2022}.

In summary, the binary kagome series T$_m$X$_n$ has been an ideal playground to initiate and deepen the exploration of kagome metal. The diverse stacking configurations and the intrinsic properties of $3d$ transition metals allow for the realization of a broad array of electronic and magnetic phenomena, providing a fertile ground for further exploration of topological and correlated phases in kagome lattices.

\subsubsection{\label{sec:chapIVA1}Fe-based: FeSn, Fe$_3$Sn$_2$, FeGe}

\paragraph{Fe$_3$Sn$_2$:}

The first kagome metal to have its electronic band structure extensively studied is Fe$_3$Sn$_2$. This material has attracted significant interest owing to its intriguing electronic, magnetic, and transport properties, all stemming from its geometrically frustrated kagome lattice. Crystals are typically prepared via chemical vapor transport methods \cite{b_malaman_structure_1976}.

Establishing the groundwork for understanding the magnetic properties of Fe$_3$Sn$_2$ \cite{b_malaman_magnetic_1978}, neutron diffraction and magnetization measurements identified the compounds as ferromagnetic, with a notable spin rotation between 250 K and 60 K. The magnetic properties of Fe$_3$Sn$_2$ were later investigated using Mössbauer spectroscopy, revealing spin direction changes across a wide temperature range \cite{g_le_caer_mossbauer_1978, g_caer_magnetic_1979}. Subsequent explorations of the the non-collinear static spin structures and re-entrant spin glass behavior were crucial for understanding the role of spin chirality in the AHE \cite{l_fenner_non-collinearity_2009}. This giant AHE was later attributed to the material's unique frustrated kagome-bilayer structure, while an unconventional scaling law in Hall resistivity was also discovered, pointing towards intrinsic mechanisms over extrinsic phenomena \cite{t_kida_giant_2011}.

\begin{figure}[!t]
\centering
\includegraphics[width=\columnwidth,angle=0,clip=true]{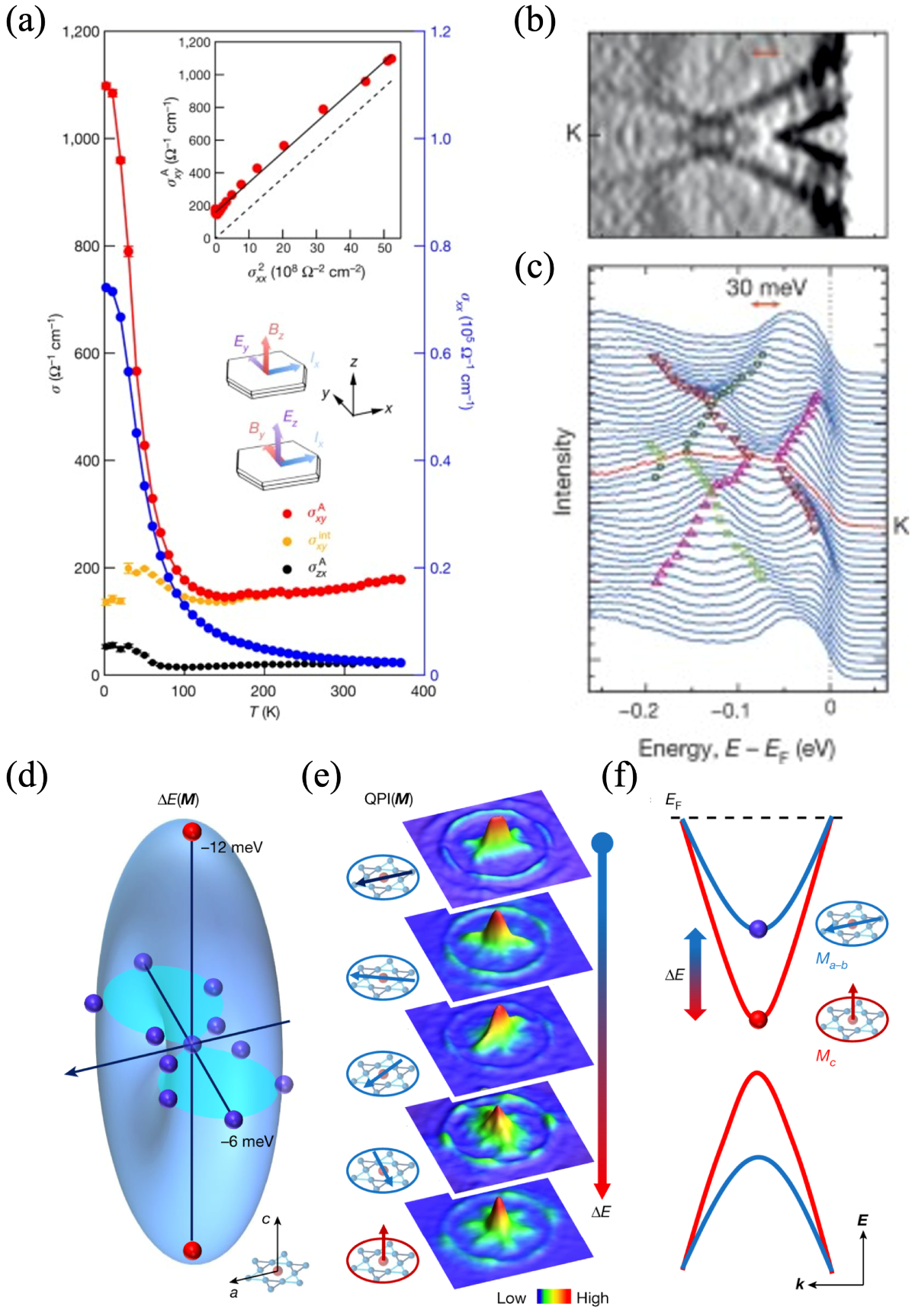}
\caption{(a) Anomalous Hall transport in kagome ferromagnet \ce{Fe3Sn2}. (b-c) Pair of massive Dirac fermions at the K point. (d) Magnetization-dependent energy shift of the Dirac fermions. (e) Quasiparticle interference pattern as a function of magnetization direction. (f) Magnetization-controlled mass gap at the Dirac point. Panels adapted from \cite{ye_massive_2018, yin_giant_2018}.}
\label{Fe3Sn2}
\end{figure}

Exploration of the electronic structure generated the discovery of massive Dirac fermions near the Fermi level using ARPES (Fig.~\ref{Fe3Sn2}(a-c)) and STM (Fig.~\ref{Fe3Sn2}(d-f)) \cite{ye_massive_2018, yin_giant_2018}. These studies highlighted the rich electronic structure of Fe$_3$Sn$_2$ and its potential for topological phenomena. A subsequent STM study provided spectroscopic evidence of the presence of flatbands and explored their connection to electron correlation and magnetic ordering \cite{zhiyong_lin_flatbands_2018}. Early theoretical works using DFT calculations predicted Weyl nodes and topological transitions, significantly advancing the theoretical framework surrounding Fe$_3$Sn$_2$ \cite{m_yao_switchable_2018,biswas_spin-reorientation-induced_2020}. More recent work reported a new type of electron band formation at low-temperatures, derived from Sn $p$ orbitals and possibly due to electron fractionalization \cite{s_ekahana_anomalous_2022}. These studies firmly established the material as a fertile ground for exploring topological quantum states. 

Additional research has focused on more exotic magnetic properties of Fe$_3$Sn$_2$.  Magnetic skyrmions were observed, highlighting potential spintronics applications due to their room-temperature stability and contributions to the topological Hall effect \cite{p_wu_evidence_2021, hou_observation_2017}. Progress in thin film synthesis has further opened new avenues for exploration.The synthesis of Fe$_3$Sn$_2$ thin films was reported using sputter deposition \cite{kacho_mumtaz_ali_khan_intrinsic_2022} and molecular beam epitaxy \cite{shuyu_cheng_atomic_2022, Ren2022Fe3Sn2STM-MBE}. This development has enabled further investigation of magnetotransport properties and device integration, paving the way for practical applications in nanostructured devices. The interplay of topological and AHEs was investigated in epitaxial films, revealing a rich interplay between electronic topology and magnetic ordering \cite{dongyao_zhang_anomalous_2022,q_du_topological_2022}.  Research, exemplified by~\textcite{hang_li_coherent_2023,manli_zhu_two-dimensional_2024}, continues to explore and harness the unique properties of Fe$_3$Sn$_2$, driving advancements in spintronics and quantum materials.

\paragraph{FeSn:}

The kagome metal FeSn, with its unique lattice structure, exhibits a fascinating interplay of electronic and magnetic phenomena. Bulk single crystals can be grown via crucible-based Sn-flux methods \cite{l_haggstrom_studies_1975}. Early investigations primarily focused on FeSn's magnetic behavior where neutron diffraction studies revealed antiferromagnetic ordering below 365 K \cite{k_yamaguchi_neutron_1967}.  Concurrently, Mössbauer spectroscopy provided insights into hyperfine interactions and magnetic configurations \cite{hisao_yamamoto_mossbauer_1966,s_ligenza_mossbauer_1971,s_ligenza_mossbauer_1972,s_k_kulshreshtha_anisotropic_1981}. These foundational works established FeSn as a compelling kagome antiferromagnet.

\begin{figure}[!t]
\centering
\includegraphics[width=\columnwidth,angle=0,clip=true]{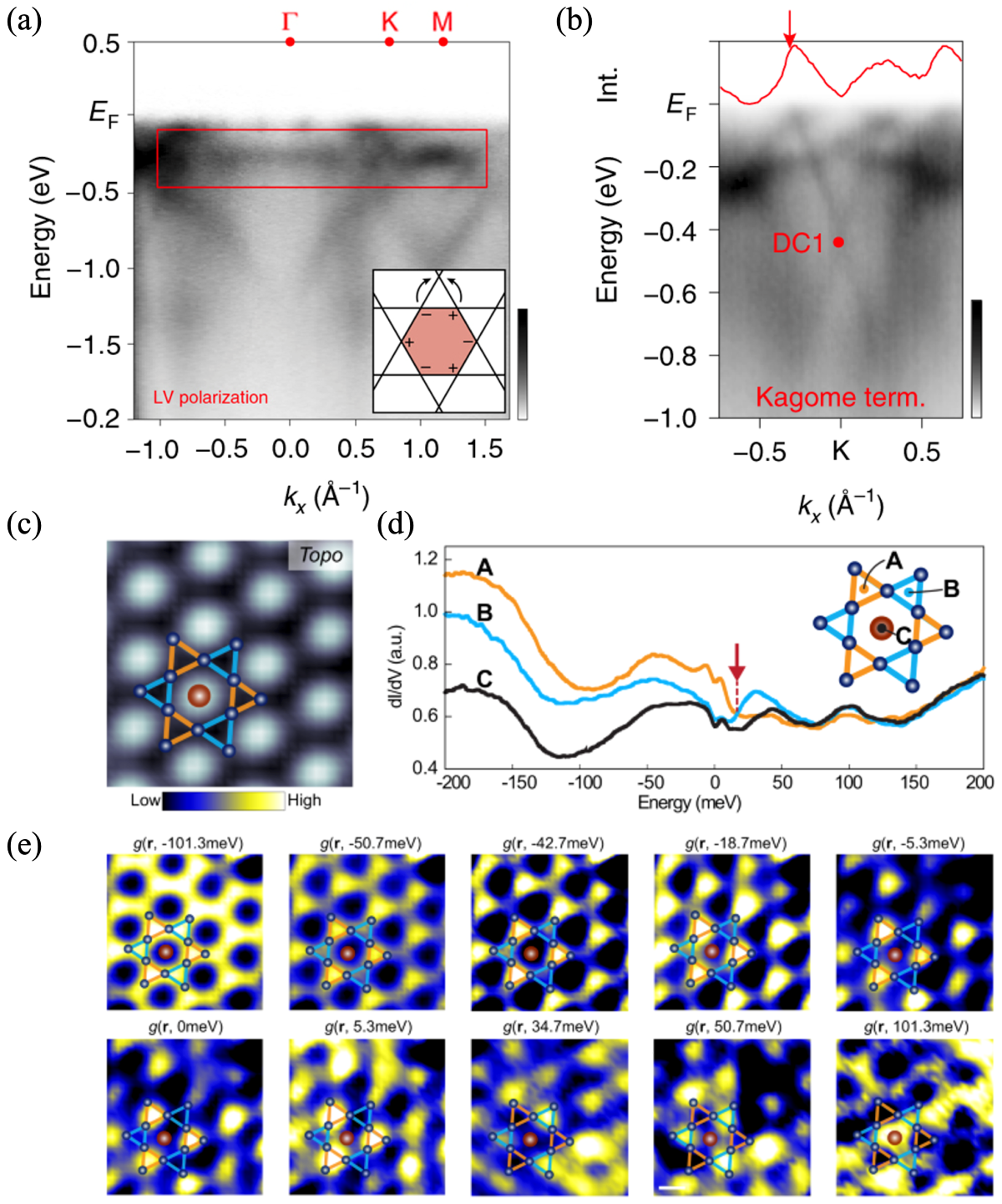}
\caption{(a-b) Flat band and Dirac fermions (right) in FeSn. (c) Topographic STM image of the kagome termination of FeSn. (d) Tunnelling conductance spectra from three representative sites A, B, and C. (e) Energy-dependent differential conductance maps. Panels adapted from \cite{kang_dirac_2020,zhang2023visualizing}}
\label{FeSn}
\end{figure}

Characterization of FeSn's electronic and band structure properties was performed by measuring its magnetic, electronic and thermal transport, and thermodynamic properties \cite{b_sales_electronic_2019, kang_dirac_2020}. DFT calculations were used to unveil a three-dimensional electronic structure with Dirac nodal lines, highlighting the potential for topological phases intertwined with magnetic order \cite{b_sales_electronic_2019}. ARPES experiments provide evidence for the presence of Dirac fermions in FeSn (Fig.~\ref{FeSn}(a-b)), solidifying its status as a platform for studying topological phenomena \cite{kang_dirac_2020}. Additional studies further elucidated the properties of these Dirac states, demonstrating their bulk nature and connection to two-dimensional surface Weyl fermions \cite{lin_dirac_2020}.

There is renewed interest in the magnetic properties and spin dynamics of FeSn. \textit{Ab initio} studies employed linear spin wave theory and density functional perturbation theory to analyze spin fluctuations, revealing Landau damping effects on magnons \cite{yi-fan_zhang_ab_2022}. Further neutron scattering work demonstrated the potential for manipulating magnetic states in FeSn to unlock exotic excitations like Dirac magnons \cite{s_do_damped_2021}.

Investigations into the transport properties of FeSn have yielded intriguing results. Contrary to expectations for quasi-2D systems, the material exhibits significant anisotropic conductivity. This anisotropy, observed in early studies on bulk \cite{b_sales_electronic_2019} and epitaxial thin film samples \cite{h_inoue_molecular_2019}, challenges conventional interpretations of Hall effects linked to Berry curvature.  An additional study quantified higher-order nonlinear AHEs, connecting these response to the quantum geometry of this system \cite{soumya_sankar_experimental_2023}. Further optical spectroscopy work emphasized the role of Sn layers in contributing to the optical anisotropy of FeSn, challenging simple 2D models \cite{j_ebad-allah_optical_2024}. This study, along with investigations of magneto-elastic couplings and lattice constants \cite{y_tao_investigating_2023}, underscores the intricate interplay of correlation effects in FeSn.

STM experiments have explored FeSn in both single crystal \cite{Li2022FeSn_STM, Multer2023FeSn_STM} and thin film forms \cite{Li2022FeSn_STM, zhang2023visualizing, Pham2024FeSn_STM} (Fig.~\ref{FeSn}(c-e)). Spin-polarized STM imaging visualized layered antiferromagnetic ordering at the surface of FeSn \cite{Li2022FeSn_STM}. An external magnetic field was found to further modify electronic properties of individual defects \cite{Li2022FeSn_STM} and tune local directionality of the electronic structure \cite{zhang2023visualizing, Multer2023FeSn_STM}. Spectroscopic signatures of electronic bands were detected as peaks in STM differential conductance d$I$/d$V$ spectra \cite{zhang2023visualizing, Multer2023FeSn_STM}.

\paragraph{FeGe:}

FeGe is a kagome lattice metal exhibiting A-type AFM \cite{watanabe1966neutron} and single crystals can be grown via halogen vapor transport methods \cite{j_b_forsyth_low-temperature_1978, mi1967crystal}.  It has become a key system for investigating the interplay of CDWs, electronic correlations, and magnetism within kagome topological systems. Its unique features, such as Dirac cones, VHS, and flat electronic bands, create a complex landscape of quantum phases influenced by lattice geometry, spin-phonon coupling, and nontrivial band topology. FeGe therefore stands out as an important platform for studying strongly correlated topological systems.

\begin{figure}[!t]
\centering
\includegraphics[width=\columnwidth,angle=0,clip=true]{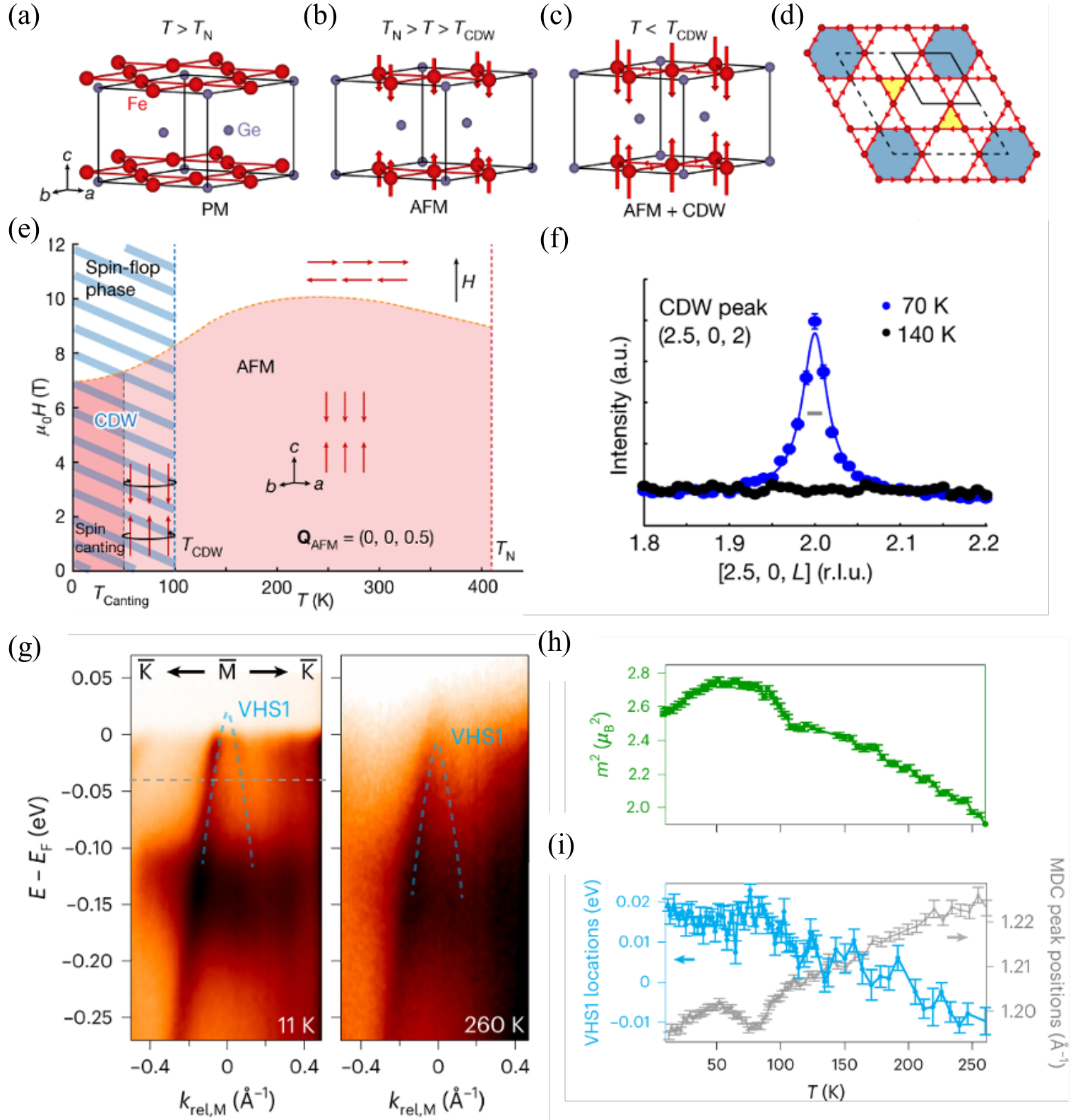}
\caption{(a-c) Symmetry breaking sequence in FeGe. (d) Schematic of the in-plane flux phase. (e) Temperature-field phase diagram of FeGe. (f) Momentum scan across the CDW peak and temperature dependence of the magnetic Bragg peak. (g) Temperature-dependent VHS from ARPES data. (h) Ordered magnetic moment measured by neutron scattering. (i) Temperature dependence of VHS binding energy. Panels adapted from \cite{teng_discovery_2022,xiaokun_teng_magnetism_2022}.}
\label{FeGe}
\end{figure}

Specifically, FeGe is host to a diverse array of phenomena including high-temperature magnetic order, unconventional CDW states, and strong electronic correlations—an interplay, that is not readily found in most other kagome materials  \cite{teng_discovery_2022, shaohui_yi_polarized_2024, c_setty_electron_2022, xiaokun_teng_magnetism_2022, yin_discovery_2022}. The interest in FeGe stems from its distinct A-type antiferromagnetic (AFM) order ($T_{N} \approx$ 410 K), coupled with an unconventional CDW phase ($T_{CDW} \approx$ 100–110 K), spin-canting transitions, and topological electronic behavior dominated by VHS \cite{teng_discovery_2022, yilin_wang_enhanced_2023, zhisheng_zhao_photoemission_2023} (Fig.~\ref{FeGe}(a-f)).

FeGe possesses electronic bands that are characteristic of the kagome lattice, namely Dirac cones, flat bands, and VHS \cite{xiaokun_teng_magnetism_2022, c_setty_electron_2022} (Fig.~\ref{FeGe}(g-i)). These features are prominent near the Fermi level, where electronic correlation effects amplify their topological and transport signatures. ARPES has revealed key intricacies of the band structure of FeGe, notably spin-split VHS below $T_{N}$, arising from magnetic exchange interactions \cite{teng_discovery_2022, xiaokun_teng_magnetism_2022}. DFT calculations with dynamical mean-field theory (DFT+DMFT) emphasize the importance of SOC and Hund's coupling in modulating these topological features \cite{c_setty_electron_2022, yilin_wang_enhanced_2023}. The coexistence of flat bands and Dirac cones in FeGe enables Berry curvature effects, driving topological phenomena like the AHE \cite{h_miao_signature_2023, yin_discovery_2022}. Spectral weight redistribution due to spin and charge ordering underscores the vital role of these kagome-derived features in shaping the electronic properties of FeGe.

The Fermi level's proximity to VHS makes FeGe highly sensitive to external conditions like pressure, strain, and doping \cite{xikai_wen_unconventional_2024}. Studies under high pressure show that $T_{CDW}$ increases rather than decreases, contrasting with conventional CDW materials and suggesting a unique correlation-driven origin for CDWs in FeGe \cite{xikai_wen_unconventional_2024, a_korshunov_pressure_2024}. Annealing and doping studies (e.g., FeGe$_{1-x}$Sb$_x$) demonstrate that modifications to lattice symmetry and bonding distances significantly influence band topology and CDW stability \cite{mason_klemm_vacancy-induced_2024, jiale_huang_fege1xsbx_2023}.

CDWs in FeGe have unconventional origins, driven by electronic correlations and accompanied by structural modulation involving Ge dimerization \cite{yilin_wang_enhanced_2023, zhisheng_zhao_photoemission_2023, chenfei_shi_charge_2024}. Unlike conventional CDWs triggered by Fermi surface nesting, FeGe’s CDW emerges from mechanisms like spin-phonon coupling and localized distortions in its kagome lattice geometry \cite{s_shao_charge_2022, yilin_wang_enhanced_2023}. 


The antiferromagnetic ground state of FeGe is intricately linked with its CDW behavior. Below $T_{CDW}$, Ge dimerization acts as an order parameter, enhancing ferromagnetic alignment within the kagome planes and indirectly influencing electronic transport phenomena like the AHE \cite{h_miao_signature_2023, yin_discovery_2022, yilin_wang_enhanced_2023, binhua_zhang_triple-well_2023}. Theoretical models suggest that spin-phonon coupling mechanisms amplify CDW instability through magnetically enhanced spectral weight redistribution near VHS \cite{h_miao_signature_2023, yilin_wang_enhanced_2023}.


Tuning CDW properties through external parameters such as pressure, annealing, and disorder has emerged as a powerful route to explore emergent CDW phases in FeGe. High-pressure experiments reveal the stabilization of quasi-long-range CDW order with superlattice shifts from $2 \times 2 \times 2$ to $\sqrt{3} \times \sqrt{3} \times 6$ \cite{xikai_wen_unconventional_2024, a_korshunov_pressure_2024}, while annealing-induced Ge-site disorder disrupts long-range CDWs but enables flexible control of magnetic coupling \cite{xue-ke_wu_annealing-tunable_2023, chenfei_shi_charge_2024, hengxin_tan_disordered_2024}. Diffraction studies further point to dimerization-driven bond-orientation order as a precursor of CDW formation \cite{d_subires_frustrated_2024}. More broadly, defect engineering through Ge-site vacancy disorder or Sb doping modulates resistivity and suppresses magnetic ordering, offering a handle to tune correlation effects in FeGe’s kagome lattice \cite{mason_klemm_vacancy-induced_2024, hengxin_tan_disordered_2024, jiale_huang_fege1xsbx_2023}.

 Below $\approx$ 60 K, spin canting induces transitions to a double-cone magnetic structure, accompanied by complex spin excitations \cite{teng_discovery_2022, xiaokun_teng_magnetism_2022, lebing_chen_competing_2023}.
The nature of this transition is however under debate, with recent inelastic neutron scattering measurements suggesting that the incommensurate magnetic structure below $\approx$ 60 K could arise from Fermi surface nesting and taking the form of an itinerant SDW~\cite{lebing_chen_competing_2023,mason_klemm_vacancy-induced_2024,oh2025disentangling}.  Regardless of the origin, the antiferromagnetic phase introduces magnetic exchange splittings, directly influencing the band structure and enabling temperature-dependent modulation of electronic properties \cite{xiaokun_teng_magnetism_2022, h_miao_signature_2023}. Magnetic interactions further stabilize CDWs, with spectral signatures evident in spin-polarized ARPES and scattering studies \cite{yilin_wang_enhanced_2023, ziyuan_chen_discovery_2023}. Magnetic anisotropies, studied via neutron scattering, reveal strong interactions between itinerant and localized magnetic moments \cite{lebing_chen_competing_2023}.

Electronic transport in FeGe is highly sensitive to the interplay of CDWs, magnetism, and electronic correlations. Quantum oscillations detected via the de Haas–van Alphen (dHvA) effect reveal reconstructed Fermi surfaces in the CDW phase, suggesting new transport channels mediated by the kagome lattice symmetry \cite{kaixin_tang_evidence_2024}. Anomalous Hall conductivities (AHE) are frequently reported below $T_{N}$, arising from enhanced Berry curvature near the Dirac points and flat bands \cite{yin_discovery_2022, zhisheng_zhao_photoemission_2023}.


FeGe is also a rare example where symmetry-ascending structural transitions are reported at temperatures below the onset of CDW ~\cite{wu2024symmetry}. This unusual tendency can manifest itself as long as an extremely weak structural instability coexists and competes with the CDW and magnetic orders, rendering FeGe a highly relevant platform for the study of intertwined orders.
Its distinct interplay of magnetism, electronic correlations, and topological excitations sets FeGe apart as a paradigmatic material for studying multi-degrees-of-freedom coupling in kagome systems \cite{teng_discovery_2022, c_setty_electron_2022}, and outstanding questions include determining the precise origin of CDW states under extreme conditions (pressure and doping) and the role of Hund's coupling in driving non-Fermi-liquid behavior. Recent functional renormalization group (FRG) analyses also hint at potential SC in FeGe under engineered conditions, opening an avenue for exploring quantum criticality in kagome magnets \cite{pietro_m_bonetti_competing_2024}.

\paragraph{Fe$_3$Sn:}

Here we discuss the relatively less explored binary Fe$_3$Sn compound, which, despite its collinear magnetic order, also exhibits a large anomalous Nernst effect (ANE) \cite{taishi_chen_large_2022}. This phenomenon parallels the ANE seen in the noncollinear antiferromagnet Mn$_3$Sn (discussed in Sect. \ref{sec:chapIVA3}). The large ANE in Fe$_3$Sn is attributed to a nodal plane that creates a flat, hexagonal-shaped electronic band structure with enhanced Berry curvature near the Fermi energy, as clarified by theoretical analysis. The observed ANE reaches up to $3\,\mu V K^{-1}$ above room temperature, making this material a candidate for developing flexible film thermopiles and heat current sensors. A numerical study of the properties of these 3:1 kagome compounds underscores the potential for high-throughput screening methods to explore a broader range of complex magnetic materials, including both collinear and noncollinear systems \cite{bouaziz_spin_2025}.

\subsubsection{\label{sec:chapIVA2}Co-based: CoSn}

\begin{figure}[!t]
\centering
\includegraphics[width=\columnwidth,angle=0,clip=true]{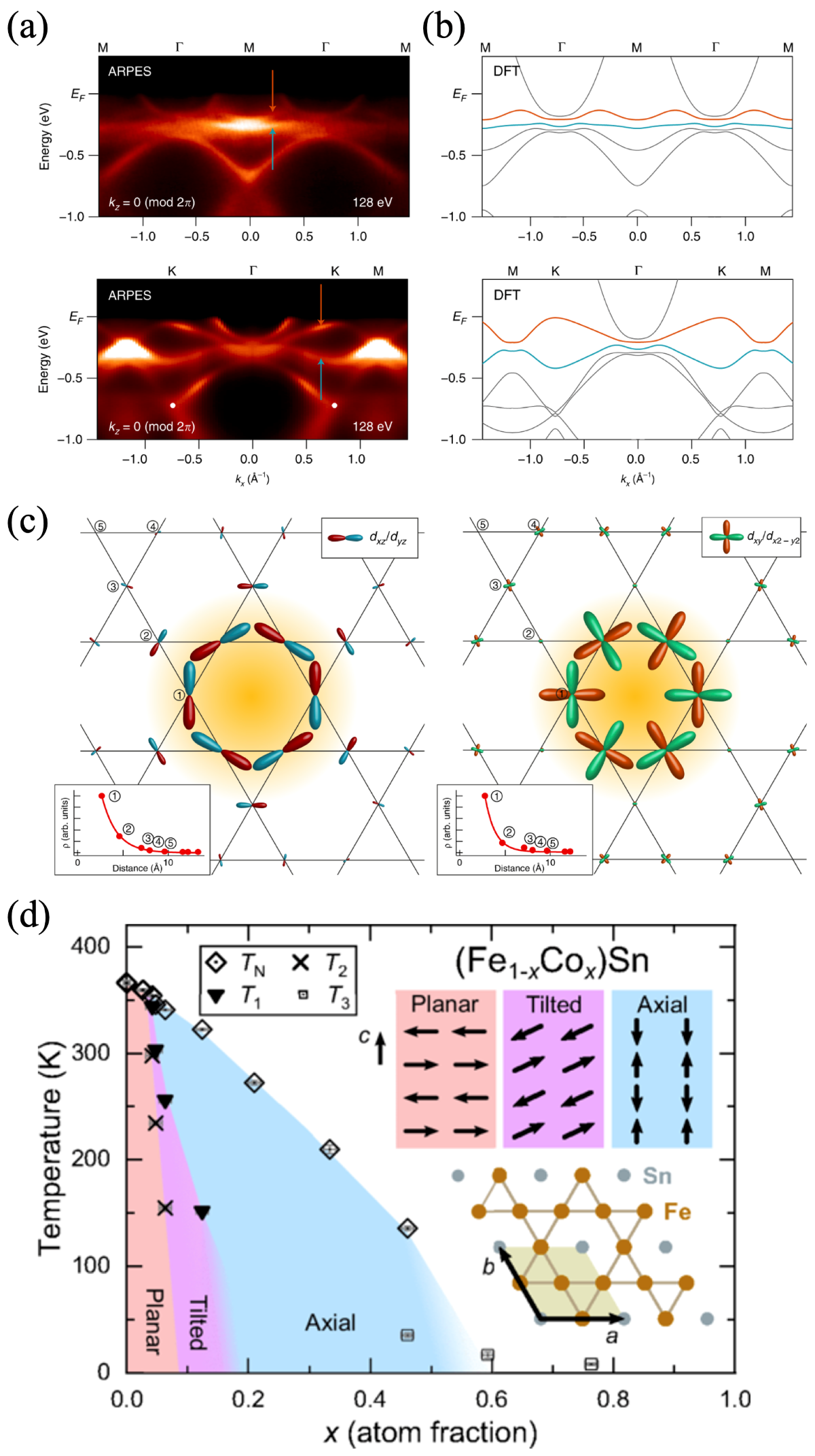}
\caption{(a-b) ARPES data (left) and DFT calculations (right) of the electronic bands in CoSn, showing evidence for the presence of a double flat band. (c) Real-space wavefunctions underlying the flat bands derived from out-of-plane (left) and in-plane (right) orbitals. (d) Magnetic phase diagram of mixed-composition kagome compound (Fe$_{1-x}$Co$_x$)Sn. Panels adapted from \cite{kang_topological_2020,w_meier_reorientation_2019}.}
\label{CoSn}
\end{figure}

CoSn is a key material for studying flat-band physics, Dirac fermions, symmetry-breaking, and unique transport properties \cite{kang_topological_2020, liu_orbital-selective_2020}. Its nonmagnetic kagome lattice and simple Fermi surface make it ideal for understanding intrinsic electronic band properties of the 2D kagome lattice without the complexities of lattice stacking or spin physics. Bulk crystals can be grown via crucible-based Sn-flux methods \cite{larsson1996single}, and CoSn crystallizes in a hexagonal structure (space group P6/mmm), with cobalt atoms forming kagome layers interspersed with tin layers. Its electronic structure displays key kagome physics features: flat bands near the Fermi level (confirmed by ARPES) \cite{kang_topological_2020, liu_orbital-selective_2020} (Fig.~\ref{CoSn}(a-b)), Dirac dispersions at BZ corners with SOC induced gaps leading to massive Dirac fermions \cite{kang_topological_2020, liu_orbital-selective_2020}, and orbital-selective features ($d_{xy}$, $d_{z2}$ hybridization) revealed by DFT and TB models \cite{liu_orbital-selective_2020, wan_temperature_2022}. Importantly, the flat bands in CoSn arise from different orbital manifolds and could be directly linked to the compact localized states predicted from 2D kagome toy models \cite{kang_topological_2020} (Fig.~\ref{CoSn}(c)).

These flat bands are highly tunable. Strain or doping with Fe, In, or Ni shifts their position relative to the Fermi level. Fe and In doping bring flat bands closer to the Fermi energy, promoting magnetic responses and instabilities, while Ni doping suppresses these effects \cite{shuyu_cheng_epitaxial_2023, sales_tuning_2021, sales_chemical_2022}. Theoretical models suggest strain could induce transitions to topologically nontrivial states with Weyl fermions or enhanced correlation phenomena \cite{kang_topological_2020, shuyu_cheng_epitaxial_2023, mojarro_tuning_2024}.

Recent studies show rotational symmetry breaking (``nematicity") in CoSn at ~225 K, driven by flat-band thermal excitations \cite{nathan_c_drucker_incipient_2024}. This is unique as it occurs without magnetic order, suggesting that the flat bands in CoSn cause electronic instabilities leading to nematic transitions. While intrinsic CoSn is nonmagnetic, doping introduces localized magnetic states and creates a rich magnetic phase diagram \cite{w_meier_reorientation_2019} (Fig.~\ref{CoSn}(d)). Fe-doping induces a spin glass state and ferromagnetic fluctuations linked to flat bands near the Fermi level \cite{sales_tuning_2021, sales_chemical_2022, kritika_vijay_tunable_2023}. For instance, Co$_{1-x}$Fe$_x$Sn displays tunable magnetoresistance due to band realignment and localized flat states \cite{kritika_vijay_tunable_2023}. Theoretical work proposes CoSn as a platform for interaction-driven quantum order, including CDWs (potentially induced by correlations near VHS and flat bands under strain or pressure) \cite{mojarro_tuning_2024} and unconventional SC (suggested by weak correlation effects and kagome physics) \cite{liu_orbital-selective_2020}.

CoSn also exhibits significant transport anisotropy, with in-plane resistivity much higher than that of the out-of-plane due to flat-band localization and large effective masses \cite{huang_flat-band-induced_2022, j_zhang_flat_2021}. These differences are tied to kagome geometry, particularly flat-band dispersion and scattering near VHS. Despite being nonmagnetic, CoSn shows flat-band-driven negative magnetoresistance in external fields, attributed to enhanced scattering and ferromagnetic correlations \cite{j_zhang_flat_2021}. Temperature-dependent transport and spectroscopy studies further reveal strong phonon-electron interactions near flat bands, affecting quasiparticle velocities, scattering rates, and resistivity \cite{huang_flat-band-induced_2022,j_yin_fermionboson_2020}.

STM has visualized the localized real-space kagome flat-band states in CoSn \cite{caiyun_chen_visualizing_2023}, serving as a fingerprint of ideal kagome physics and a path to explore strongly correlated phases. While CoSn is topologically trivial, small SOC gaps suggest the proximity to a topological phase. Strain or doping could potentially realize Weyl or Dirac semimetal phases, as suggested theoretically \cite{kang_topological_2020,shuyu_cheng_epitaxial_2023,sales_tuning_2021}.

The development of thin films of CoSn offers added control over its properties, where, for instance, ARPES studies reveal enhanced SOC effects and sharper flat bands compared to bulk samples \cite{shuyu_cheng_epitaxial_2023}. Additionally, strain engineering can shift energy levels, modulate Fermi surface topology, and explore emergent phases \cite{shuyu_cheng_epitaxial_2023,mojarro_tuning_2024}.

Compared to magnetic kagome metals like FeSn, CoSn lacks intrinsic magnetism, simplifying its electronic structure \cite{kang_topological_2020, liu_orbital-selective_2020}. Unlike superconducting AV$_3$Sb$_5$ materials, CoSn displays weaker correlations and no ambient CDWs or superconducting phases \cite{mojarro_tuning_2024}. Its simplicity makes CoSn a benchmark for kagome physics, highlighting the interplay between flat bands, orbital differentiation, and Dirac features.

Despite substantial progress, questions remain: How can CoSn be tuned into a robust topological regime? What mechanisms underlie the reported nematic order, and how is it connected to flat-band physics? Addressing these requires systematic doping experiments, thin-film engineering, and advanced spectroscopies. Further research on tunability and correlated phases holds the potential to unlock more exotic properties of this kagome system.

\subsubsection{\label{sec:chapIVA3}Mn-based: Mn$_3$Sn, Mn$_3$Ge}
\begin{figure}[t]
\centering
\includegraphics[width=\columnwidth,angle=0,clip=true]{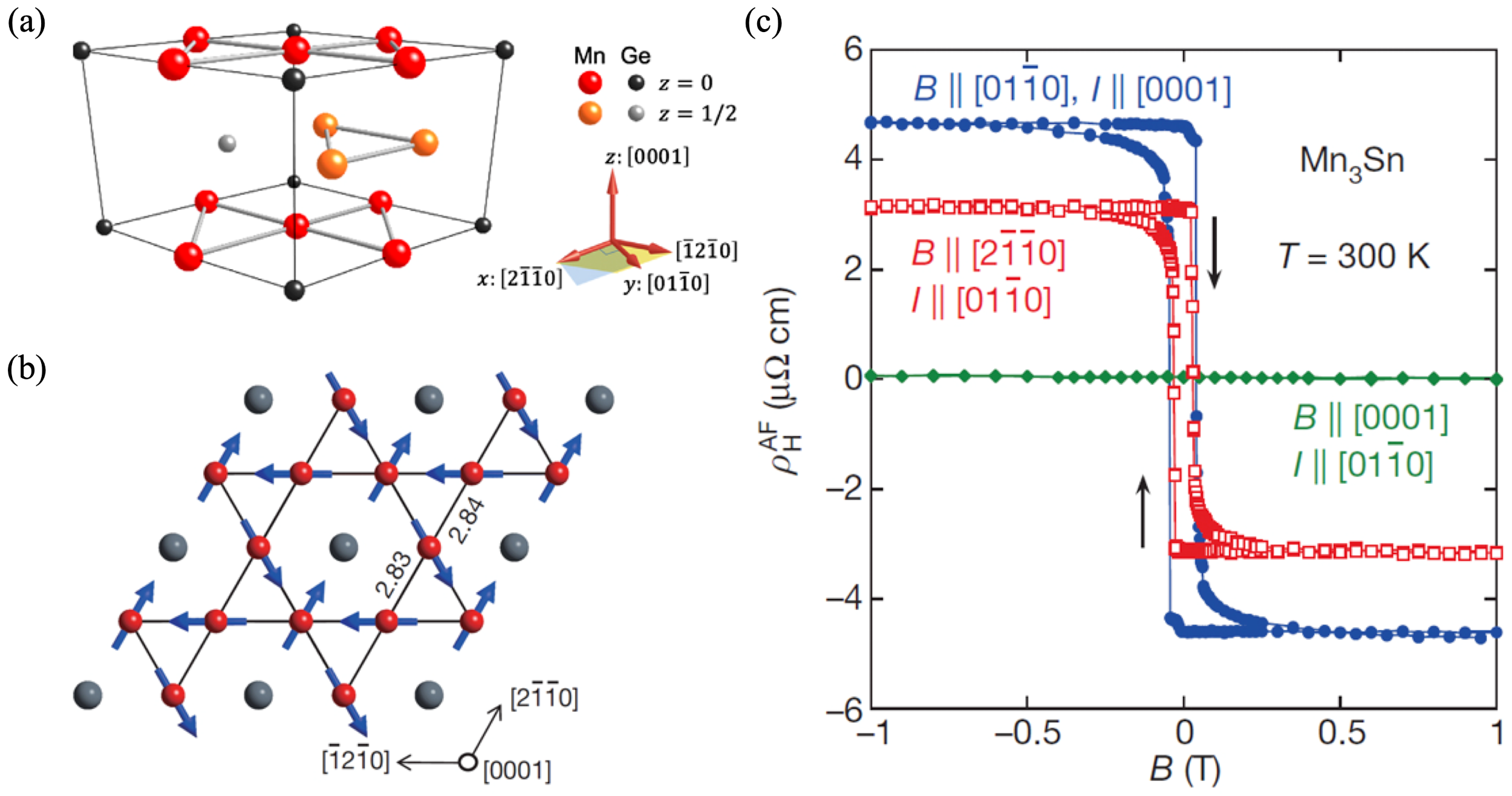}
\caption{(a) Crystal structure of \ce{Mn3Ge}, isostructural to \ce{Mn3Sn} (from \cite{kiyohara2016giant}). (b) Magnetic ordering in the kagome plane for \ce{Mn3Sn} and (c) Anomalous Hall response (from \cite{nakatsuji2015large}).}  
\label{Mn_structure}
\end{figure}

The Mn-based family of binary kagome compounds supports unusual magnetically ordered states.  Both \ce{Mn3Sn} and \ce{Mn3Ge} crystallize in the \ce{Ni3Sn}-type structure with hexagonal space group $P6_{3}/mmc$ (see Fig.~\ref{Mn_structure}(a)) and crystals can be grown via Bridgman techniques \cite{tomiyoshi1982polarized, kren1975study}. It is known that these structures are stabilized by excess Mn, which randomly occupies the Sn or Ge sites \cite{kren1975study, yamada1988magnetic}.  Both systems realize an exotic antiferromagnetic state which exhibits an anti-chiral spin texture (see Fig. \ref{Mn_structure}(b)) \cite{tomiyoshi1982polarized, nagamiya1982triangular}.  For \ce{Mn3Sn} this occurs at $T_{N} \approx 430$~K while for \ce{Mn3Ge} this is at $T_{N} \approx 380$~K.  The latter retains this magnetic structure to the lowest temperatures whereas the former has a transition to a non-coplanar magnetic phase below approximately 50~K \cite{nakatsuji2015large, nayak2016large, kiyohara2016giant}.  The Mn atoms host a total moment of approximately 3 $\mu_B$; however a small tilting of the in-plane moments results in a net ferromagnetic moment of order 0.001 $\mu_B$.  

A striking property of these systems is a large anomalous Hall response observed for magnetic fields in the kagome plane (see Fig.~\ref{Mn_structure}(c)) despite the system only having a very weak ferromagnetic component to its magnetic order \cite{nakatsuji2015large, nayak2016large, kiyohara2016giant}.  This is captured by calculations of Berry curvature contributions to the anomalous Hall response \cite{kubler2014non}.  The presence of Weyl nodes in these materials has been investigated theoretically  \cite{yang2017topological} and experimentally \cite{kuroda2017evidence}.

Several extensions have followed the above discoveries, including the observation of large responses in other anomalous transport responses including large Nernst \cite{ikhlas2017large} and thermal Hall \cite{li2017anomalous} effects.  Given its large anomalous Hall response but weak stray and demagnetizing field owing to its largely antiferromagnetic nature (as well as $T_{N}$ above room temperature), these systems have attracted significant attention for their possible utility in spintronics. Some of these properties can be rationalized from their connection to antiferromagentic Weyl systems \cite{liu2017anomalous}.  

\begin{figure}[!t]
\centering
\includegraphics[width=\columnwidth,angle=0,clip=true]{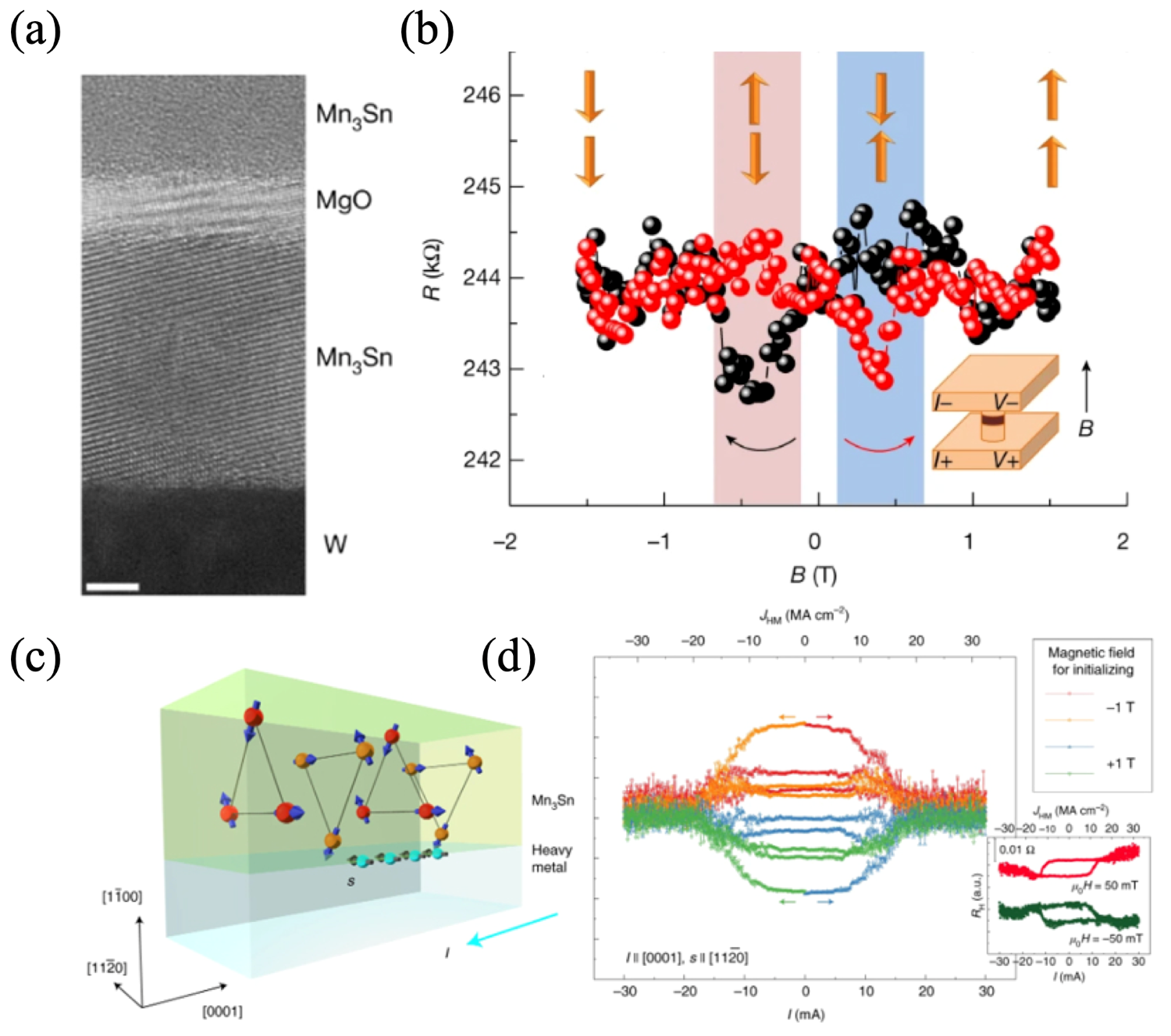}
\caption{(a) TMR device geometry and (b) response for \ce{Mn3Sn} (from \cite{chen2023octupole}).  (c) Kagome-heavy metal heterostructure and (d) spin-orbit torque switching for \ce{Mn3Sn} (from \cite{takeuchi2021chiral})), with initializing field shown upper inset and bipolar switching shown lower inset.}  
\label{Mn_spintronics}
\end{figure}

These compounds have been studied in a number of spintronic device geometries, and an early study confirmed the presence of spin Hall effects using NiFe electrodes \cite{kimata2019magnetic}. More recently, tunneling magnetoresistance (TMR) device structures using MgO barriers (see Fig. \ref{Mn_spintronics}(a)) demonstrated a tunneling resistance response at room temperature (see Fig. \ref{Mn_spintronics}(b)) \cite{chen2023octupole}.  In addition to this, spin-orbit torques have been demonstrated using conventional heterostructures of \ce{Mn3Sn} and heavy element metals (see Fig. \ref{Mn_spintronics}(c,d)) \cite{takeuchi2021chiral, higo2022perpendicular}.  This was extended in recent work on Si/\ce{SiO2}/\ce{Mn3Sn}/AlO$_{x}$ structures using all-electrical inputs \cite{deng2023all}.  The development of thin films of these materials has played a crucial role in these efforts \cite{higo2022thin}, and further development of spintronic devices based on these and related materials is a forefront area of research.  

\subsubsection{\label{sec:chapIVA4}Ni-based: Ni$_3$In}
\begin{figure}[!t]
\centering
\includegraphics[width=\columnwidth,angle=0,clip=true]{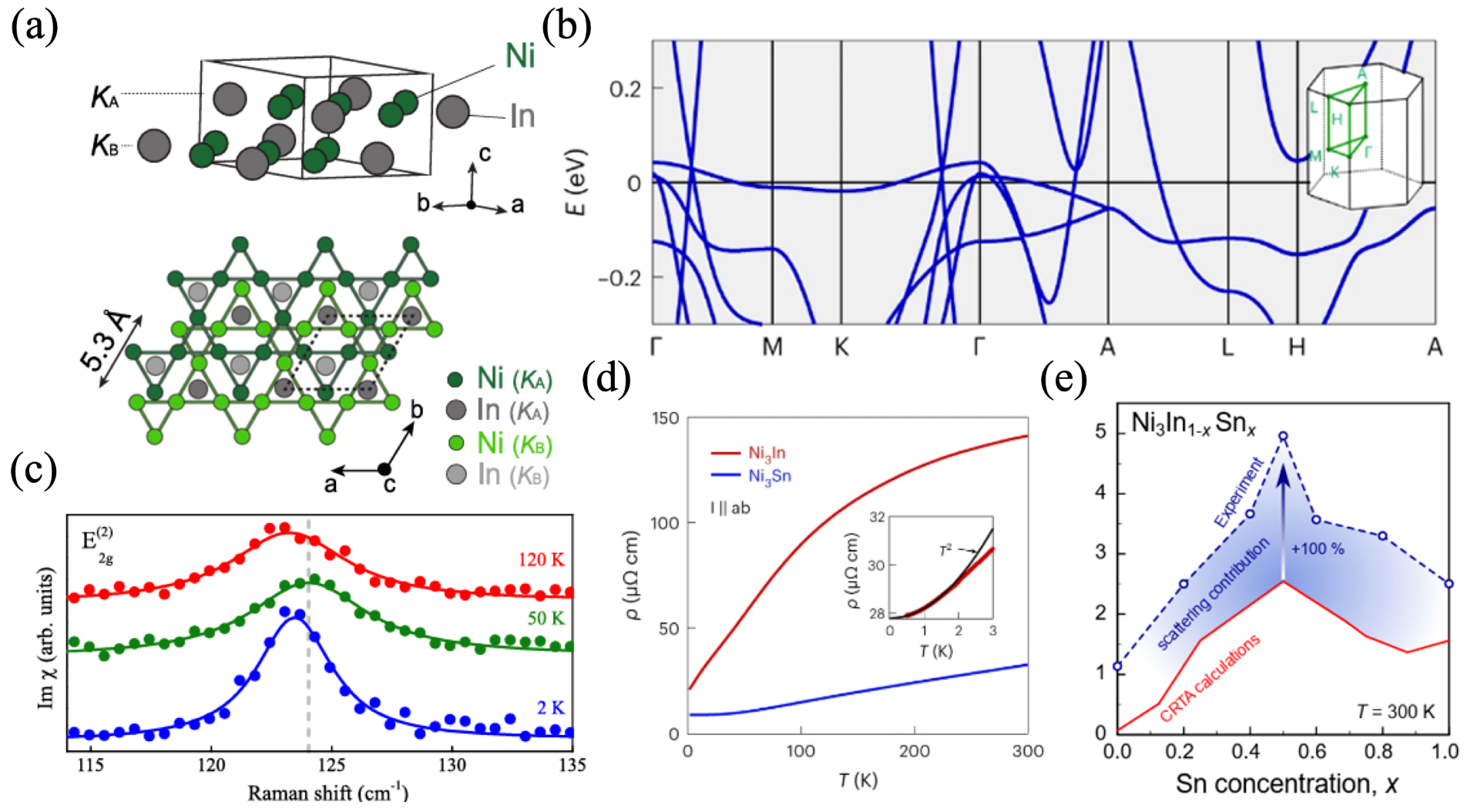}
\caption{(a) Crystal structure and (b) DFT electronic structure of \ce{Ni3In} (from \cite{han2024molecular} and \cite{ye_hopping_2024}, respectively). (c) Temperature dependence of Raman response as a function of temperature in single crystals of \ce{Ni3In} (from \cite{gim2023fingerprints}). (d) Electrical transport of \ce{Ni3In} and \ce{Ni3Sn} bulk single crystals (from \cite{ye_hopping_2024}). (e) Concentration dependence of calculated and observed power factor in polycrystalline \ce{Ni3(In,Sn)} alloys (from \cite{garmroudi2024high}). }  
\label{ni3in}
\end{figure}

The binary Ni-based kagome metal \ce{Ni3In} has been shown to support correlated electronic states.  The system crystallizes in the  \ce{Ni3Sn}-type structure with hexagonal space group $P6_{3}/mmc$ (see Fig.~\ref{ni3in}(a) \cite{baranova1966electron} and crystals can be grown via halogen catalyzed reaction and chemical vapor transport \cite{ye_hopping_2024}.  Recent studies of bulk single crystals \cite{gim2023fingerprints, ye_hopping_2024}, thin films \cite{han2024molecular}, and polycrystalline materials \cite{garmroudi2024high} have been reported.  Of central interest in this system is the expectation from DFT that (before interactions are considered) a partially flat band associated with the Ni ions appears at the Fermi level (see Fig.~\ref{ni3in}(b)).  

This electronic configuration appears to give rise to correlated electron phenomena reminiscent of heavy electron systems including $f$-electron Kondo systems \cite{checkelsky2024flat}.  Studies of Raman scattering exhibit a low temperature renormalization of frequency and lineshape that recalls that of the onset of Kondo screening (see Fig. \ref{ni3in}(c)) \cite{gim2023fingerprints}.  Electrical transport reveals a $T$-linear behavior below 100~K and points towards a strongly correlated Fermi liquid below 2~K (see Fig. \ref{ni3in}(d)) \cite{ye_hopping_2024, han2024molecular}. By applying magnetic field and pressure, this can be changed to a more conventional transport response \cite{ye_hopping_2024}.  Moreover, conventional metalicity is observed for the isostructural system \ce{Ni3Sn}, wherein the Fermi level is shifted away from the bands with quenched kinetic energy, suggesting the role of the flat band in generating strong electronic correlations. The consequences of correlation effects for band topology in these systems remains the subject of ongoing study.  

These band features have also been identified as a possible source of a large power factor $PF$ with potential relevance for metallic thermoelectrics \cite{garmroudi2023high}.  As shown in Fig. \ref{ni3in}(e), studies of alloys of \ce{Ni3In} and \ce{Ni3In} reveal a maximum in $PF$, consistent with the role of assymetric scattering across the Fermi level between the flat band states and highly dispersive bands \cite{garmroudi2024high}.  The design of such metallic thermoelectrics in kagome and other flat band metals (\textit{e.g.}, pyrochlores metals \cite{wakefield2023three}) is an emerging field of study.

\subsubsection{\label{sec:chapIVA5}Nb-based: Nb$_3$X$_8$}

The Nb$_3$X$_8$ family (X = Cl, Br, or I) comprises layered 2D materials with each unit consisting of two Nb$_3$X$_8$ monolayers stacked along the c-axis through weak van der Waals (vdW) interactions. Crystals can be produced via high-temperature sintering \cite{miller1995solid} or via halide liquid flux techniques \cite{haraguchi2017magnetic}, and within each monolayer, a breathing kagome lattice of Nb$_3$ trimers forms, where the intracluster Nb-Nb distance is significantly shorter than the intercluster distance. Strong metal-metal bonding within each trimer results in a localized spin $S=1/2$, primarily occupying the Nb $^2 a_1$ orbital \cite{yaoji_wang_quantum_2023,pasco_tunable_2019,gao_discovery_2023}. DFT calculations predict a metallic state; however, transport and ARPES measurements consistently indicate that Nb$_3$X$_8$ is an insulator \cite{gao_discovery_2023, sun_observation_2022,yoon_anomalous_2020}. This discrepancy may be attributed to strong electron correlations, suggesting that Nb$_3$Cl$_8$ and Nb$_3$Br$_8$ are Mott insulators rather than conventional band insulators, with the Fermi level within the Mott gap \cite{gao_discovery_2023,s_grytsiuk_ab_2019,zhang_mottness_2023}. 

Magnetic susceptibility measurements reveal a first-order phase transition in both Nb$_3$Cl$_8$ and Nb$_3$Br$_8$, from a high-temperature paramagnetic ($\alpha$-phase) to a low-temperature nonmagnetic ($\beta$-phase), accompanied by a structural transition. The transition temperature is $\approx 90$~K for Nb$_3$Cl$_8$ and $\approx 380$~K for Nb$_3$Br$_8$ \cite{pasco_tunable_2019}. At high temperatures, both adopt the $P3\bar{m}1$ space group in the $\alpha$-phase. In the $\beta$-phase, Nb$_3$Br$_8$ has a well-established $R3\bar{m}$ symmetry \cite{pasco_tunable_2019}, while recent evidence increasingly supports a similar assignment for Nb$_3$Cl$_8$ \cite{jeff_raman_2023,kim_terahertz_2023}. The structural transition is attributed to a mechanical shift of layers, altering the stacking order while retaining the same monolayer structure in both phases. 

Weak magnetic interactions between trimer clusters are strongly affected by charge fluctuations within Nb-Nb and Nb-Cl bonds, reflecting the interplay between structural and electronic degrees of freedom \cite{nakamura_charge_2024}. The nonmagnetic ground state has been proposed to arise from a spin-singlet configuration with a thermal gap to an excited triplet state, rather than charge disproportionation ([Nb$_3$]$^{7+}$-[Nb$_3$]$^{9+}$), as the core-level spectra indicate equivalent Nb valence states \cite{gao_discovery_2023}. Additionally, anomalous magnetic susceptibility enhancements around 10–20 K have been observed in specific samples, likely due to defect-induced spins or residual high-temperature phases retained during synthesis \cite{pasco_tunable_2019}. These sample-specific effects suggest an extrinsic rather than intrinsic origin. Notably, pressure-induced metallization in Nb$_3$Cl$_8$ is coupled with a structural phase transition and pronounced changes in its band structure \cite{shan_pressure-induced_2023}.

While flat bands have been identified in several kagome metals, the lack of exfoliatable and semiconducting kagome materials significantly limits their integration into device applications. In a breathing kagome lattice, the absence of inversion symmetry opens a gap in the Dirac cone, resulting in a semiconducting ground state, while flat bands persist due to the protection of mirror reflection symmetry \cite{sun_observation_2022}. Unlike flat bands in an ideal kagome lattice, those in the trigonally distorted lattice of Nb$_3$X$_8$ are highly isolated. Recent ARPES experiments have observed flat bands across various breathing kagome systems, including Nb$_3$Cl$_8$, Nb$_3$Br$_8$, Nb$_3$I$_8$, and Nb$_3$TeCl$_7$ \cite{gao_discovery_2023,sun_observation_2022,sabin_regmi_observation_2023,sabin_regmi_spectroscopic_2022,zhang_topological_2023}. These observations align well with theoretical band calculations, confirming that the flat bands originate from the breathing kagome lattice of niobium atoms, predominantly involving niobium $d$-orbital character. Furthermore, energy splitting in the lower Hubbard band of bilayer $\beta$-Nb$_3$Cl$_8$ has been attributed to enhanced interlayer hybridization, resulting in the formation of bonding-antibonding states in the $\beta$-phase, further underscoring the role of Mott physics in breathing kagome systems \cite{pasco_tunable_2019,s_grytsiuk_ab_2019,zhang_mottness_2023}. 

\begin{figure*}[t]
\centering
\includegraphics[width=\textwidth,angle=0,clip=true]{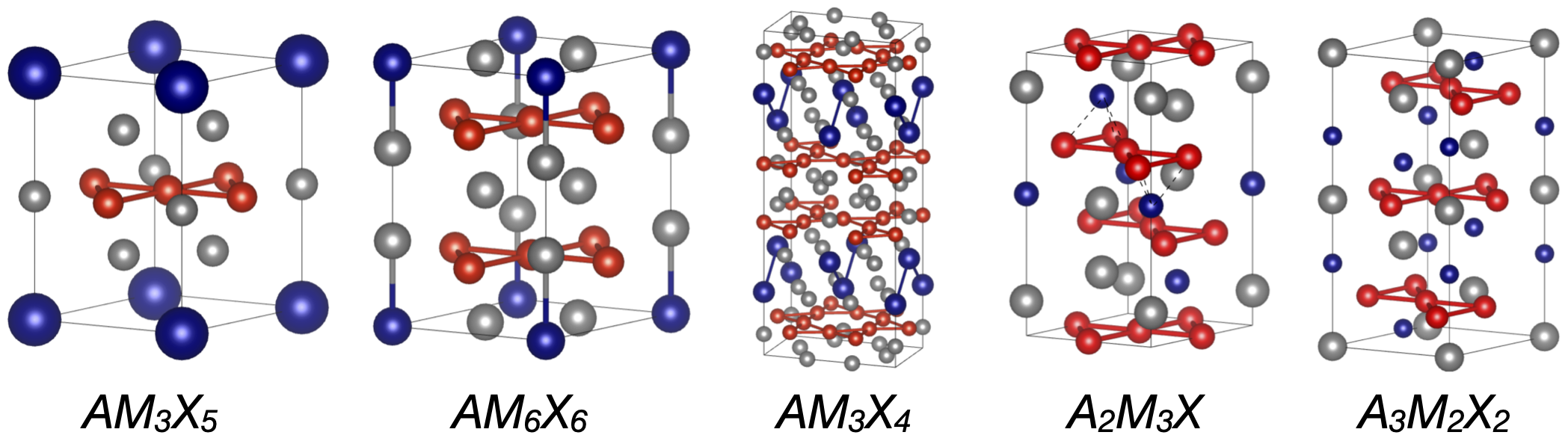}
\caption{Crystal structures of common ternary families possessing isolated kagome networks. Kagome networks and the constituent atoms are shown as red spheres connected by red bonds.  The out-of-plane bonds highlighted in the bilayer AM$_6$X$_6$ structure illustrate the atoms driving the chain-instability underlying the CDW-like state in select variants.  Blue spheres and bonds shown in the AM$_3$X$_4$ structure illustrate the zig-zag network of interleaving A-site ions.  Dashed lines in the ternary C15 Laves AM$_3$X structure show the native pyrochlore network broken up by the ordered incorporation of X-site ions.  
} 
\label{ternarystructures}
\end{figure*}

The magnetic order in the Nb$_3$X$_8$ family is predicted to depend on the number of layers in thin film samples, with unique phenomena emerging as the system approaches the 2D limit \cite{conte_layer-dependent_2020}. The van der Waals nature of Nb$_3$X$_8$ enables easy exfoliation into monolayers, providing a rare opportunity to explore its magnetic properties, such as monolayer ferromagnetism, all-in antiferromagnetic spin structures, and quantum spin-liquid states \cite{yaoji_wang_quantum_2023,peng_intrinsic_2020}. More importantly, the intrinsic breaking of both time-reversal and inversion symmetries positions Nb$_3$I$_8$ as a remarkable 2D anomalous valley Hall material, capable of exhibiting spontaneous valley polarization without the need for external tuning \cite{peng_intrinsic_2020}. In ferroelectric bilayer Nb$_3$I$_8$ with A-type antiferromagnetic coupling, the coupling of electric polarization to spin, valley, and layer degrees of freedom enables the realization of a spin-valley-layer polarized AHE \cite{feng_nonvolatile_2023}. Furthermore, mechanical exfoliation experiments on Nb$_3$Cl$_8$ have successfully yielded few- and single-layer samples with excellent ambient stability and enhanced conductivity \cite{yoon_anomalous_2020,sabin_regmi_observation_2023}. These exceptional properties, along with the novel low-dimensional magnetic behaviors, establish the Nb$_3$X$_8$ family as a versatile platform for exploring fundamental physics and advancing device applications.


\subsection{\label{sec:chapIVB}Ternary}

The incorporation of a third element into the crystal structures of kagome metals affords an additional degree of tunability and chemical diversity not achievable in simpler binary compounds.  Layered kagome planes can be better isolated via layers of interleaving ions, rendering reduced out-of-plane hopping and more two-dimensional electronic properties, and additional modes of symmetry breaking can be interfaced with the kagome network.  Key examples are illustrated in Fig. \ref{ternarystructures}.  For instance, magnetic layers can be proximitized with kagome planes or alternate units that undergo bond/charge order coupled to the kagome network, all within the same structural unit.  There is a broad phase space of ternary structure types that host kagome networks and allow for this tunability, much of it unexplored. In this section we highlight some of the most commonly studied variants to date.  

\subsubsection{\label{sec:chapIVB1}M$_3$A$_2$X$_2$ compounds}
\begin{figure}[!t]
\centering
\includegraphics[width=\columnwidth,angle=0,clip=true]{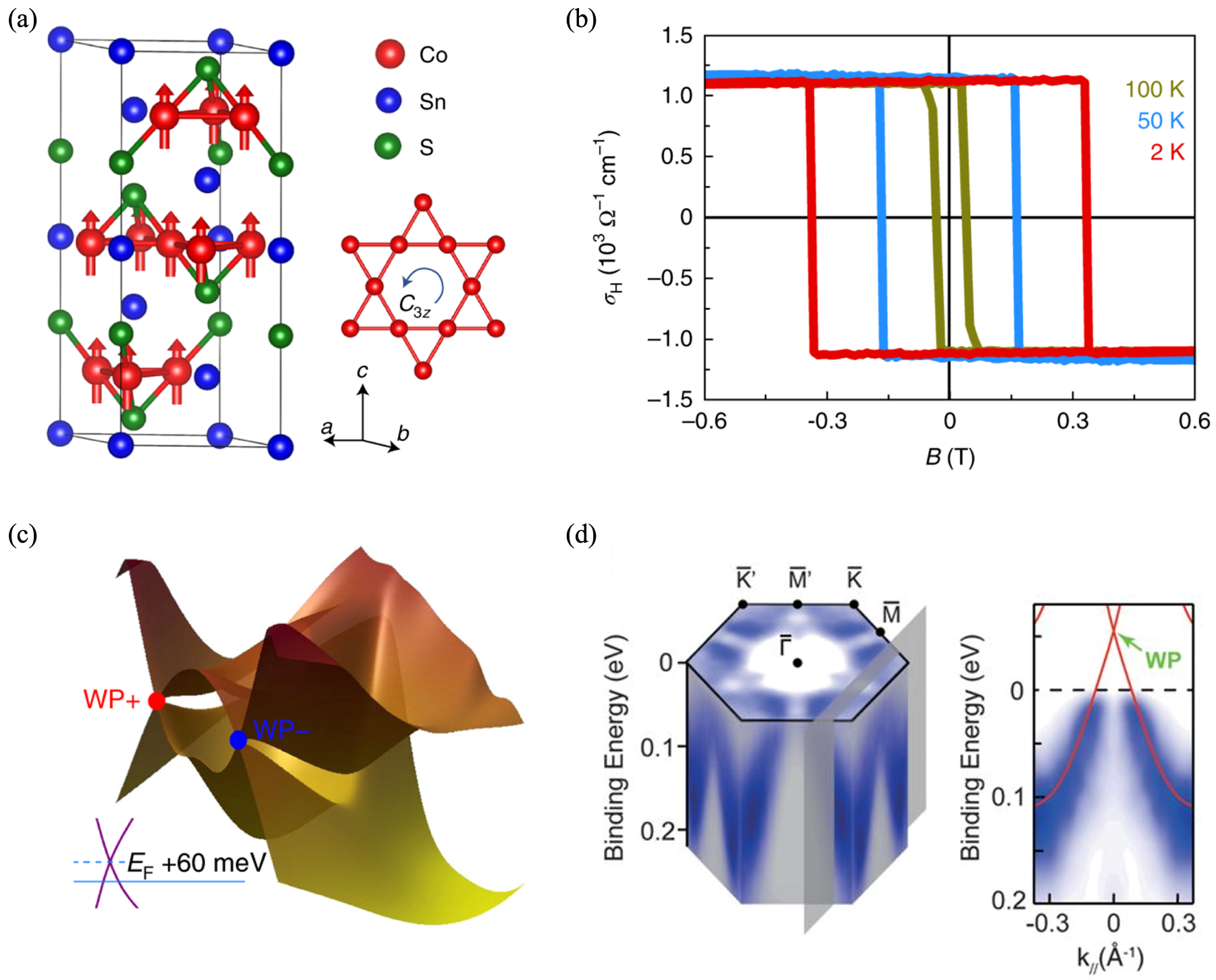}
\caption{(a) crystal structure of layered kagome metal Co$_3$Sn$_2$S$_2$. (b) giant anomalous Hall conductivity. (c) low-energy electronic structure featuring a pair of Weyl points. (d) Fermi surface and low-energy bands showing Weyl point right above $E_F$. Panels adapted from \cite{liu_giant_2018,defa_liu_magnetic_2019}.}
\label{Co3Sn2S2}
\end{figure}

The M$_3$A$_2$X$_2$ or shandite crystal structure consists of kagome planes of M-site transition metal ions coordinated with main group A-site ions and X-site chalcogens (Fig.~\ref{Co3Sn2S2}(a)).  Bulk crystals can be grown typically via Bridgman \cite{m_holder_photoemission_2009} or sintering techniques with mineralizers \cite{guowei_li_surface_2019}. The kagome layers are stacked in an ABC sequence and the triangles of the kagome network have alternately one X-site ion above and below them.  There are two unique A-site positions in the cell and one rests within the kagome planes.  The most widely studied material in this class in terms of its electronic properties is the half metal ferromagnet Co$_3$Sn$_2$S$_2$. Initial work \cite{p_vaqueiro_powder_2009} reported ferromagnetic order at a Curie temperature $T_C \approx 180$~K and metallic behavior, with magnetic moments localized within the Co kagome lattice. 

Work exploring the electronic structure of Co$_3$Sn$_2$S$_2$ identified the compound as an intrinsic magnetic Weyl semimetal. Liu \textit{et al.} observed a giant AHE (Fig.~\ref{Co3Sn2S2}(b), attributed to Berry curvature concentrated around Weyl points near the Fermi energy (Fig.~\ref{Co3Sn2S2}(c).  DFT calculations predicted six Weyl points within 60 meV of the Fermi level, and experiments confirmed a large anomalous Hall conductivity ($\sigma_{xy}$ up to 1100 $\Omega^{-1}$ cm$^{-1}$) \cite{liu_giant_2018}, with a modulation of the anomalous Hall angle up to a magnitude of $25^{\circ}$~\cite{yang2025modulation}.  \textcite{wang_large_2018} further confirmed the presence of intrinsic ferromagnetism (moment $\approx$0.3 $\mu_B$ per Co atom) and Weyl points near the Fermi energy using ARPES and DFT.  This study provided direct evidence of band crossings and highlighted the connection between topology and magnetism as the source of the exceptional transport properties.  In the same year, \textcite{yang_giant_2020} reported a giant ANE with a significant Nernst conductivity up to 10 $\mu$V/K at low temperatures, further emphasizing the role of Berry curvature in anomalous transverse thermoelectric responses.

Focusing on the surface,  \textcite{noam_morali_fermi-arc_2019} used ARPES and STM to observe diverse surface terminations and confirm the connectivity of bulk Weyl cones via Fermi arcs, while \textcite{qiunan_xu_topological_2017} resolved topological surface states linked to Weyl nodes and SOC.  The character of the states at the Fermi level, however, varies depending on the surface termination probed, leading to discrepancies in the literature \cite{Mazzola2023}. \textcite{defa_liu_magnetic_2019} mapped the bulk band structure and topology in the magnetic Weyl semimetal phase, linking the giant AHE and anomalous Hall angle to Berry curvature contributions from Weyl nodes  (Fig.~\ref{Co3Sn2S2}(d)).  

Interestingly, \textcite{jia-xin_yin_negative_2019} discovered ``negative-flat band magnetism", with a negative effective $g$ factor, driven by the kagome lattice and dominated by SOC effects.  This study proposed that SOC lifts degeneracies in flat bands, creating an orbital moment that suppresses ferromagnetism.  The magnetoresistance (MR) and planar Hall effect (PHE) were conencted to topological Berry curvature in Weyl nodes \cite{shama_observation_2020} linking PHE anomalies to magnetic field rotation. \textcite{sandeep_howlader_domain_2020} visualized temperature-dependent domain structures using MFM, demonstrating strong pinning of domain walls below 130~K .

Additional studies focusing on correlations between magnetic phases, doping, and Berry-phase physics brought advances in understanding the magnetic and electronic properties of Co$_3$Sn$_2$S$_2$. A neutron diffraction and DFT investigation detailed the magnetic exchange couplings, predicting a Weyl-dominated intermediate temperature regime  with 120° AFM orders \cite{qiang_zhang_unusual_2021}. Other neutron and magnetization studies instead identified an intermediate crossover driven by domain wall nucleation \cite{j_soh_magnetic_2022}. Yanagi et al. explored hole- and electron-doped variants, revealing that doping induces band parity inversions and enhances anomalous Nernst conductivity \cite{y_yanagi_first-principles_2020}. Shen et al. reported extrinsic AHE enhancement via Fe and Ni doping, introducing Berry curvature hotspots \cite{jianlei_shen_33_2020, thakur_intrinsic_2020}.

Other investigations have emphasized experimental tunability. \textcite{f_schilberth_generation_2024} showed how magnetization reorientation can generate new Weyl nodes or nodal lines, modulating Berry curvature response. \textcite{abhirup_roy_karmakar_giant_2022} found that uniaxial strain enhances thermal Hall and Hall conductivity by realigning band topology.  \textcite{miuko_tanaka_topological_2020} demonstrated fabrication of Co$_3$Sn$_2$S$_2$ thin films using CVT synthesis, achieving quantum-limit mobility, and \textcite{hongjun_gao_synthesis_2023} synthesized ultra-pure crystals with record-setting 10,490 cm$^2$/V$\cdot$s mobility and 2500\% MR.  Recently, in fact, there are also observations of chiral phonons (see Sec.~\ref{sec:chapIIC}) rising from the coupling between the electronic topology and the magnetic order~\cite{yang2025inherent,che2025magnetic}.


Other than Co$_3$Sn$_2$S$_2$, a number of related shandite kagome systems have been explored. Substituting Sn by In and Co by Ni provides access to nonmagnetic counterparts such as Ni$_3$In$_2$S$_2$~\cite{fang2023record}, which was recently shown to host Dirac nodal lines~\cite{zhang2022endless}. Intermediate substitution in Co$_3$Sn$_{2-x}$In$_x$S$_2$ tunes the magnetic order and electronic structure, while Co$_3$In$_2$S$_2$ itself shows a field-induced butterﬂy-like anisotropic MR~\cite{pielnhofer2014half,corps2015interplay,weihrich2004magnetischer,lv2025field}. Beyond the Co- and Ni-based shandites, compounds incorporating Rh or Pd have also been investigated from first-principles methods, although their properties remain far less explored~\cite{buiarelli2025first}. These materials enlarge the accessible parameter space of the kagome shandite family, spanning from magnetic Weyl semimetals to nonmagnetic nodal-line systems, and suggest that further investigation of the less studied members may reveal yet unexplored topological and correlated phases.

\subsubsection{\label{sec:chapIVB2}AM$_3$X$_5$ compounds}
AM$_3$X$_5$ compounds crystallize in a layered structure with kagome nets formed via the M-site (M=V, Ti, Cr) ion which are octahedrally coordinated via the X-site (X=Sb, Bi). M$_3$X$_5$ blocks are separated by triangular lattice layers of A-site (A=K, Rb, Cs) ions, forming a chemically and electronically two-dimensional structure.  X-site ions occupy two sites in the lattice.  The first is in the kagome plane, within the voids of the hexagons of the M-site kagome network, and the second is within a honeycomb network above and below the kagome planes. While the interlayer bonding is ionic, the lattice remains highly exfoliable, and layers as thin as two unit cells have been achieved \cite{song2023anomalous} as well as some degree of gate tunability \cite{zheng2023electrically}.    

\paragraph{AV$_3$Sb$_5$:}

AV$_3$Sb$_5$ (A=K, Rb, Cs) compounds were reported as a new structure type, featuring a V-based kagome sublattice with kagome-derived saddle points close to $E_F$ \cite{ortiz2019new}. Bulk crystals can be grown via crucible-based self-flux techniques \cite{PhysRevLett.125.247002}, where a high-temperature CDW transition was observed in electrical transport, heat capacity and magnetic susceptibility followed by a low-temperature transition into a superconducting state \cite{PhysRevLett.125.247002, Yin_2021, PhysRevMaterials.5.034801} as shown in Fig. \ref{AVSTransport}.  

\begin{figure}
\centering
\includegraphics[width=\columnwidth,angle=0,clip=true]{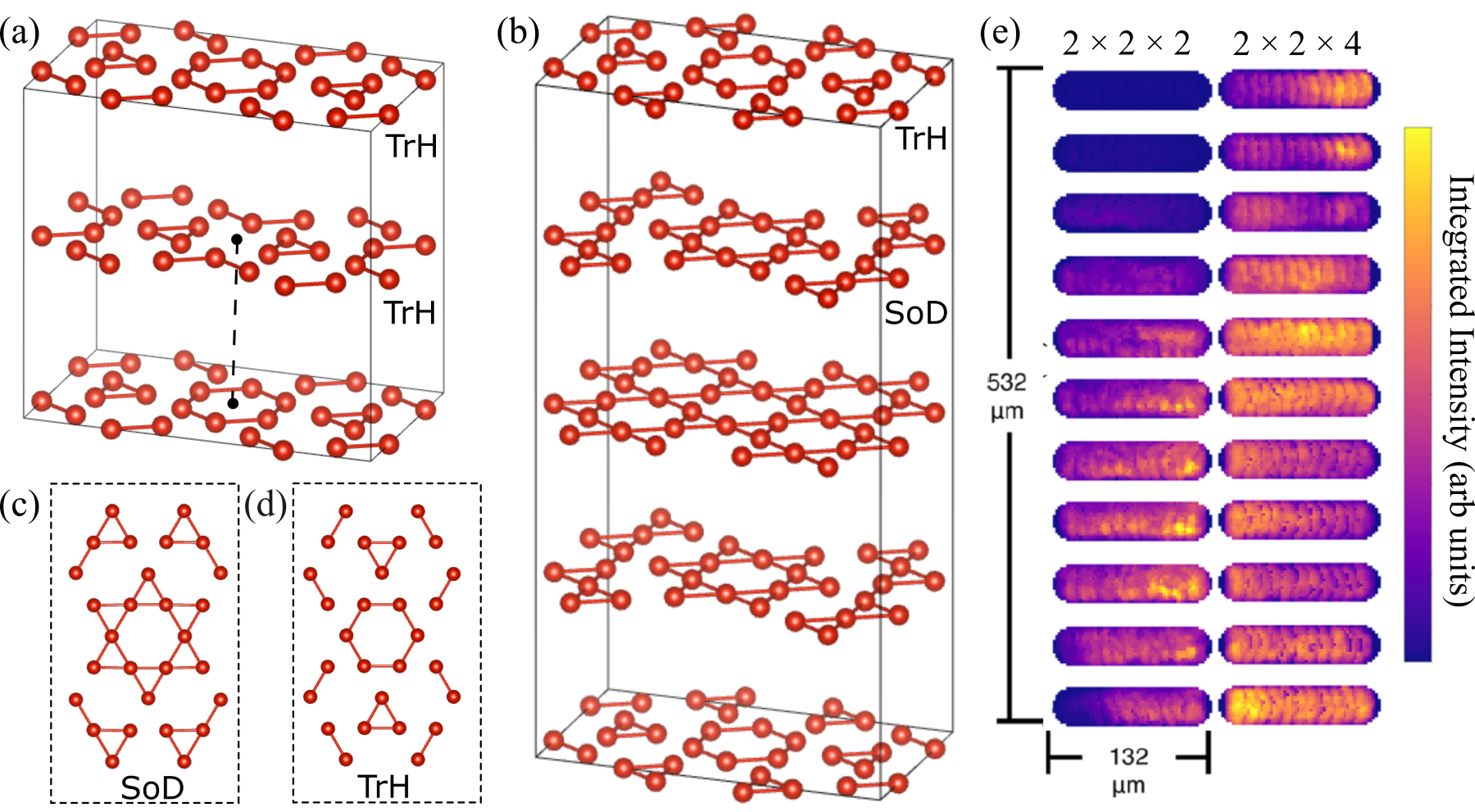}
\caption{(a) Staggered tri-hexagonal structure that comprises the $2 \times 2 \times 2$ reconstructed cell in the CDW state of RbV$_3$Sb$_5$ and KV$_3$Sb$_5$. (b) Average structure of the $2 \times 2 \times 4$ reconstructed cell in the CDW state of CsV$_3$Sb$_5$. (c) The star-of-David pattern of distortion for a single kagome layer (d) the tri-hexagonal pattern of distortion for a single kagome layer (e) Real space mapping of the $2 \times 2 \times 2$ and $2 \times 2 \times 4$ cells in a CsV$_3$Sb$_5$ crystal, showing phase separation and competition between the two states.  Data are reproduced from \cite{PhysRevMaterials.8.093601}.}
\label{CDWStructure}
\end{figure}

\begin{figure}
\centering
\includegraphics[width=\columnwidth,angle=0,clip=true]{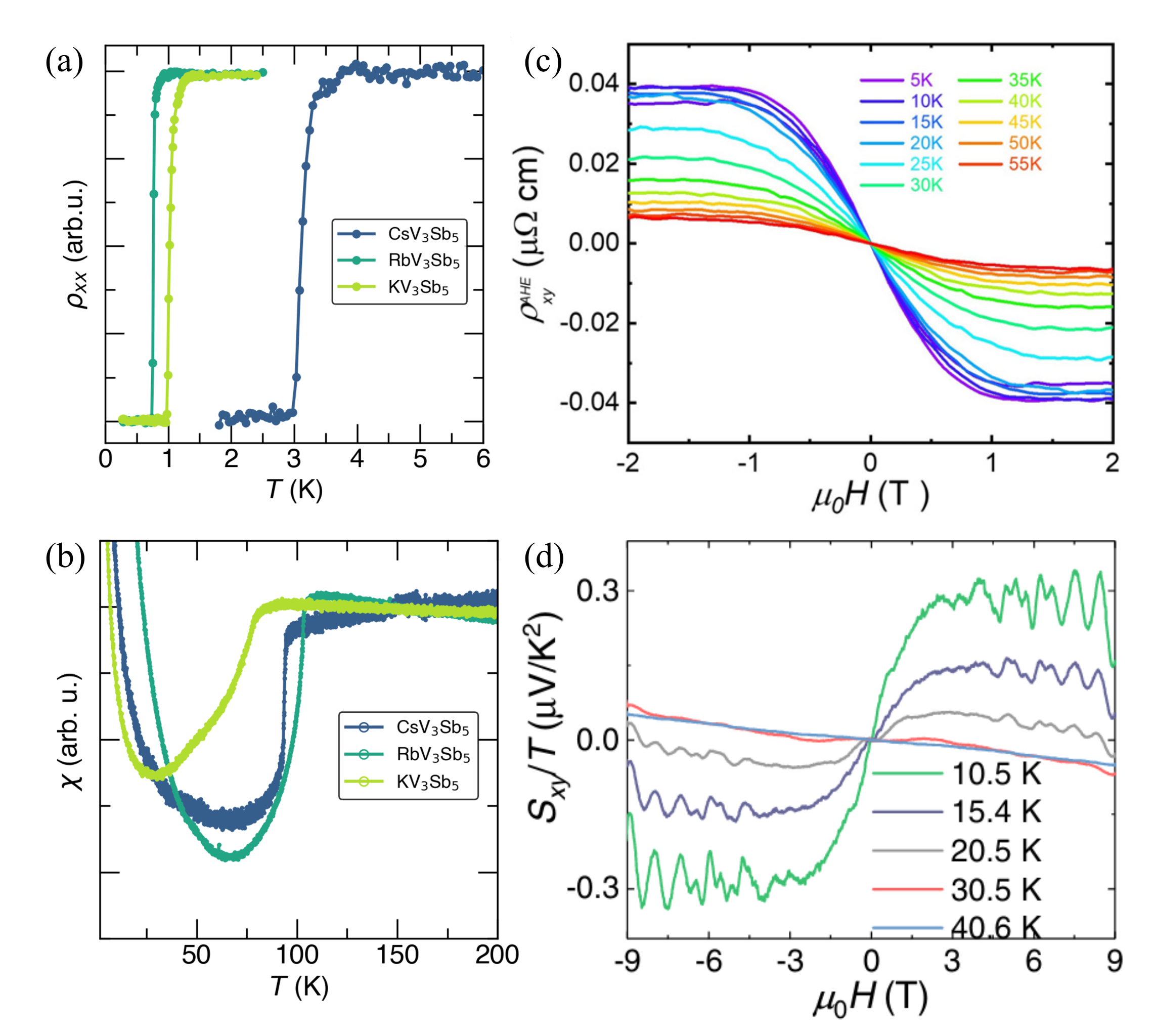}
\caption{(a) Resistivity data showing the superconducting transitions for AV$_3$Sb$_5$ compounds. (b) Magnetization data showing the CDW transitions for AV$_3$Sb$_5$ compounds. (c) Magnetotransport data showing the onset of the AHE below the CDW transition for CsV$_3$Sb$_5$.  Figure is reproduced from \cite{PhysRevB.104.L041103} (d) ANE data suggesting the onset of broken TRS below 30 K.  Figure is reproduced from \cite{PhysRevB.105.L201109}.}
\label{AVSTransport}
\end{figure}

\textit{Charge density wave state:} The CDW order that forms is three-dimensional in nature \cite{PhysRevX.11.031026, PhysRevX.11.031050, PhysRevX.11.041030, hongen_zhu_electronic_2023,li_coexistence_2022} and forms a staggered tri-hexagonal (inverse star-of-David) pattern of order that is shifted by half a unit cell from layer to layer along the $c$-axis \cite{PhysRevLett.127.046401} as shown in Fig. \ref{CDWStructure}.  The distortion and variants of it can be parameterized by three modes along high-symmetry directions in the BZ \cite{PhysRevB.104.214513}. 

The importance of the $e-ph$ interaction in forming the CDW state is suggested by a number of experimental results. High-resolution ARPES on KV$_3$Sb$_5$ observed a CDW-induced Fermi surface reconstruction, associated band folding and gap opening at the boundary of both pristine and reconstructed BZs, with reported typical signatures of $e-ph$ coupling such as kinks~\cite{luo2022electronic}. Broadband optical spectroscopy observed phonon anomalies above and below the CDW transition in KV$_3$Sb$_5$, indicating a strong coupling of phonons to the underlying electronic structure~\cite{ece2022optical}. Neutron scattering experiments also reveal that the CDW order in CsV$_3$Sb$_5$ is linked to a static lattice distortion and a sudden hardening of the B$_{3u}$ longitudinal optical phonon mode at the BZ boundary, further highlighting the significant role of a wave vector-dependent $e-ph$ coupling ~\cite{xie2022electron}.

One notable exception to the standard $2\times2\times2$ enlargement of the unit cell is the presence of a coexisting  $2\times2\times4$ CDW order in CsV$_3$Sb$_5$ that seemingly also possesses an interleaved star-of-David character distortion \cite{l_kautzsch_structural_2022, PhysRevResearch.5.L012032, PhysRevB.105.195136, ning2024dynamical}.   Regardless of this detail, all three V-based 135 variants show broken rotational symmetry in the CDW state as the crystal symmetry is lower from hexagonal to orthorhombic, though this symmetry breaking is largely driven by shifts in the interlayer phasing of breathing modes between kagome planes versus broken rotational symmetry in any one kagome layer, which is seemingly preserved.  Between the onset of the high-temperature CDW transition and low-temperature SC state, a host of unconventional phenomena have been reported in V-based 135 materials, and more extensive, focused reviews can be found in~\textcite{s_d_wilson_av3sb5_2023} and~\textcite{k_jiang_kagome_2021}.  In the following sections, we highlight some of the key phenomenology reported to date.

As discussed in \ref{sec:chapIIIA}, the possibility of flux order on the kagome lattice is a major route of inquiry regarding whether TRS is broken in the CDW state \cite{fernandes2025loopcurrentorderkagomelooking}.  This can occur due to the presence of \textit{p}-type VHS at $E_F$ and theoretical models have been put forward utilizing this fact \cite{PhysRevLett.132.146501, Zhou2022pockets, lin2021complex}.  Electrical and thermal magnetotransport studies identified the onset of an AHE \cite{yang_giant_2020, PhysRevB.104.L041103, Wang_2023_RVS} and an ANE \cite{PhysRevB.105.L201109, PhysRevB.105.205104} below the CDW transitions of the V-based 135 compounds. There is however no spontaneous (zero-field) component to the AHE; instead, at low fields, a strong nonlinear Hall response is observed. The origin of this is debated.  It can potentially arise, for instance, due to TRS breaking, Berry curvature effects from partial gapping of topological electronic bands \cite{liang2018anomalous}, or from sharp velocity changes along Fermi contours in the CDW-reconstructed band structure \cite{PhysRevB.110.024512}.

Further magnetotransport data identified the onset of a chiral transport term at low temperatures (within the CDW state) implying either a field-tunable structural chirality or the presence of a TRS breaking order parameter \cite{Guo2022chiralTransport}.  Coincident with the onset of the CDW, $\mu$SR studies report the onset of an unusual source of depolarization, suggestive of the appearance of a weak magnetic moment \cite{mielke2022time,PhysRevResearch.4.023244, Guguchia2023}.  Supporting this, tuning fork resonator measurements report the formation of a magnetic state below 30 K within the CDW state of CsV$_3$Sb$_5$ \cite{gui2025probing} and a superconducting diode effect is observed in the absence of an external magnetic field \cite{le2024superconducting}.  Similarly, STM measurements report a field-tuned pattern of CDW order \cite{jiang_unconventional_2021}, which will be discussed at greater length later in this section.  

Optical measurements have also extensively characterized the CDW transition. Frequency-dependent optical conductivity measurements have quantified the band gap and parameterized the relative role of electronic correlations across the 135 family \cite{ece2022optical, PhysRevB.104.045130, PhysRevB.105.245123}, and coherent phonon and Raman spectroscopy identified the evolution of lattice modes through the CDW transition \cite{PhysRevMaterials.5.L111801, PhysRevB.104.165110, PhysRevB.105.155106, PhysRevResearch.4.023215}.  Intriguingly, optical studies resolved the onset of circular dichroism and a subdomain structure in the CDW state, consistent with the appearance of magnetic domains \cite{xu_three-state_2022}.  This result is, however, debated.  It is seemingly at odds with polar Kerr measurements using Sagnac interferometry that fail to resolve signs of \textbf{q}=0 magnetic order \cite{PhysRevLett.131.016901, PhysRevMaterials.8.014202}.

Magnetic torque measurements also seemingly report the onset of TRS breaking, though at temperatures slightly above the onset of the CDW transition \cite{asaba2024evidence}, inviting further studies exploring fluctuation phenomena above the CDW transition.  Select x-ray studies have identified short-range order or fluctuations in this regime \cite{subires2023order, chen2022charge}, though the transition itself is weakly first-order. However, \textcite{scheurer2025spin} have recently argued that the torque response above the CDW temperature cannot be explained in terms of conventional CDW fluctuations or intraband magnetic order. Instead, an interband ordering mechanism, enabled by symmetry-allowed interband SOC, could naturally account for the observed odd-parity TRS-breaking nematic torque signal and associated hysteresis. In this picture, the experimental observation is interpreted as a crossover rather than a true phase transition, challenging the view that nematic or fluctuation-driven physics alone underlies the torque response.   

\textit{Electronic structure:} As discussed as an example earlier in Sect. \ref{sec:chapIIA2}, the electronic structure of V-based 135 compounds contains a multiorbital mixture of V d-states and Sb p-states crossing $E_F$.  The key features are generally believed to be the p-type VHS (derived from the V $d$-states) that are close to or cross the Fermi level and an electron-like pocket at the $\Gamma$-point derived from the Sb $p_z$ states.  Bulk quantum oscillation measurements identify several orbits with a nontrivial Berry phase in the CDW state \cite{PhysRevB.105.024508, PhysRevLett.127.207002} and analysis of the band structure classifies these materials as Z$_2$ topological metals with a protected surface state \cite{PhysRevLett.125.247002, PhysRevLett.127.046401}.

ARPES studies, in particular, have been instrumental in establishing the electronic band structure of AV$_3$Sb$_5$ kagome metals and its role within the rich landscape of symmetry-broken phases. Reviews by \textcite{yong_hu_electronic_2023} and \textcite{yigui_zhong_photoemission_2024} compile the growing evidence of multi-phase coexistence and ARPES-based spectral features relevant to kagome SC, and here we highlight a few key points. Early ARPES studies reported a momentum-dependent CDW gap, showing that the electronic states near the $M$-points were particularly affected, whereas those near the $\Gamma$-point remained gapless \cite{k_nakayama_multiple_2021,wang_distinctive_2021,cho_emergence_2021}. Subsequent works focused on the mapping of low-energy band features, uncovering multiple VHS near the Fermi level and identifying $m$-type (saddle point) and $p$-type (symmetric) VHS \cite{yong_hu_rich_2022,kang2022twofold}.

Using angle- and photon-energy-dependent ARPES, it was further shown that CsV$_3$Sb$_5$ exhibits $k_z$-dependent VHS, indicating a three-dimensional (3D) nature of the electronic structure. These studies shed initial light on the link between VHS and CDW formation by demonstrating Fermi surface nesting, a hallmark of electronic instability. However, a study of KV$_3$Sb$_5$ further revealed energy gaps due to CDW-induced band folding and propose electron-phonon coupling as a primary driver of CDW transitions \cite{luo2022electronic}. The growing focus on electronic reconstruction led to a study of the CDW-induced peak-dip-hump structures in CsV$_3$Sb$_5$ \cite{lou_charge-density-wave-induced_2022}. ARPES data in this work indicated that the CDW phase strongly modulated the low-energy excitations, reminiscent of physics seen in strongly correlated superconductors. Collectively, these studies established that CDW ordering reorganizes the electronic structure of kagome metals at the Fermi surface, modifying both the DOS and Fermi surface topology.

\textit{Scanning tunneling microscopy:} A great deal of electronic complexity has been revealed by STM measurements of 135 compounds. A variety of surface terminations and reconstructions contribute to this complexity \cite{zhao2021cascade,chen2021roton,PhysRevX.11.031026, Wang2021STM-CVS,Xu2021STM, Li2022STM_RSB} with a large number of possibilities born from the mobile alkali atoms on the surface \cite{zhao2021cascade, Yu2022RVS_STM, Han2023STM_RVS, Nie2022nematicity, meng2023STM_RVS}.  In general, STM experiments have primarily focused on studying the Sb surface termination after sweeping away these mobile $A$-site ions.


The 2 $\times$ 2 CDW order in the kagome plane is detectable in STM images on both $A$ \cite{PhysRevX.11.031026, Nie2022nematicity,Li2022STM_RSB} and Sb \cite{jiang_unconventional_2021,zhao2021cascade, PhysRevX.11.031026} surfaces. While the CDW state clearly breaks the translation symmetry of the lattice, which rotation symmetries are broken by this order remain debated. The results can be broadly categorized into two pictures. The first involves breaking of all reflection symmetries in the kagome plane, resulting in an inequivalence of the three CDW peaks in STM data and giving rise to a chiral CDW state \cite{jiang_unconventional_2021, Shumiya2021STM}. The second CDW picture, instead maintains a single in-plane reflection symmetry, while the other two are still broken \cite{Li2022STM_RSB, Nie2022nematicity}. Experimentally, the distinction is whether two of the three inequivalent $M$-point peaks in the Fourier transform of the CDW pattern are identical or whether all three are distinct.  It is debated whether the apparent inequivalence of all three CDW peaks can be explained by STM tip anisotropy \cite{Li2022STM_RSB} or if strain can locally modulate the intensity of CDW peaks \cite{Xing2024opticalM, Wang2023strainSTM}.  Temperature-dependent STM measurements revealed that the symmetry breaking signal persists up to at least 60 K \cite{Li2023STM_35K}, and three 120$^{\circ}$-rotated domains \cite{Li2022STM_RSB,Nie2022nematicity} are imaged. 
 This is consistent with the orthorhombic twinning and micro-sized domain formation imaged by optical birefringence measurements \cite{xu_three-state_2022}. 

Signatures of broken-TRS within the $2\times2$ CDW phase were first reported in STM measurements. Evidence here comes from examining the $M$-point intensities in Fourier-transformed STM data as a function of magnetic field, where $B$ is applied and then reversed along the interlayer $c$-axis. The reversal of the magnetic field is reported to change the relative intensities of the CDW peaks in some experiments, resulting in an apparent switch in the chirality of the CDW order \cite{jiang_unconventional_2021, Xing2024opticalM}. Equivalent experiments performed by other groups, however, report no field-induced change to the CDW signal \cite{Li2022STM_RSB, Li2022hai-huwen, Candelora2024}. Discrepancies between the measurements potentially arise from one of two origins.  The first is that the apparent change in the CDW intensities under different magnetic fields results from an experimental artifact from unconsidered, microscopic changes of the STM tip probe (i.e. reconfiguration of the atoms at the tip apex) between the data sets \cite{Li2022STM_RSB, Candelora2025}. The second potential origin is that small amounts of strain may locally suppress TRS and spatially inhomogeneous strain fields may generate the differing behaviors seen between STM measurements.  Resolving this discrepancy via future STM experiments on strain-free samples attached without epoxy \cite{Guo2022chiralTransport} and with a meticulous examination of experimental artifacts \cite{Candelora2024, Candelora2025} is highly desirable.

\begin{figure}[!t]
\centering
\includegraphics[width=\columnwidth,angle=0,clip=true]{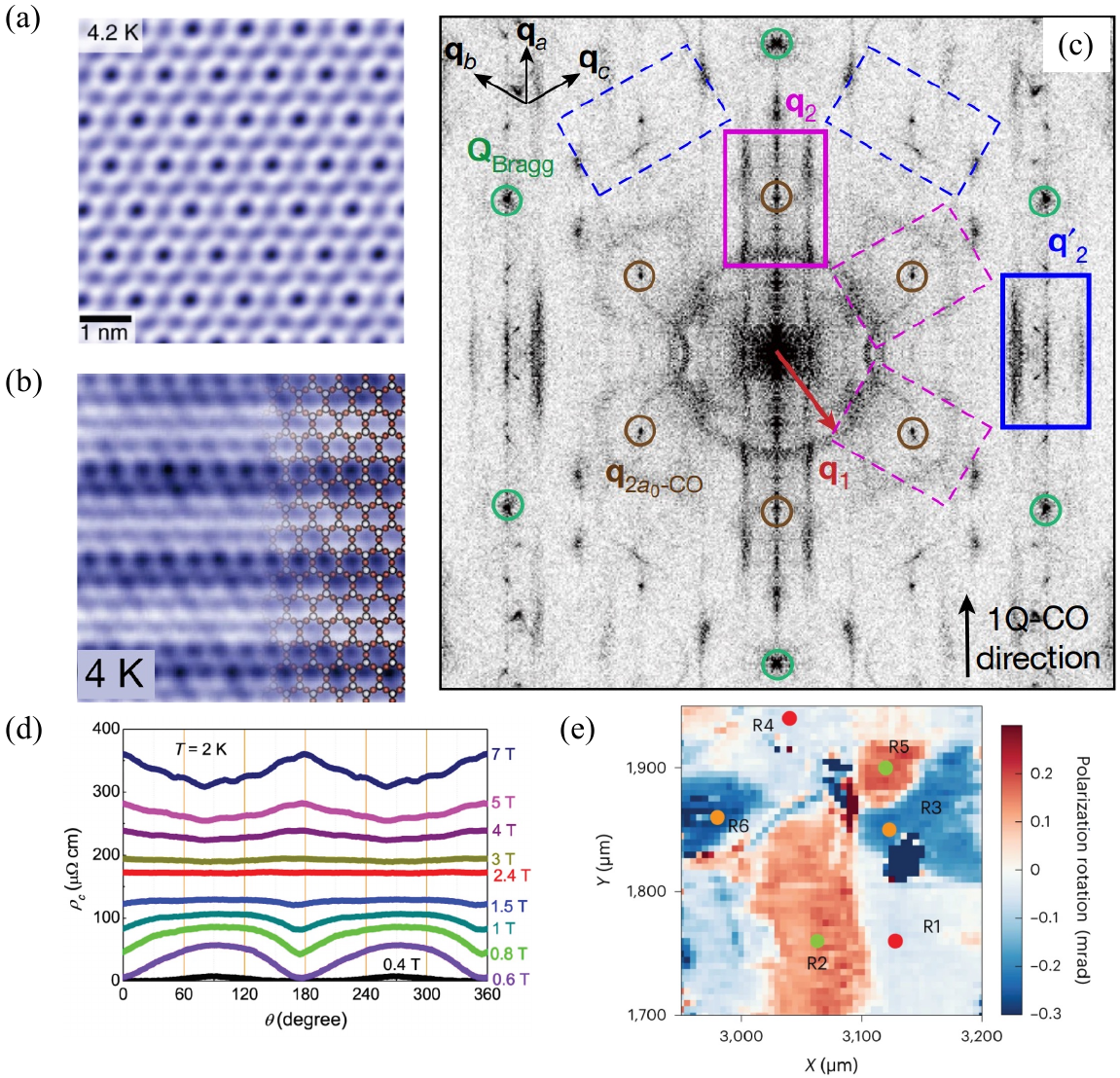}
\caption{(a) STM topograph of KV$_3$Sb$_5$ showing a $2 \times 2$ CDW \cite{jiang_unconventional_2021}. (b) STM topograph of CV$_3$Sb$_5$ at 4.5 K showing additional 4 x 1 charge modulations \cite{zhao2021cascade}. (c) Fourier transform of STM d$I$/d$V$ map of CsV$_3$Sb$_5$ showing $C_2$-symmetric QPI from V-bands, marked with $q_2$ and $q_2^{\prime}$ \cite{zhao2021cascade}. (d) Angular dependence of $c$-axis resistivity in CsV$_3$Sb$_5$ measured at different in-plane magnetic fields \cite{Xiang2021two-foldCVS}. (e) Spatial mapping of polarization angle at 6 K in a RbV$_3$Sb$_5$ sample showing three types of equivalent domains rotated with respect to one another \cite{xu_three-state_2022}.
}
\label{AVSb135}
\end{figure}

STM measurements have also played a key role in identifying charge correlations that form beyond the $2\times2$ in-plane CDW order (Fig. \ref{AVSb135}).  Upon cooling in the CDW state, experiments reveal a unidirectional 4$a_0$ charge order that forms on the Sb surface termination of CsV$_3$Sb$_{5}$ (below about 60 K) ~\cite{zhao2021cascade, Wang2021STM-CVS, Wang2021chiralCDW_STM,Li2022hai-huwen} and also in RbV$_3$Sb$_{5}$ ~\cite{Shumiya2021STM}; however it is notably absent, or relatively rare, on the equivalent Sb surface of KV$_3$Sb$_{5}$ ~\cite{jiang_unconventional_2021, Li2022STM_RSB, PhysRevX.13.031030}. One hypothesis is that this stripe order stems from a set of small Fermi pockets, arising due to electronic band reconstruction upon entering the $2\times2$ CDW state ~\cite{Zhou2022pockets,PhysRevX.13.031030}. While the 4$a_0$ charge stripe order is absent as a static order parameter in scattering experiments ~\cite{Li2022conjoinedCDW}, its well-defined onset below 60 K ~\cite{zhao2021cascade} coincides with the temperature scale of lattice anomalies observed in bulk single crystals of CsV$_3$Sb$_{5}$.  These include Raman spectroscopy ~\cite{PhysRevB.105.155106, PhysRevResearch.4.023215} and time-resolved optical reflectivity ~\cite{PhysRevMaterials.5.L111801}, suggesting that it is a manifestation of an energy scale present in the bulk. 

Complementing ARPES measurements on 135 compounds, spectroscopic-imaging STM (SI-STM) was used to measure electronic band structure at the nanoscale. SI-STM mapping detected features associated with both the Sb-derived bands and the V-derived bands \cite{zhao2021cascade, chen2021roton, wu2023unidirectional, PhysRevX.13.031030, Li2023STM_35K}, largely consistent with DFT calculations and ARPES measurements \cite{ortiz2019new, PhysRevLett.125.247002}. The signal from the V-derived bands shows pronounced unidirectional features, while the Sb-derived signal remains rotationally symmetric \cite{zhao2021cascade, Li2023STM_35K, wu2023unidirectional, PhysRevX.13.031030}. Electron–phonon coupling is reported to renormalize the Fermi velocity of the in-plane vanadium bands, but preferentially along the symmetry-breaking direction, making the low-energy dispersion highly non-equivalent along the three lattice directions \cite{wu2023unidirectional}.

SI-STM measurements of CsV$_3$Sb$_{5}$ and KV$_3$Sb$_{5}$ reveal an emergence of quantum interference of quasiparticles below $\approx$35 K, a key signature for the formation of a coherent electronic state  \cite{Li2023STM_35K}. These quasiparticles, associated with V-derived kagome bands, display a pronounced unidirectional feature in reciprocal space that strengthens as the superconducting state is approached. This temperature is substantially below the formation of the 4$a_0$ stripes \cite{zhao2021cascade} or the $2\times2$ CDW phase \cite{PhysRevLett.125.247002}. This additional temperature scale has been confirmed in subsequent STM experiments \cite{wu2023unidirectional,Zhang2024KVS_40K_RSB} as well as by multiple other probes reviwewed in \textcite{s_d_wilson_av3sb5_2023} such as nuclear magnetic resonance (NMR) measurements \cite{Luo2022_2, Song2022}, suggesting the formation of electronic coherence correlated with a structural crossover in the system.

\textit{Superconducting state:} Turning now to the SC state, SC in $A$V$_3$Sb$_5$ compounds has been studied by a large number of probes, including tunnel diode oscillator \cite{Duan2021}, scanning SQUID \cite{kaczmarek2024direct}, ARPES \cite{Zhong2023}, heat capacity and thermal transport \cite{hossain2025unconventional}, point-contact tunneling \cite{PhysRevB.104.174507}, irradiation studies \cite{Roppongi2023, zhang2023nodeless}, $\mu SR$ \cite{gupta2022microscopic, Guguchia2023}, and NMR \cite{Mu_2021}.  These compounds host highly anisotropic quasi-two-dimensional SC states in the clean limit \cite{Duan2021}.  Broadly, the picture is one of a nodeless, multigap, singlet SC state whose smaller gap is rather anisotropic and whose larger gap is isotropic \cite{Roppongi2023,grant2024superconducting}. STM experiments generally show a V-shaped superconducting gap, with non-zero conductance at zero energy, suggestive of residual ungapped Fermi surface \cite{chen2021roton, PhysRevX.11.031026, Xu2021STM}. Some STM experiments also resolve multiple superconducting gaps in differential conductance spectra \cite{Xu2021STM}. $\mu SR$-measurements in select compounds claim to resolve a nodal gap \cite{Guguchia2023}; however this differs from conclusions of tunnel diode oscillator measurements. Recent work by \textcite{akifumi_mine_direct_2024} report ultra-high-resolution laser-based ARPES to directly probe the anisotropic superconducting gap in CsV$_3$Sb$_5$. Strong orbital selectivity was observed in the gap structure: V-based 3d orbitals exhibited an 80\% gap anisotropy, whereas Sb-derived bands retained an isotropic gap. This provides strong evidence that SC is intertwined with CDW-driven Fermi surface reconstruction, possibly pointing toward an unconventional pairing mechanism. 

Constraining interpretations of this combined data, theoretical treatment of a kagome metal with \textit{p}-type VHS at $E_F$ \cite{PhysRevB.108.144508, dai2024existence} showed that traditional BCS-like metrics of a Hebel-Slichter peak in NMR \cite{Mu_2021} and resilience to nonmagnetic disorder upon irradiation \cite{Roppongi2023} fail to exclude sign-changing, singlet SC order parameters.  This aggregate picture suggests the pairing symmetry to be either $s^{++}$, $s^{+-}$, or $d+id$, with $d+id$ favored if TRS is broken in the SC state.  Curiously, the signature of TRS reported by $\mu SR$ measurements below the CDW transition, reappears below the SC transition once long-range CDW order is suppressed via hydrostatic pressure \cite{Guguchia2023}.  One additional aspect of research into the SC state of $A$V$_3$Sb$_5$ is the presence of unconventional phenomena in the fluctuation regime of the SC state, such as an unusal extended vortex regime \cite{PhysRevB.109.144507} and the proposal of a composite, 6e-pairing condensate \cite{PhysRevX.14.021025, PhysRevB.108.214516}. 

Studies of the surface-dependent electronic structure include a report of topological surface states and flat bands in CsV$_3$Sb$_5$ \cite{hu_topological_2022}, germane to connate models of topological superconductivity \cite{PhysRevLett.100.096407}. Additionally, similar work shows that SC emerges within the CDW-modified structure, with partial CDW gap suppression and enhanced low-energy spectral weight \cite{t_kato_surface-termination-dependent_2023}. Different surface terminations drastically affect the electronic structure, with some terminations suppressing CDW order locally.\cite{l_huai__surface-induced_2022, t_kato_polarity-dependent_2022}.  A recent STM study has shown that manipulation of surface alkali ions can locally fault the CDW stacking and suppress the CDW gap, resulting in a higher $T_c$ surface SC state that potentially couples to the Z$_2$ topological surface state \cite{han2024emergent}.

\textit{Impact of pressure and strain:} A number of studies have also explored the impact of hydrostatic pressure and strain on the electronic properties of V-based 135 compounds.  Hydrostatic pressure has been shown to rapidly suppress the CDW transition while enhancing the SC transition, which eventually forms a dome-like feature \cite{PhysRevLett.126.247001, PhysRevB.103.L220504}.  At even higher pressures (beyond the initial suppression of SC), SC eventually reappears.  The disappearance and reemergence of SC correlates to the presence of Sb-states at the $E_F$ and suggests their importance for both SC and the CDW order \cite{PhysRevB.107.174107,PhysRevB.107.205131, PhysRevB.105.235145}.  

CsV$_3$Sb$_5$, distinct from its Rb and K-based cousins, shows the formation of a ``double-dome" SC state in the low-pressure regime \cite{PhysRevLett.126.247001, Yu2021}.  This has been linked to the presence of a distinct CDW phase for CsV$_3$Sb$_5$ (discussed earlier) and an evolution into a competing CDW phase \cite{PhysRevLett.133.236503}. Externally applied in-plane strain induces a modest suppression of the CDW state and a slight enhancement of the $T_c$ \cite{PhysRevB.104.144506}. STM experiments reveal unusually high decoupling rate between the direction of electronic unidirectionality and anisotropic strain, revealing weak smectoelastic coupling in the 2 $\times$ 2 CDW phase of kagome superconductors \cite{Wang2023strainSTM}. The 4$a_0$ charge-stripe order appears to be easily suppressed by strain induced by buckling in the sample \cite{Wang2023strainSTM}. Both the impact of in-plane uniaxial strain and nominally hydrostatic pressure are largely driven by changes in the $c$-axis, either via the Poisson ratio or by anisotropic compressibility of the lattice.   

\begin{figure}[!t]
\centering
\includegraphics[width=\columnwidth,angle=0,clip=true]{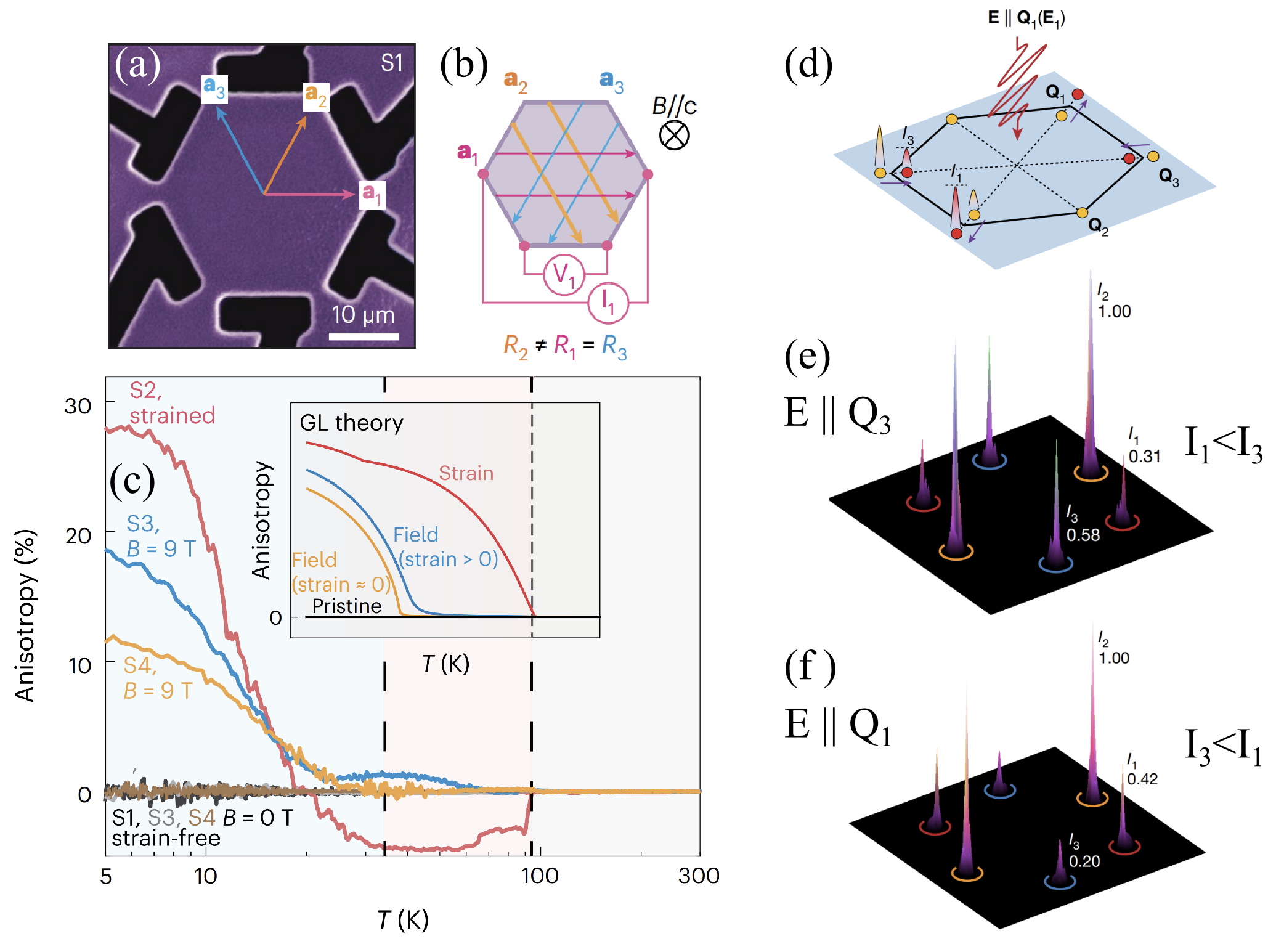}
\caption{(a) and (b) show a CsV$_3$Sb$_5$ crystal fabricated into a device for mitigating strain and measuring in-plane transport anisotropy and the corresponding measurement geometry.  (c) The resulting transport anisotropy measured with and without extrinsic strain, showing negligible in-plane anisotropy when strain is largely suppressed.  An out-of-plane magnetic field drives the reappearnce of in-plane anisotropy in ``strain-free" crystals.  Figures are from Guo et al. \cite{Guo24}.  (d)  Measurement geometry for STM measurements conducted under optical illumination from Xing et al. \cite{Xing2024opticalM}.  (e) and (f) show the change in intensities of the Fourier transformed pattern of CDW peaks driven via the alignment of an in-plane optical polarization, signifying a shift in the pattern of CDW order. 
}
\label{piezomagnetic}
\end{figure}

Related to strain, one line of inquiry is the potential impact of extrinsic strain (\textit{e.g.} due to differential cooling between a crystal and its sample mount) on the measured properties of $A$V$_3$Sb$_5$ compounds. 
Small amounts of strain at the order of 0.1 \% can suppress the TRS-breaking/chiral signal in transport experiments of nearly strain-free samples \cite{Guo2022chiralTransport}. Experiments further report that an applied out-of-plane magnetic field \cite{Guo24,Xing2024opticalM} or light pulse with an in-plane polarization in select directions can drive an in-plane transport anisotropy or modulate the in-plane propagation vectors of the CDW state \cite{Xing2024opticalM} as shown in Fig. \ref{piezomagnetic}. These measurements suggest the presence of an intrinsic piezomagnetic order parameter and a TRS-broken CDW state in $A$V$_3$Sb$_5$---one whose signatures in optics, magnetotransport, and STM measurements would be highly sensitive to the presence of extrinsic global or local strain fields imposed within the crystal under study. The magnetic field-induced structural and electronic response in STM measurements, however, has subsequently been questioned and alternatively attributed to experimental artifacts \cite{Candelora2025}.

\textit{Chemical doping:} A tremendous amount of studies have also been performed exploring the impact of tuning $E_F$ in $A$V$_3$Sb$_5$ via chemical substitution or doping.  Hole-doping (Sn, Ti) \cite{PhysRevMaterials.6.L041801, liu2023doping} and electron-doping (Te, Cr) \cite{capa2023electron,PhysRevB.106.235151} studies reveal a CDW order parameter that is rapidly suppressed upon tuning the proximity of the VHS relative to the Fermi level. ARPES studies directly reveal that doping the vanadium site shifts the VHS positions, affecting the CDW stability and enhancing SC \cite{kato_fermiology_2022}.  Comparing with the known evolution of the CDW upon doping, the microscopic nature of the CDW state and its relationship to the electronic structure and band folding can be resolved \cite{kang_charge_2023}. The impact of carrier-doping is largely orbitally selective, with a large change in the Sb $p_z$ pocket's filling \cite{PhysRevB.104.205129}, and, similar to pressure studies, SC seemingly vanishes coincident with the disappearance of this pocket.  Paralleling the phenomenology of pressure studies, CsV$_3$Sb$_5$ uniquely develops a double-dome SC feature \cite{PhysRevMaterials.6.L041801} with signatures of a nearby competing pattern of charge correlations \cite{Kautzsch2023}.  A nematic quantum critical point has been proposed to account for this unusal evolution of $T_c$ across the phase diagram \cite{Sur-Nematic-NC-2023}.  Isoelectronic substitution on the V-site similarly has a strong impact on the CDW and SC transitions \cite{liu2022evolution, PhysRevB.105.L180507} while isoelectronic substitution on the alkali-site largely interpolates CDW and SC behaviors between the 135 end members \cite{PhysRevMaterials.7.014801}. 

\textit{Nematicity and PDW order:} Moving beyond conventional CDW descriptions, electronic nematicity in KV$_3$Sb$_5$ was identified using ARPES, demonstrating that the CDW phase breaks $C_3$ symmetry by forming a unidirectional electronic modulation \cite{jiang_observation_2023}.  It was similarly found that the Fermi surface exhibits small reconstructed pockets corresponding to an intertwined charge-stripe order in AV$_3$Sb$_5$ compounds \cite{PhysRevX.13.031030}. This work suggested that beyond a simple CDW, the Fermiology of AV$_3$Sb$_5$ might support additional orders, including a potential PDW coexisting with SC.  At low temperature, STM studies report the observation of an additional 3q CDW-like wave vector linked to the local modulation of the superfluid density, suggesting a PDW instability \cite{chen2021roton, Deng-Josephson-Nature-2024}.


A host of experimental studies have revealed a rich landscape of competing and cooperating phases in $A$V$_3$Sb$_5$ materials, but fundamental questions remain unanswered. Chief among them is the potential presence and precise role of TRS breaking in both the CDW and SC states, and the underlying SC pairing mechanism. The relative role of electron-phonon coupling versus Fermi surface nesting effects in driving both CDW and SC order remain debated~\cite{ritz2023superconductivity}.  Electron-phonon coupling, initially thought small \cite{PhysRevLett.127.046401}, has later been shown to be appreciable and nonequilibrium studies have shown lattice vibrations couple kagome band features \cite{yigui_zhong_unveiling_2024}. The role of extrinsic strain in many of the discrepancies reported between experimental studies using the same technique and between different experimental probes (each reporting the existence or absence of TRS) is a key issue that must be resolved. \textit{In situ} strain-tuning techniques as well as careful strain mapping measurements carried out in combination with optical, STM, magnetotransport, and x-ray scattering measurements will be crucial for solving this issue.


\paragraph{CsCr$_3$Sb$_5$:}

CsCr$_3$Sb$_5$ is an isostructural cousin of CsV$_3$Sb$_5$ \cite{liuCr135}, with a kagome net composed of Cr atoms. This metastable phase can be grown in bulk crystal form via a high-temperature, crucible-based self-flux growth method \cite{liuCr135}.  Unlike members of the $A$V$_3$Sb$_5$ family, it exhibits an electronic flat band favorably placed near Fermi level \cite{liuCr135, Guo2024Cr135_flatband, Li2024cr135_flatband, xieCr135, Peng2024cr135}, and it is also believed to be magnetic and more strongly correlated \cite{giorgioNV}. At ambient conditions, it undergoes a 4$a_0$ structural transition at 55 K observed by XRD measurements, also reported to be accompanied by a magnetic transition \cite{liuCr135}. Under pressure, this phase transition evolves into two separate transitions at different temperatures, interpreted as spin density or CDWs. Both of these density wave transitions are gradually suppressed as SC ultimately emerges above 3.65 GPa, peaking at $T_c \approx$ 6.4 K \cite{liuCr135}.  It is worth noting that at intermediate pressures, both SC and magnetism are believed to coexist \cite{liuCr135}. The nature of emergent SC, however, remains to be determined.  Recent STM experiments image a similar 4$a_0$ density wave on the surface to those in the V-based 135 compounds, and further reveal the emergence of another unidirectional charge density wave below 45 K \cite{Cheng2025Cr135Frieze}. This lower-temperature density wave breaks all mirror symmetries, but preserves the mirror-glide symmetry \cite{Cheng2025Cr135Frieze}, and it may be potentially connected to an altermagnetic state \cite{Huang2025Cr135alt}. First principles and RPA calculations suggest that antiferromagnetic fluctuations are crucial for mediating SC in this system \cite{Wu2025cr135_theory}. There is also little known about the nature of magnetism, with theoretical proposals for 4 × 2 altermagnetic spin-density-wave state and other energetically favorable phases close by \cite{Xu2023altermagnetism_Cr135}.


\paragraph{ATi$_3$Bi$_5$:}

   $A$Ti$_3$Bi$_5$ ($A$=Cs, Rb) crystallizes in the same structure as its $A$V$_3$Sb$_5$ counterpart, but with a kagome net composed of Ti atoms instead of V \cite{dominik_werhahn_kagome_2022, Yang2022CTB135_discovery2}.  It is synthesized via similar high-temperature crucible-based self-flux growth techniques. Similar to $A$V$_3$Sb$_5$, it is also non-magnetic \cite{dominik_werhahn_kagome_2022, Yang2024CTB135_STM}. While some studies show that CsTi$_3$Bi$_5$ is superconducting below $T_c \approx$ 4.8 K \cite{Yang2024CTB135_STM}, SC has not been observed in other studies \cite{dominik_werhahn_kagome_2022, Wang_2023CTB135}. The discrepancy may lie in the small superconducting volume fraction of impurity phases and low critical magnetic fields that necessitate screening Earth's magnetic fields \cite{Yang2024CTB135_STM}. The superconducting $T_c$ of 4.8 K is substantially higher than the theoretically predicted $\approx$ 2 K in both Cs and Rb variants \cite{Yi2023CTB_theory}. 
  
  In contrast to $A$V$_3$Sb$_5$, experiments on $A$Ti$_3$Bi$_5$ reveal no obvious anomalies in magnetization, resistivity and heat capacity \cite{dominik_werhahn_kagome_2022, Wang_2023CTB135, Chen2023CTB135} indicative of a CDW state. Phonon dispersion calculations also show no imaginary phonons \cite{Li2023CTB_nematicity, Jiang2023CTB135nematicity}, further supporting the absence of a CDW instability. This in turn provides an opportunity to explore electronic phenomena in this class of non-magnetic kagome metals in the absence of translational symmetry breaking CDW. 

\begin{figure*}[!ht]
\centering
\includegraphics[width=\textwidth,angle=0,clip=true]{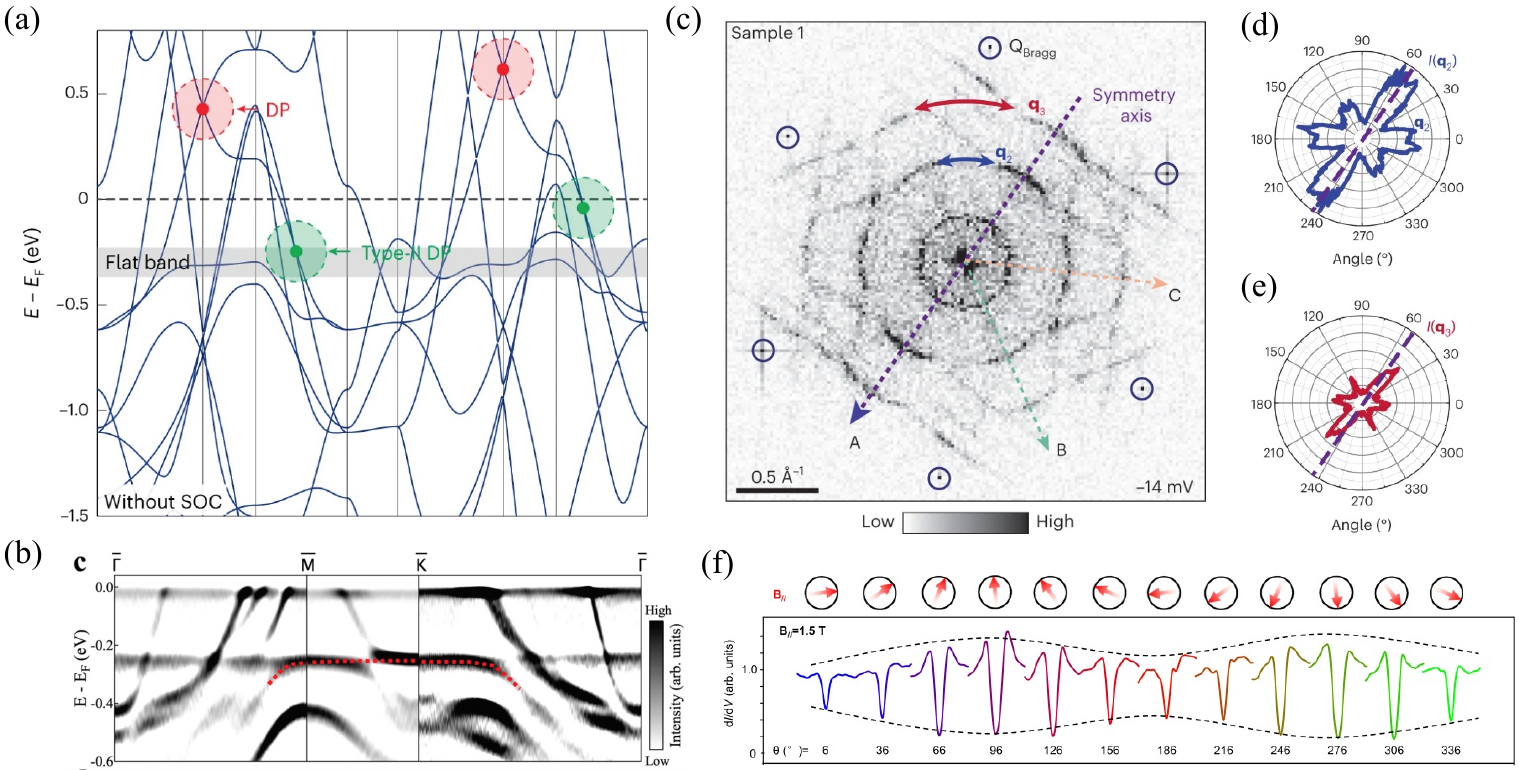}
\caption{(a) DFT-calculated electronic structure of RbTi$_3$Bi$_5$ without SOC \cite{Hu2023CTB135_ARPES}. Flat band, Dirac point and Type-II Dirac nodal lines are denoted. (b) Second derivative image with respect to energy obtained from raw ARPES data of CsTi$_3$Bi$_5$ along $\Gamma$-M-K-$\Gamma$ direction \cite{Yang2023CTB135_ARPES}. Flat band is highlighted by dotted red lines. (c) Fourier transform of normalized STM differential conductance map at -14 meV showing two-fold symmetric scattering and interference of electrons on the surface of CsTi$_3$Bi$_5$ \cite{Li2023CTB_nematicity}. (d,e) Angle-dependent FT amplitudes of q$_2$ and q$_3$ scattering wave vectors from (c), both plotted in polar coordinates on the same scale \cite{Li2023CTB_nematicity}. (f) A panorama of d$I$/d$V$ spectra showing a superconducting gap on the surface of CsTi$_3$Bi$_5$ at different in-plane magnetic ﬁeld orientations exhibiting two-fold rotational symmetry \cite{Yang2024CTB135_STM}.}
\label{CTB-135}
\end{figure*}

From the Fermiology perspective, $A$Ti$_3$Bi$_5$ hosts VHS at M points, but these are positioned well above the Fermi level \cite{Hu2023CTB135_ARPES,Jiang2023CTB135nematicity,Wang_2023CTB135, Yang2023CTB135_ARPES, Liu2023CTB135_ARPES} unlike the equivalent features in $A$V$_3$Sb$_5$ that appear in closer proximity to the Fermi level. It may be possible to shift these VHS closer to Fermi level under pressure \cite{Yi2023CTB_theory}. ARPES experiments reveal a partial flat band is located several hundred meV below the Fermi level \cite{Hu2023CTB135_ARPES, Jiang2023CTB135nematicity, Yang2023CTB135_ARPES} (Fig.\ \ref{CTB-135}a,b), which was also inferred from infrared spectroscopy \cite{Cao2024IFspectroscopy}. 

First-principle calculations predict that CsTi$_3$Bi$_5$ has robust topological surface states \cite{Wang_2023CTB135}, some of which could be further tuned by pressure \cite{Yi2023CTB_theory, Wu2024CTB135_theory_pressure}. Topological surface states are visualized in ARPES in RbTi$_3$Bi$_5$ \cite{Hu2023CTB135_ARPES, Jiang2023CTB135nematicity} and CsTi$_3$Bi$_5$ \cite{Yang2023CTB135_ARPES}, and the topological nature of certain Fermi surfaces has also been inferred from quantum oscillations \cite{Dong2024CTB135_QO}. Dirac nodal lines are also detected in CsTi$_3$Bi$_5$ \cite{Yang2023CTB135_ARPES} and RbTi$_3$Bi$_5$ \cite{Jiang2023CTB135nematicity, Hu2023CTB135_ARPES}. Overall, the electronic band structure extracted from ARPES \cite{Hu2023CTB135_ARPES,Wang_2023CTB135, Jiang2023CTB135nematicity, Yang2023CTB135_ARPES, Liu2023CTB135_ARPES} and STM \cite{Li2023CTB_nematicity, Yang2024CTB135_STM}, and Fermi surface determined from quantum oscillations \cite{Rehfuss2024CTB135_QO} shows a reasonable agreement with DFT calculations, suggesting only a modest degree of electronic correlations in the system. 

Rotation symmetry breaking, or electronic nematicity, is reported in this family of materials using several complementary probes. Notably, it emerges without any accompanying CDW states, in sharp contrast to $A$V$_3$Sb$_5$. Spectroscopic imaging STM measurements revealed electronic rotation symmetry breaking in Fourier transforms of differential conductance maps (Fig.\ \ref{CTB-135}c) \cite{Li2023CTB_nematicity, Yang2024CTB135_STM}. In particular, by comparing the wave lengths and amplitudes of scattering wave vectors along different directions in reciprocal space, STM experiments detected electronic anisotropy that breaks the sixfold symmetry of the lattice \cite{Li2023CTB_nematicity, Yang2024CTB135_STM} (Fig.\ \ref{CTB-135}d,e), arising from both in-plane and out-of-plane titanium-derived $d$ orbitals \cite{Li2023CTB_nematicity}.  Polarization-dependent ARPES report orbital-selective electronic nematicity in RbTi$_3$Bi$_5$ \cite{Hu2023CTB135_ARPES} and CsTi$_3$Bi$_5$ \cite{Bigi2024CTB135}. Autocorrelation of ARPES Fermi surface of RbTi$_3$Bi$_5$ is also consistent with rotation symmetry breaking, reported to persist to 200 K \cite{Jiang2023CTB135nematicity}, and a comparable onset temperature of 150 K was inferred from anomalous phonon line shapes in infrared spectroscopy experiments \cite{Wenzel2025CTB135}. 

The origin of nematicity has been hypothesized to be an electron correlation driven Pomeranchuk instability \cite{Bigi2024CTB135}. Interestingly, an anomalous spin-optical helical effect suggestive of loop currents circulation has also been reported in CsTi$_3$Bi$_5$~\cite{mazzola2025anomalous}.  When SC is observed, the superconducting gap size under in-plane magnetic field also exhibits a two-fold symmetry, consistent with the parent state electronic nematicity \cite{Yang2024CTB135_STM} (Fig.\ \ref{CTB-135}f).
\textcite{Hu_2024CTB135review} provides a more in-depth review specific to this family of materials.

\subsubsection{\label{sec:chapIVB3}AM$_6$X$_6$ compounds}
Kagome metals with the chemical formula AM$_6$X$_6$, often referred to as 166 compounds, form with a broad array of elements, where $A$- and $M$-sites are occupied predominantly with rare-earth ($R$=Y, Sc, La-Lu) and transition metal ions and the $X$-site ions are commonly Sn or Ge. In the stoichiometric limit, the structures most commonly studied are those that form in the hexagonal P6/mmm HfFe$_6$Ge$_6$-type cell, and these can be thought of as a modification of the CoSn-type cell due to the addition of a triangular lattice $A$-site plane between the kagome planes.  The result is a bilayer kagome lattice interleaved with a triangular lattice of $A$-site ions, and it is one where the $X$-site ions (nominally within the kagome plane in 11-type structures) are now displaced above and below the kagome plane.  Studies of 166 materials can be roughly separated into those with magnetic and nonmagnetic kagome sites. AM$_6$X$_6$ compounds with $M$=Fe, Mn, Cr are the most common magnetic variants while those with $M$=V, Nb have nonmagnetic sites.  In the following discussion, we focus first on the magnetic $M$-site variants and then return to the nonmagnetic $M$-site members.   

\paragraph{RMn$_6$Sn$_6$:}

Strongly magnetic members of the 166 family have been studied extensively as hosts of noncollinear magnetic ground states born via a number of mechanisms. Interlayer frustration between kagome planes of magnetic $M$-site variants has long been known to promote the formation of double-flat and other spiral states \cite{rosenfeld2008double}, and tuning the relative interactions between the kagome planes as well as coupling magnetic $A$-site moments can lead to a host of complex magnetic phase diagrams \cite{malaman1999magnetic, riberolles2024new, PhysRevB.111.054410, SCHOBINGERPAPAMANTELLOS199792}.  We will not focus on the diversity of these states here and instead focus on phenomena derived from electronic instabilities and interactions endemic to the kagome networks and band structures of 166 compounds. 

An early focus of studies of magnetic Mn-based 166 compounds was their unconventional magnetotransport properties, and single crystals can be grown via Sn-based flux methods \cite{clatterbuck}.  When the Dirac crossings arising from the kagome network are close to the Fermi level, an out-of-plane ordering of the magnetic moments can generate a large Chern gap and a large AHE is predicted to result. Strong SOC can similarly gap the Dirac point, but the appeal of a field-switchable response requires the presence of tunable out-of-plane moments.  

Absent anisotropies from interstitial magnetic $R$-site ions, the $M$-moments orient ferromagnetically within the basal planes modulated along the $c$-axis in a helical ground state in the case of $M$=Mn.  Specifically, when the R-sites are nonmagnetic such as in $R$Mn$_6$Sn$_6$ ($R$=Sc, Y, Lu), an incommensurate, spiral state forms within the Mn kagome network with moments aligning in the kagome plane and winding along the interplane direction \cite{venturini1996incommensurate}.  An in-plane field can induce an out-of-plane component to the magnetization and a noncollinear spin texture with real-space spin chirality.  For instance, an in-plane field in YMn$_6$Sn$_6$ has been shown to induce a transverse conical spiral state induced by thermal fluctuations \cite{doi:10.1126/sciadv.abe2680}, leading to the formation of a strong topological Hall effect.  Similar effects have been reported in ScMn$_6$Sn$_6$ \cite{zhang2022magnetic} and LuMn$_6$Sn$_6$ \cite{mozaffari2025diversemagneticphasediagram}. By tuning the out-of-plane magnetic field, STM/S experiments reported an anomalous momentum-dependent $g$-factor associated with the gapped Dirac point \cite{Li2022manipualtions_STM}.  In-plane magnetic field experiments further show additional tunability of the $g$-factor strongly dependent on the in-plane crystalline direction \cite{Jia2024YMS166_STM}.

Adding anisotropic moments via magnetic lanthanide $R$-site ions can induce a ferrimagnetic $c$-axis aligned moment for strongly uniaxial Tb$^{3+}$ \cite{ELIDRISSI1991143}, as well as out-of-plane canting for Ho$^{3+}$, and Dy$^{3+}$.  For $M$=Mn, all of these ions induce some degree of collinear spin reorientation transitions upon cooling as the in-plane anisotropies of the stronger exchange-coupled Mn-moments give way to the anisotropies of the lanthanide moment network \cite{PhysRevX.12.021043}. The filling of the kagome band structure places Dirac points and saddle points just below and above $E_f$ \cite{li2021dirac}.  Unconventional magnetotransport effects are also reported in $R$Mn$_6$Sn$_6$ ($R$=Er, Tm) where an in-plane helical magnetic state persists \cite{PhysRevB.104.L161115,PhysRevB.106.125107}. 

For TbMn$_6$Sn$_6$ in particular, the ground state forms a $c$-axis aligned ferrimagnetic state of antialigned Mn and Tb moments, generating a sizable 6 $\mu_B$ $c$-axis aligned moment. This ordered state, combined with the Dirac crossing close to $E_F$ led to the report of a Chern-gapped state, intrinsic anomalous Hall conductivity, and topological edge-states mapped via STM \cite{yin2020quantum}.  Quantum oscillation measurements and Nernst effect data reveal a quasi-two dimensional orbit with a nontrivial Berry phase as well as an anomalous thermal Hall signal \cite{xu2022topological}.  The origin of the anomalous Hall conductivity as deriving primarily from the quasi-2D Dirac band features, however, is debated \cite{PhysRevB.110.115134, PhysRevB.108.045132}.

\paragraph{R(Fe,Cr)$_6$(Sn,Ge)$_6$:}

Other variants with potentially magnetic $M$-sites include $R$Fe$_6$(Ge,Sn)$_6$ and $R$Cr$_6$(Ge,Sn)$_6$ \cite{10.1063/1.372607, SCHOBINGERPAPAMANTELLOS199792}, though these compounds often suffer from vacancies and off-stoichiometries within the Ge/Sn sublattices.  Bulk crystals are typically grown via Sn- or other metal-based flux methods \cite{Avila_2005}.  Compounds such as YFe$_6$Sn$_6$ form in both disordered YCo$_6$Ge$_6$-type and vacancy-ordered HoFe$_6$Sn$_6$-type structures, and magnetic $R$-site variants form various mixtures of these phases \cite{ELIDRISSI19911331}.  A similar case occurs with $M$=Cr variants \cite{SCHOBINGERPAPAMANTELLOS199792}, though some reports claim YCr$_6$Ge$_6$ forms in the fully ordered structure where elements of the Cr-based kagome band structure were investigated \cite{wang2020experimental}.  The structurally ordered YCr$_6$Ge$_6$ variant is characterized as a Pauli paramagnet with moderate correlation effects in transport and a partially-flat band predicted to form just below $E_F$\cite{ishii2013ycr6ge6}. There is considerable opportunity in the future exploration of these compounds if processing routes forming stoichiometric crystals can be developed.  This is interesting, in particular due to their ability to mix magnetic $R$-site networks with magnetic kagome ($M$-site) networks with variable electronic filling.    

\paragraph{RV$_6$Sn$_6$:}

An ordered 166 structure also forms with the nonmagnetic $M$-site ion V, where an array of fully ordered $R$V$_6$Sn$_6$ ($R$=Sc, Y, Gd-Lu) compounds are known to stabilize \cite{ROMAKA20118862} and bulk crystals can again be grown via Sn-based flux methods \cite{pokharel_electronic_2021}.  The band structures of these compounds, similar to their Mn-based cousins, have Dirac crossings and saddle points very close to $E_F$ \cite{PhysRevLett.127.266401, hu_tunable_2022}, and, for nonmagnetic $R$-sites, they can be categorized as topological $\mathbb{Z}_2$ metals \cite{pokharel_electronic_2021}.  

Magnetism within the $R$-site network is determined by a combination of the frustrated triangular lattice motif, RKKY coupling, and anisotropies imposed by the spin-orbit entangled wave function of the lanthanide ground state multiplet.  Varying the character of the magnetic $R$-site can tune the magnetic anisotropy and the nature of the order \cite{lee2022anisotropic,zhang2022electronic}.  While (Y,Lu)V$_6$Sn$_6$ are nonmagnetic, YbV$_6$Sn$_6$ has an easy-plane anisotropy and exhibits signs of heavy fermion behavior in its low-temperature electronic properties \cite{PhysRevB.107.205151}, while (Tm,Er)V$_6$Sn$_6$ also possesses easy-plane anisotropy and antiferromagnetic order \cite{PhysRevResearch.6.043291}. (Ho,Dy)V$_6$Sn$_6$ show weak easy-axis anisotropy and ferromagnetism, while TbV$_6$Sn$_6$ is strongly uniaxial with Tb$^{3+}$ moments oriented out of the plane \cite{pokharel2022highly,rosenberg2022uniaxial}. GdV$_6$Sn$_6$, in contrast, has a nearly isotropic spin-only state that forms an incommensurate helical modulation along the interplane direction \cite{PhysRevB.108.035134}.  Larger $R$-site ions such as Sm result in the formation a disordered 166 variant \cite{PhysRevMaterials.7.054403}.

In strongly uniaxial TbV$_6$Sn$_6$, strong SOC and the resulting mass gap of Dirac bands at the Fermi level generates strong spin-Berry curvature effects and large orbital magnetic moments, which were detected via their Zeeman coupling to a magnetic field in QPI measurements \cite{li2024spin}. Spin-resolved ARPES also provides evidence for finite spin Berry curvature contributions at the center of the BZ, where the electronic flat band detaches from the Dirac band due to SOC \cite{di2023flat}. Even when magnetism is absent, such as in YV$_6$Sn$_6$, quantum oscillation measurements similarly detect a nontrivial Berry phase and light electron masses assigned to multiple Dirac crossings near $E_f$ in addition to heavy orbits attributed to nearby VHS \cite{PhysRevB.110.035119}.  Similar effects are seen in GdV$_6$Sn$_6$ \cite{PhysRevB.109.235145} as well as recently reported as nonmagnetic variants (Ti,Zr,Hf)V$_6$Sn$_6$ \cite{PhysRevB.109.155117}.  

Despite VHS existing close to $E_f$ across the broader family of $R$V$_6$Sn$_6$ compounds and a qualitatively similar filling of the kagome sublattice to $A$V$_3$Sb$_5$ compounds, no purely electronic phase transitions manifest in the
absence of magnetic order. Due to the lack of a
 dominant steric role, neither CDW order nor SC emerge down to 50 mK in stoichiometric (undoped compounds). However, one notable exception occurs when the size of the $R$-site ion is decreased below a critical threshold as in the case of ScV$_6$Sn$_6$, which we discuss next.

\paragraph{ScV$_6$Sn$_6$:}
Similar to YV$_6$Sn$_6$, ScV$_6$Sn$_6$ can be grown via a Sn-based flux method and does not show magnetic ordering or SC down to 80 mK \cite{Arachchige2022Sc166}. It hosts a similar band structure to YV$_6$Sn$_6$, and quantum oscillation experiments suggest that the electron Dirac band located at the K-point is characterized by a finite Berry phase, inferring a topologically nontrivial Fermi surface \cite{Yi2024QO, Zheng24LuLi, Shrestha2023QO}. Topological Dirac surface states have also been observed near $\Gamma$-point in ARPES measurements \cite{Hu2024MingShi_Sc166}.

\begin{figure}[!t]
\centering
\includegraphics[width=\columnwidth,angle=0,clip=true]{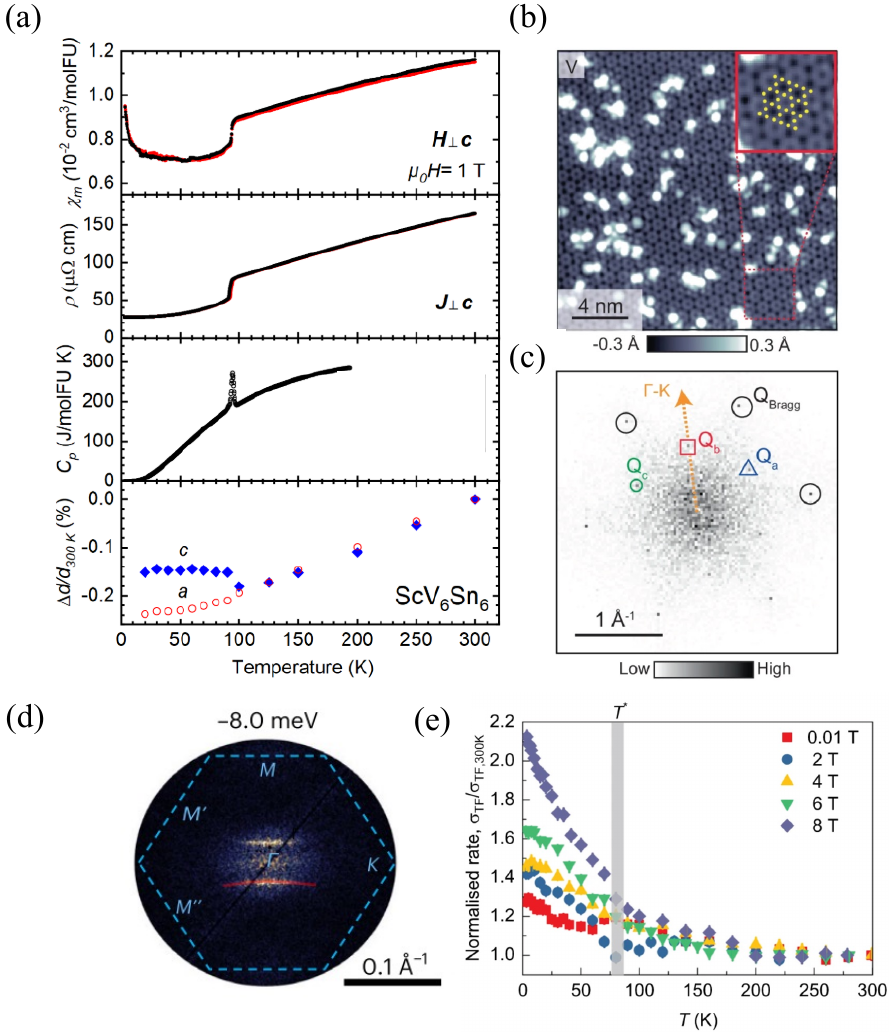}
\caption{(a) Temperature dependence of magnetic susceptibility on cooling (black) and warming (red) of bulk single crystals of ScV$_6$Sn$_6$ (top panel), temperature dependence of resistivity in the $ab$-plane measured on cooling (black) and warming (red) (second panel), specific heat capacity (third panel), and relative change in lattice parameters versus temperature (bottom panel). \cite{Arachchige2022Sc166}. (b-c) STM topograph and associated Fourier transform showing the CDW peaks (enclosed by triangles). \cite{Cheng2024STM_Sc166}. (d) Fourier transform of STM d$I$/d$V$ map showing unidirectional electron scattering and interference signature. \cite{Jiang2024Sc166_nematicity} (e) Temperature dependence of the high transverse field muon spin relaxation rate in $\mu$SR normalized to the value at 300 K, measured under different $c$-axis magnetic fields \cite{Guguchia2023usR_Sc166}.
 } 
\label{Sc-166}
\end{figure}

Notably, ScV$_6$Sn$_6$ is unique in that it hosts a first-order phase transition into a CDW-like phase below $T^*\approx $ 92 K with the $\overline{K}$ q=($\frac{1}{3}, \frac{1}{3}, \frac{1}{3}$) wave vector \cite{Arachchige2022Sc166}.  This corresponds to an enlargement of the unit cell by $\sqrt{3}\times \sqrt{3} \times3$ and a lowering of the lattice symmetry to an R32 cell.  The transition is reflected in concomitant discontinuities in magnetization, resistivity and heat capacity data \cite{Arachchige2022Sc166} (Fig.\ \ref{Sc-166}a). The absence of a gradual spectral gap opening from optical spectroscopy \cite{Hu2023optical_spec_Sc166} and abrupt spectral weight redistribution at $T^*$ from ARPES \cite{Cheng2024STM_Sc166} further indicate the first-order nature of the transition. Lattice distortions accompanying the CDW-like transition come primarily from a chain-like instability driven by the motion of the out-of-plane Sc and Sn$_1$ atoms, each with up to 0.16 $\AA$ out-of-plane displacements. The kagome net V atoms, however, show substantially weaker displacements \cite{Arachchige2022Sc166}, consistent with \textit{ab initio} calculations \cite{Tan2023Binghai}.  

Understanding charge ordering in ScV$_6$Sn$_6$ is the subject of a number of theoretical works \cite{cao2023competing,Tan2023Binghai, Subedi2024theorySc166, Liu2024theorySc166, Wang2024Sc166_theory}. Strong anharmonic effects \cite{Wang2024Sc166_theory}, order-disorder transitions \cite{Liu2024theorySc166}, and an interplay between electron-phonon coupling soft-phonon fluctuations of a locally flat H-mode are a few examples \cite{PhysRevB.111.054113}. In particular, the ($\frac{1}{3}, \frac{1}{3}, \frac{1}{2}$) CDW wave vector initially appeared to be energetically favorable from theory \cite{cao2023competing, Tan2023Binghai, Liu2024theorySc166}, which contrasts the experimentally observed q=($\frac{1}{3}, \frac{1}{3}, \frac{1}{3}$) distortion \cite{Arachchige2022Sc166}. Both elastic and inelastic x-ray scattering experiments instead reveal short-range charge order and phonon softening along q=($\frac{1}{3}, \frac{1}{3}, \frac{1}{2}$) at temperatures much higher than $T^*$ \cite{Korshunov2023flat_phonon_Sc166, cao2023competing, Pokharel2023Sc166}. 

The apparent onset temperature of this short-range charge order varies between reports due to differences in experimental resolution, but signatures are reported as high as room temperature. Additionally, weak short-range order in some cases persists below $T^*$, potentially pinned by variable levels of disorder in different samples \cite{Pokharel2023Sc166}. The fluctuations associated with the build up of short-range ($\frac{1}{3}, \frac{1}{3}, \frac{1}{2}$) order couples to charge carriers, where transport data suggest the presence of a pseudo-gap phase above $T^*$ \cite{DeStefano2023Sc166}. 

In general, phonons and lattice effects are believed to play the primary role in the formation of the CDW-like phase in ScV$_6$Sn$_6$ \cite{tuniz2023dynamics, Hu2024MingShi_Sc166, Korshunov2023flat_phonon_Sc166, Lee2024Sc166_arpes}, and any purely electronic instabilities are likely to play a secondary role. This notion is supported by ARPES measurements, which observe similar VHS near the Fermi level in both ScV$_6$Sn$_6$ and other 166 systems \cite{Cheng2024STM_Sc166, Lee2024Sc166_arpes, Cheng2024Zahid_Sc166, Hu2024MingShi_Sc166}, whereas a bulk CDW is only reported in ScV$_6$Sn$_6$. As such, any potential Fermi surface nesting is believed to only have a marginal effect on the CDW formation \cite{Cheng2024STM_Sc166, Lee2024Sc166_arpes,yuPRB2024}, and instead, steric effects and dynamics associated with an inherent out-of-plane chain instability of the Sn-Sc-Sn-Sn ions drive its unconventional behavior.  A real space model of frustration of this chain instability, one whose in-plane correlations are frustrated via steric effects mediated via the kagome network, was shown to capture and quantitatively model the short-range ($\frac{1}{3}, \frac{1}{3}, \frac{1}{2}$) charge correlations using a 2D Ising model \cite{PhysRevB.110.L140304}.

Supporting the notion of a primarily one-dimensional lattice instability, the CDW-like correlations in ScV$_6$Sn$_6$ are highly sensitive to disorder effects within the chains. Substituting Sc for larger $R$ ions, such as Lu or Y rapidly suppresses both long-range and short-range CDW order \cite{Meier2023Sc166_doping, Pokharel2023Sc166}, which are attributed to the smallness of the Sc atoms compared to other possible $R$ atoms \cite{Tan2023Binghai} and the impact of Sn-ion rattling along the chain axis. Small amounts of substitution at the V site \cite{Yi_Sc166doping} or the use of high pressure \cite{Yi_Sc166doping, Gu2023Sc166} also suppress the CDW transition temperature. A strong coupling to lattice disorder is also reflected in the varied transition temperatures $T^*$ reported of nominally stoichiometric crystals.

Of all of the stoichiometric $R$V$_6$Sn$_6$ compounds studied thus far, only the Sc variant exhibits a bulk CDW phase. Theory suggests that some $R$V$_6$Sn$_6$ systems may exhibit surface charge orders ($\sqrt{3}$ x $\sqrt{3}$ or 2 x 2) even in the absence of bulk CDW phase on kagome surface terminations \cite{Tan2024surfaceCDW_166}. The ($\frac{1}{3}, \frac{1}{3}$) bulk, in-plane CDW wave vector is consistent with STM measurements of ScV$_6$Sn$_6$, which observe periodic $\sqrt{3}$ x $\sqrt{3}$ modulations, rotated by 30 degrees with respect to the atomic lattice \cite{Cheng2024STM_Sc166, Kundu2024AbhaySc166, Jiang2024Sc166_nematicity} (Fig.\ \ref{Sc-166}b,c). The out-of-plane periodicity was further resolved by imaging the charge order across different terraces \cite{Cheng2024STM_Sc166, Kundu2024AbhaySc166}. The CDW is, however, only observed on some surface terminations \cite{Cheng2024STM_Sc166, Jiang2024Sc166_nematicity}, and there are conflicting reports in the literature reporting the presence \cite{Cheng2024STM_Sc166, Kundu2024AbhaySc166} or absence \cite{Jiang2024Sc166_nematicity} of charge order on the V termination. There are similarly dissonant reports on the honeycomb Sn$_2$ termination \cite{Cheng2024STM_Sc166, Jiang2024Sc166_nematicity} and the triangular Sn$_1$/ScSn$_3$ termination \cite{Jiang2024Sc166_nematicity, Kundu2024AbhaySc166}. The multitude of inequivalent surface terminations possible in 166 systems (up to 6 different ones) complicate identification of surfaces imaged by STM. 

STM experiments generally observe no spectral gap opening at E$_f$ \cite{Jiang2024Sc166_nematicity, Kundu2024AbhaySc166}, albeit a small partial gap of about 20 meV has been reported in d$I$/d$V$ spectra on one surface termination \cite{Cheng2024STM_Sc166}. Some ARPES \cite{Lee2024Sc166_arpes} and optical spectroscopy \cite{Kim2023Sc166_optics} experiments suggest the largest gap opening occurs for $k_z \approx \pi$ along the A-L direction of 260-270 meV. STM d$I$/d$V$ maps suggest the existence of another spectral gap at positive energies at the order of 50 meV, based on charge modulations that exhibit an abrupt phase shift near that energy \cite{Cheng2024STM_Sc166}.

Beyond translation symmetry breaking induced by the CDW, additional rotational symmetries are reported to be broken in this system, potentially via a secondary instability of the V-sublattice. Signatures of rotational symmetry breaking in the electronic structure are reported by STM experiments \cite{Jiang2024Sc166_nematicity}. First, spectroscopic-imaging STM detected a deformation of the low-energy electronic states on the V termination where the CDW signal is absent \cite{Jiang2024Sc166_nematicity} (Fig.\ \ref{Sc-166}d). Second, anisotropy in the intensity of atomic Bragg peaks in STM topographs is also observed on the same termination below about 70 K \cite{Jiang2024Sc166_nematicity}. No structural transition is observed in the temperature range, suggesting the electronic nature of electronic nematicity observed. No rotation symmetry breaking is detected in the intensities of the three in-plane CDW peaks \cite{Cheng2024STM_Sc166, Jiang2024Sc166_nematicity}, in sharp contrast to $A$V$_3$Sb$_5$ \cite{jiang_unconventional_2021, Li2022STM_RSB}. Electronic nematicity observed in ScV$_6$Sn$_6$ by STM is phenomenologically similar to that previously reported in CsTi$_3$Bi$_5$ \cite{Li2023CTB_nematicity}, where Fermi surface deformation, anisotropic spectral weight and inequivalent atomic Bragg peaks have also been reported. It has been suggested that favorably positioned VHSs near the Fermi level play a role in the emergence of electronic nematicity in ScV$_6$Sn$_6$ \cite{Jiang2024Sc166_nematicity}. 

Adding to the above, TRS is suggested to be broken below $T^*$.  Here the enhancement of the internal field width within the CDW-like state sensed by the $\mu$SR spectroscopy suggests TRS breaking from ``hidden" magnetism \cite{Guguchia2023usR_Sc166} (Fig.\ \ref{Sc-166}e). Anomalous Hall-like behavior up to the CDW phase transition further was also suggested to hint at TRS breaking \cite{Yi2024QO, Mozaffari2024mandrus}. It is worth mentioning that $^{51}V$ NMR measurements have not observed any signatures of underlying magnetism, possibly because the signals are below the resolution of the experiment of 10$^{-2}$ $\mu_B$ \cite{Guehne2024NMR_SC166}, or due to nuclear fields and alternative Fermiology effects accounting for the $\mu$sR and AHE results

\paragraph{RNb$_6$Sn$_6$:}
A new class of nonmagnetic $M$-site 166 variants was recently reported in $R$Nb$_6$Sn$_6$ with $R$=Y,Ce-Lu \cite{b_r_ortiz_stability_2024} with crystals grown via a high-temperature self-flux growth technique.  These compounds form a nonmagnetic Nb-based kagome net also with K-point Dirac crossings at $E_F$ and M-point saddle points nearly at and above $E_F$.  A rather rich spectrum of magnetic states are reported upon varying the $R$-site ion, where Gd, Tb, and Dy stabilize a variety of intermediate plateau states. Crucially, $R$=Lu seemingly crosses the steric barrier necessary to stabilize an out-of-plane chain CDW-like instability in LuNb$_6$Sn$_6$.  Below $T^{*}\approx$ 68 K, transport, heat capacity, and magnetization record a transition into a lower symmetry $\overline{K}$-type distortion with q=(1/3, 1/3, 1/3).  This is identical to the distortion observed in ScV$_6$Sn$_6$, and the distortion is frustrated with short-range charge correlations observed along q=(1/3, 1/3, 1/2) with phonon softening likely to occur along that wave vector as well.

\paragraph{CsCr$_6$Sb$_6$:
Another member of the 166 family is the recently discovered CsCr$_6$Sb$_6$ \cite{Song2025Cr-166, Liu2025Cr-166}, which consists of Cr-based kagome networks and whose crystals can be grown via a high-temperature self-flux growth technique. ARPES and theoretical calculations reveal flat bands at the Fermi level that are composed of Cr 3$d$ electrons \cite{Song2025Cr-166}. The material exhibits magnetic frustration and Kondo insulating behavior at low temperature, enabling the investigation of Kondo physics in kagome systems \cite{Song2025Cr-166}. As the thickness of the material is reduced to a few layers, its bulk magnetic frustration leads to A-type antiferromagnetic order and anomalous Hall response dependent on the parity of the number of layers \cite{Song2025Cr-166}.} 

\subsubsection{\label{sec:chapIVB4}AM$_3$X$_4$ compounds}
$A$M$_3$X$_4$ compounds form with a formula unit where $A$ can be either a trivalent/divalent rare-earth ion or a divalent alkali earth ion such as Ca. The $M$-site can be nonmagnetic V or Ti with the anion $X$ comprised of either Sb or Bi respectively and crystals are typically grown via using high-temperature metal-flux methods \cite{ortiz2023ybv, Ovchinnikov2020, Ortiz2023,Chen2024}.  The structure is formed from units of buckled kagome bilayers that are spaced from one another via zig-zag chains of the $A$-site ions in an orthorhombic cell. The magnetic $A$-site variants often form ferromagnetic ground states, though Ce, Gd, and Tb order antiferromagnetically in $A$Ti$_3$Bi$_4$ \cite{Ortiz2023, motoyama2018magnetic, guo20241, ortiz2024intricate}.  Low-field helical states may exist for the Eu$^{2+}$ variants, and a number of magnetic plateau states have been reported in antiferromagnetic members when the field is oriented along the chain-axis.

While band features reflective of the kagome networks such as flat bands, saddle points, and Dirac crossings have been identified in ARPES investigations of these compounds \cite{sakhya2024diverse, Ortiz2023, ortiz2023ybv}, electronic instabilities within the V-kagome network are typically not observed.  One notable exception is the recent STM report of a CDW intertwined with magnetic order in the GdTi$_3$Bi$_4$ compound \cite{han2025discoveryunconventionalchargespinintertwineddensity}. Studies of this compound and related variants are ongoing.  

\subsubsection{\label{sec:chapIVB5} Other compounds}
\paragraph{A$_2$M$_3$X$_2$:}
Another subset of kagome intermetallic compounds can be formed within ternary variants of three-dimensional Laves phases.  C14 and C15 Laves phases have the chemical formula AM$_2$ where the $M$-site ions form kagome nets stacked with equidistant triangular lattice networks. The in-plane translational phasing between these stacked layers generates alternating face- and vertex-sharing tetrahedra in the C14 case and an extended pyrochlore network in the C15 case. By adding a third main group $X$ ion that chemically orders within the triangular lattice network, an ordered ternary Laves phase can be formed, comprised of chemically-unique kagome layers, often distorted by a breathing mode. As one example, the prototypical C15 Laves compounds MgCu$_2$ forms a pyrochlore network of corner-sharing Cu tetrahedra.  A third element such as Si can be incorporated to form Mg$_2$Cu$_3$Si where Si occupies one tetrahedral Cu site in an ordered fashion.  This results in planes of Cu kagome nets stacked with triangular nets of Si ions embedded within a lattice of Mg.  

Many ordered Laves phases compounds are superconductors, such as Mg$_2$Ir$_3$Si \cite{kudo2020superconductivity2} and Li$_2$IrSi$_3$ \cite{hirai2014superconductivity} with Ir- and Si-based kagome networks respectively. The former can be grown in bulk crystal form via a high-temperature melt with excess Mg \cite{kudo2020superconductivity2}.  SC also appears when there is partial ordering of the kagome lattice \cite{kudo2020superconductivity1} suggesting that it is not an instability derived exclusively from the kagome networks.  Indeed, the electronic structure of ordered Laves phases remains quite three-dimensional, although a recent report of SC in Ta$_2$V$_3$Si reports a connection between enhanced SC and the presence of kagome-derived VHS close to $E_f$ \cite{PhysRevB.108.104504}.

\paragraph{RM$_3$(Si,B)$_2$:}
A structurally distinct family of compounds with a similar chemical formula as ordered Laves phases are RM$_3$(Si,B)$_2$ where R is lanthanide ion and M is a transition metal.  These compounds possess distorted kagome nets of M-site ions seperated by M(Si,B)$_2$ planes. The structure is distinct from the ordered Laves phase description, and the M-site kagome nets are intersected by chains of R and (Si,B) ions that thread the triangles and hexagons of the kagome network, respectively.   A complicated interplay of magnetism and SC is realized across the phase diagrams of these materials \cite{KU198091,BARZ19801489}.  SC occurs in a number of members of this materials class with LaRu$_3$Si$_2$ being the most widely studied due to its higher $T_c$.  There are a large number of compounds, so we focus on the LaRu$_3$Si$_2$ variant in this review as emblematic of the phenomena possible in these materials. Crystals can be obtained via standard melt-cool techniques \cite{c_mielke_charge_2024}.  

LaRu$_3$Si$_2$ was reported to possess moderate electron-electron correlation effects via an enhanced Wilson ratio \cite{PhysRevB.84.214527} with a conventional, nodeless BCS superconducting state \cite{KISHIMOTO2004507} whose enhanced $T_c$ is attributed to a narrow band of conduction elections.  The narrow conduction band is proposed to arise from the proximity of a kagome-derived flatband close to $E_F$ \cite{PhysRevMaterials.5.034803}, and correlation effects are further evident via a high-temperature charge ordered phase that evolves upon cooling \cite{Plokhikh2024}.  Recent work reports muon spin relaxation in $\mu$SR and magnetotransport data showing an unconventional coupling to external magnetic fields upon cooling into a modified CDW state \cite{c_mielke_charge_2024}.  The impact of both the kagome derived flatbands and the nearby saddle points on this phenomenology is a topic of current experimental \cite{misawa2025enhancedsuperconductivitylaru3si2chemical, PhysRevB.107.024503} and theoretical study \cite{deng2025theorysuperconductivitylaru3si2predictions}.

\paragraph{Homologous variants:}
The phase space for ternary kagome compounds is vast, and here we highlight a few additional structural motifs.  As one example, the $A$V$_3$Sb$_5$ structure exists within a broader homologous family, ($A_{m-1}$Sb$_{2m}$)(V$_3$Sb)$_n$,  where filled V$_3$Sb kagome planes are stacked with variable modulations of alkali and antimony layers. For instance, this generates $A$V$_6$Sb$_6$ and $A$V$_8$Sb$_{12}$ type structures with bilayer and higher groupings of the kagome planes \cite{Yin_2021, PhysRevMaterials.7.115002}. While the bilayer member of this family shows SC under high pressure \cite{shi2022new}, no other member of this expanded family is reported to form CDW or SC order under ambient pressure.  

Another kagome-related structure is the partially disordered Ln$_{2–x}$Ti$_{6+x}$Bi$_9$ class of compounds \cite{ortiz2024intricate}.  The buckled Ti-based kagome layers partially connect out of the plane and low-temperature magnetic transitions are reported arising from the partial Ln-chain networks in the lattice.  Another hybrid kagome structure is exemplified by Yb$_{0.5}$Co$_3$Ge$_3$. This compound possesses partially occupied Yb and Ge sublattices \cite{weiland2020refine} and provides another example of the diversity of kagome structures possible.  Interestingly, it exhibits a CDW-like low-temperature distortion of the kagome lattice as well as Yb-based magnetic order \cite{wang2022electronic}.

\section{\label{sec:Conc}Conclusions}

\subsection{\label{sec:ConcA}Summary}

There has been tremendous progress in both experimental and theoretical aspects of kagome metals.  Previously, existing  (and often well-studied) materials have been revisited using new theoretical frameworks, and new experimental probes have been leveraged to test predictions of unconventional electronic states and phase behaviors in kagome compounds.  This is readily apparent in the notable shift in the research focus from magnetic to electronic structure in numerous classes of magnetic kagome lattices. The notion of topological bands intrinsic to the band structure and the potential for combining these features with electronic correlations has become a topic of central focus.

From the perspective of materials, both the binary and ternary kagome metals have grown into a broad and diverse class of exotic quantum materials.  The binary systems have established the relevance of the kagome band structure in real materials including Dirac crossings and flat bands, sources of strong electronic correlation, and have opened new opportunities for spintronic devices. The ternary systems, with the addition of a third element into the crystal structure, brought enhanced degree of chemical tunability and diversity, which in turn provided a richer plethora of electronic properties and symmetry breaking than their binary kagome siblings.

In terms of experimentally observed phenomena, there have now been observations of the Dirac, Van Hove, and flat band features expected for the kagome network.  Positioning these features at the Fermi level has led to a variety of phenomena, ranging from aspects of electronic topology to strong electronic correlations.  For example, narrow band-induced strange metal behavior has been observed in binary kagome metals, while electronic symmetry breaking tied to the presence of \textit{p}-type Van Hove points has been observed in ternary systems.  The interplay between these kagome-derived phase behaviors and other types of order and energy scales (e.g. local moment magnetism, structural instabilities) in both ternary and binary compounds adds an additional layer of richness to the reported phenomena.

The theoretical investigation of kagome metals has also made significant progress and has offered a deep understanding of how their lattice geometry leads to a wide spectrum of electronic and collective phenomena. At the single-particle level, theoretical models of spinless electrons reveal the emergence of electronic features rooted in the kagome lattice's intrinsic symmetry and topology. The inclusion of SOC has further enriched this landscape, enabling the realization of topologically nontrivial phases. Lattice dynamics, treated through phonon calculations, highlights the importance of soft modes and structural instabilities, setting the stage for intertwined electron-phonon and many-body effects. 

Building on this foundation, theoretical work has now advanced to treat the interaction-driven phases that we have surveyed in this review. Models of CDWs describe the nesting-driven instabilities supported by VHS, whereas mean-field and beyond-mean-field treatments have provided frameworks to understand unconventional SC, including signatures of PDWs. Treatments of electron-phonon coupling have shed light on its role in competing with electronic orders, and models incorporating local moment fluctuations and strong electron correlations have begun to describe the emergence of magnetism. Taken together, these developments not only provide solid theoretical foundations to the experimental discoveries in binary and ternary kagome metals but also suggest a rich and tunable platform for exploring novel quantum phases. In fact, understanding kagome metals means taking on the interacting electron problem in its entire complexity, treating electronic correlations and electron-phonon coupling on an equal footing. As such, kagome metals promise to play the role of an incubator and benchmark for understanding electronic quantum matter.

\subsection{\label{sec:ConcB}Future Questions}

Despite the rapid theoretical and experimental progress in understanding kagome metals, a wide array of open questions and challenges remain that demand further refinement and exploration.  
In terms of materials, much of the recent progress in these systems has leveraged the use of metallic systems with a significant number of electronic bands.  While this overcomes the challenge of introducing charge carriers previously encountered with chemical and electrostatic doping in insulating kagome systems, finding materials with simpler Fermi surfaces is of significant interest. Additionally, finding compounds with more ideal $s$-electron orbitals, better matching idealized models, is an open frontier. Many of the kagome systems to date are based on $d$-electrons with multiple orbital states present at the Fermi level. Often, the bandwidth of the kagome band structure is broader than the energetic spacing within the $d$-electron manifold. Remedying this, and finding systems with \textit{e.g.}, large crystal field splittings  and finding schemes where this would be possible with absent localization due to Coulomb interactions is an open challenge.  Finally, a broader range of effectively two-dimensional systems is a clear target.

Achieving such effectively two-dimensional kagome systems remains a crucial challenge for the field. Out-of-plane hopping strongly perturbs the ideal kagome band structure, reducing both the flat-band localization and the separation from non-kagome-derived states. To overcome this, several promising approaches could be pursued. The first involves van der Waals kagome metals, which are naturally prone to exfoliation and integration into atomically thin heterostructures. The second exploits thin-film growth and surface-engineering techniques, which enable the controlled realization of ultrathin kagome lattices or even artificial kagome layers. Both strategies open new opportunities for enhancing correlation effects, tuning topology, and engineering interactions in a manner not possible in bulk crystals.

Beyond these, while it is now established that both correlated electronic phases and topological electronic phases can be realized in kagome materials, to what degree these two phenomena can enrich one another to create fundamentally new electronic phases is an open question.
Additionally, while the kagome network is now recognized as the (chemically) most common line-graph compound, to what extent other theoretically well-known line-graph lattices can be realized experimentally (and with the potential richness of kagome metals) is an emerging area of investigation.

From the theoretical side, one of the most pressing challenges lies in developing accurate minimal models that can faithfully capture the essential band features responsible for the emergence of CDW order, SC, nematicity, and TRS breaking in AV$_3$Sb$_5$ compounds. Theories must address the complex roles played by VHS, SOC, and multi-orbital effects near the Fermi level, as well as the interplay between in-plane and out-of-plane Sb-$p$ orbitals, which remain poorly understood yet crucial to stabilizing correlated phases. The diversity of newly discovered AM$_3$X$_5$ materials, such as Ti-, Cr-, and V-based variants, further expands the parameter space of kagome systems, demonstrating the urgent need for theoretical tools that can adapt to different electronic fillings, orbital makeups, and structural distortions while remaining predictive and physically transparent.

Another frontier involves the theoretical treatment of strain and disorder. Experimental findings increasingly suggest a strong sensitivity of certain classes of kagome metals to mechanical strain, which appears to modulate TRS-breaking signatures and influence CDW domain formation. Theoretical models that incorporate the coupling of strain and structural deformation to electronic, magnetic, and orbital degrees of freedom are necessary to understand emergent piezomagnetic responses and chirality switching observed in surface-sensitive probes. Such models will be fundamental in understanding whether the exotic signatures, such as Kerr rotation or nematic anisotropy, arise intrinsically or are mediated by strain-induced symmetry breaking.

A closely related challenge is uncovering the origins and consequences of intertwined orders. Kagome metals often exhibit similar energy scales for competing or coexisting states such as SC, charge order, magnetism and nematicity. Capturing the competition and entanglement of these orders, particularly in materials near Van Hove filling or in the presence of sizable correlations, will require beyond-mean-field approaches such as renormalization group methods, in which different electron orders compete toward the final ground state instability. Furthermore, the coupling of topology with magnetism, SC, and correlation effects presents another rich area for theoretical advancement. Several kagome metals present bands with a nontrivial topology, namely Dirac cones, flat bands, and Weyl nodes, and their interplay with electronic correlations, electron-phonon interaction and magnetism is still an underexplored direction.

The exploration of unconventional SC in kagome metals is equally promising and challenging. In fact, the pairing mechanisms leading to SC in systems like AV$_3$Sb$_5$, whether conventionally mediated by phonons or involving unconventional pairing channels, remain unresolved despite intense research activity. Resolving this puzzle requires studies that integrate realistic band structures with dynamical interactions. The field must also address the modeling of kagome materials with partial flat-band filling, which opens the door to strongly correlated quantum phases such as Mott insulators, non-Fermi liquid metals, and fractional Chern insulators. Here, understanding how electronic interactions drive localization will be vital, especially in systems with minimal bandwidth and strong SOC.

Another important open question is to connect the conceptual viewpoints traditionally used for geometrically frustrated systems. Historically, frustrated lattices such as the kagome have been explored extensively within the context of Hubbard and spin models, often from the angle of strong correlations, Mott physics, and magnetic ground states. In contrast, much of the recent progress in kagome metals was pushed by Fermi liquid-based approaches that center on band structure instabilities, VHS, and weak-to-intermediate correlation effects. However, as evidence grows for strong-coupling phenomena, such as charge ordering, unconventional SC, and potential Mott-like behavior in certain regimes, it becomes important to investigate the link between these two pictures. This puts forward the necessity of theoretical frameworks that interpolate between weak-coupling Fermi liquid and non-perturbative many-body physics, particularly in systems with flat bands or partial band fillings. Developing a unified understanding that captures both itinerant and localized electron behavior in kagome systems will be fundamental to shed light onto the full range of correlated phases they can support.

Furthermore, an interesting question in kagome metals is the role of excitons and how they relate to charge and spin orders. In multiband systems with strong SOC, in particular when TRS is broken, the distinction between CDWs, SDWs, and exciton condensates becomes vague. Yet, most studies treat these orders separately. A more unified view that considers all these possibilities together could offer new insights into the nature of the symmetry-breaking phases in kagome materials.

Beyond their rich interplay of topology, correlation, and lattice frustration, kagome metals offer also a new and natural setting to explore the quantum geometry of Bloch states, captured by the quantum geometric tensor. Their flat bands, arising from compact localized states, are classified as singular flat bands \cite{rhim2021singular}. Under strong magnetic fields, these give rise to an anomalous Landau level spectrum that directly encodes the quantum metric \cite{rhim2020quantum}, with predicted consequences such as diverging orbital magnetic susceptibility and nonlinear Hall effects. Building on these insights, recent work has proposed and demonstrated a photoemission-based approach to directly measure the quantum geometric tensor in kagome systems~\cite{kang2025measurements}, establishing them as a promising class of materials to experimentally probe band geometry and its role in topological and transport phenomena.

Lastly, kagome metals potentially represent an interesting platform for new device applications that could range from spintronics to orbitronics and quantum information. Investigations that link bulk properties to nanoscale or interface phenomena, such as proximity-induced SC, spin-orbit torque effects and charge-to-spin-to-orbit interconversion, will be important for future experimental efforts that aim at constructing functional kagome-based heterostructures and quantum devices. In this context, the recent demonstration of a sizable modulation of the anomalous Hall angle in the magnetic Weyl semimetal Co$_3$Sn$_2$S$_2$ represents a significant advance. By achieving Hall angles far beyond those of conventional magnetic materials and realizing Fe-doped nanoflake devices with ultrahigh sensitivity and nanoscale magnetic field detectability, kagome metals are a highly promising platform for next-generation magnetic sensors and spintronic applications~\cite{yang2025modulation}.

In summary, while the study of kagome metals has already revealed a rich landscape of correlated and topological phenomena, the field still remains full of open questions that span materials discovery, theoretical modeling, and emergent functionalities. The search for new kagome systems continues, so too does the promise of discovering entirely new states of quantum matter, many of which may lie just beyond the current frontier.

\section*{List of Abbreviations}

\begin{longtable}{ll}
\caption{\parbox{.8\linewidth}{Alphabetical list of abbreviations.}}\\
\hline
\textbf{Abbreviation} & \textbf{Meaning} \\
\hline
\endfirsthead
\hline
\textbf{Abbreviation} & \textbf{Meaning} \\
\hline
\endhead
AFM  & Antiferromagnetism \\
AHE  & Anomalous Hall effect \\
ANE  & Anomalous Nernst effect \\
ARPES& Angle-resolved photoemission spectroscopy \\
BCS  & Bardeen–Cooper–Schrieffer \\
BZ   & Brillouin zone \\
CD   & Circular dichroism \\
CDW  & Charge density wave \\
CL   & Circular left \\
CR   & Circular right \\
DFT  & Density functional theory \\
DMFT & Dynamical mean field theory \\
DOS  & Density of states \\
EBR  & Elementary band representation \\
EDC  & Energy distribution curve \\
FFLO & Fulde–Ferrell–Larkin–Ovchinnikov \\
FRG  & Functional renormalisation group \\
MR   & Magnetoresistance \\
$\mu$SR & Muon spin spectroscopy \\
PDW  & Pair density wave \\
PES  & Photoemission spectroscopy \\
PHE  & Planar Hall effect \\
QPI  & Quasiparticle interference \\
RPA  & Random phase approximation \\
SBC  & Spin Berry curvature \\
SC   & Superconductivity \\
SDW  & Spin density wave \\
SOC  & Spin-orbit coupling \\
STM  & Scanning tunneling microscopy \\
TB   & Tight-binding \\
TMO  & Transition metal oxide \\
TMR  & Tunneling magnetoresistance \\
TRS  & Time reversal symmetry \\
VHS  & van Hove singularity \\
XRD  & X-ray diffraction \\
\hline
\end{longtable}

\section*{Acknowledgments}

The authors acknowledge Armando Consiglio, Matteo D\"{u}rrnagel and Niklas Wagner for helping with the preparation of the manuscript as well as Federico Mazzola, Mark H. Fischer, Binghai Yan, Maia Vergniory, Rafael Fernandes, Dante Kennes, Roser Valent\'i, Ryotaro Arita, Julian Ingham, Harley Scammell, Morten H. Christensen and Brian M. Andersen for discussions and critical comments.  
DDS acknowledges the Deutsche Forschungsgemeinschaft (DFG, German Research Foundation) through SFB 1170 ToCoTronics Mercator Fellowship. RT, GS, TN acknowledge support from the FOR 5249 (QUAST) by the DFG and the associated funding from the Swiss National Science Foundation (Project 200021E198011). 
RT, GS, DDS acknowledge support by the DFG Cluster of Excellence ct.qmat (Project-ID 390858490). 
RC, JGC and SDW were supported by the Air Force Office of Scientific Research under grant FA9550-22-1-0432. IZ acknowledges the support from the National Science Foundation (NSF), Division of Materials Research 2216080.


\bibliography{biblio_1_2_cleaned_subset_check_by_hand}

\end{document}